\RequirePackage{fix-cm}
\RequirePackage{amsmath}
\documentclass[smallextended]{svjour3}       
\smartqed  
\usepackage{graphicx}
\usepackage[utf8x]{inputenc}
\usepackage{amsmath}
\usepackage{latexsym}
\usepackage{amssymb}
\usepackage{braket}
\usepackage{xcolor}
\usepackage{hyperref}
\hypersetup{colorlinks,linkcolor=blue,linktoc=page,citecolor=blue,filecolor=blue,urlcolor=blue}
\usepackage{url}
\usepackage[numbers,sort&compress]{natbib}
\bibpunct{(}{)}{,}{a}{}{,}
\usepackage{threeparttable}
\usepackage{comment}
\usepackage{booktabs}
\usepackage{enumitem}

\newcommand{\xmm}{\textit{XMM-Newton}}
\newcommand{\erosita}{\textit{eROSITA}}
\newcommand{\athena}{\textit{ATHENA}}

\begin{document}

\title{Soft proton scattering at grazing incidence from X-ray mirrors: analysis of experimental data in the 
framework of the non-elastic approximation}



\titlerunning{A semi-empirical model for soft proton scattering}        

\author{Roberta Amato \and Teresa Mineo \and Antonino D'A\`i \and Sebastian Diebold \and Valentina Fioretti \and Alejandro Guzman \and Simone Lotti \and Claudio Macculi \and Silvano Molendi \and Emanuele Perinati \and Chris Tenzer \and Andrea Santangelo}


\institute{Roberta Amato \at
              Dipartimento di Fisica e Chimica - Emilio Segr\'e, Università degli Studi di Palermo, via Archirafi, 36, 90123 Palermo, Italy \\
              INAF-IASF Palermo, via Ugo La Malfa, 153, 90146 Palermo, Italy \\
              IAAT, University of T\"ubingen, Sand 1, 72076 Tübingen, Germany\\
              \email{roberta.amato@inaf.it}           
           \and Teresa Mineo \and  Antonino D'A\`i \at INAF-IASF Palermo, via Ugo La Malfa, 153, 90146 Palermo, Italy   
           \and Sebastian Diebold \and Alejandro Guzman \and Emanuele Perinati \and Chris Tenzer \and Andrea Santangelo \at IAAT, University of T\"ubingen, Sand 1, 72076 Tübingen, Germany
           \and Valentina Fioretti \at INAF-IASF Bologna, Via Piero Gobetti, 101, 40129 Bologna, Italy
           \and Simone Lotti \and Claudio Macculi \at INAF-IAPS, Via del Fosso del Cavaliere, 100, 00133 Roma, Italy
           \and Silvano Molendi \at INAF-IASF Milano, Via Alfonso Corti 12, 20133 Milano, Italy
           }

\date{Received: date / Accepted: date}

\maketitle

\begin{abstract}
Astronomical X-ray observatories with grazing incidence optics face the problem of pseudo-focusing of low energy protons from the mirrors towards the focal plane. Those protons constitute a  variable, unpredictable component  of  the  non  X-ray  background that strongly affects astronomical observations and a correct estimation of their flux at the focal plane is then essential. For this reason, we investigate how they are scattered from the mirror surfaces when impacting with grazing angles. We compare the non-elastic model of reflectivity of particles at grazing incidence proposed by \cite{Remizovich} with the few available experimental measurements of proton scattering from X-ray mirrors. We develop a semi-empirical analytical model based on the fit of those experimental data with the Remizovich solution. We conclude that the scattering probability weakly depends on the energy of the impinging protons and that the relative energy losses are necessary to correctly model the data. The model we propose assumes no dependence on the incident energy and can be implemented in particle transport simulation codes to generate, for instance, proton response matrices for specific X-ray missions. Further laboratory measurements at lower energies and on other mirror samples, such as \athena{} Silicon Pore Optics, will improve the resolution of the model and will allow us to build the proper proton response matrices for a wider sample of X-ray observatories.

\end{abstract}

\section{Introduction}
\label{section:introduction}

X-ray missions that carry on board grazing incidence telescopes and orbit outside the Earth's radiation belts, such as the \textit{Chandra X-ray Observatory} \citep{Weisskopf2000} and \xmm{} \citep{Jansen2001}, are subjected to the impact of different types of charged particles from the surrounding environment; among them,  ``soft protons" (SPs), i.e. protons with energies up to a few hundreds of keV, are of primary concern. SPs impacting on the mirrors of grazing incidence X-ray telescopes with low incidence angles are scattered and funneled towards the focal plane, where they reach the detectors, producing signals indistinguishable from the ones generated by photons and, in the worst cases, damaging them \citep{Struder2001,Turner2001}. Depending on the satellite orbit and solar activity, periods of intense particle background can last up to several hours \citep{XMMbkg-4}, thus 
affecting the performance and reliability of scientific observations and the overall duty cycle. For instance, the \xmm\ observing time is reduced by $\sim$30-40\%, due to the proton flares \citep{XMMbkg-3}.

Since SPs affect the scientific performance of the detectors and the overall sensitivity of X-ray missions, it is important to have a correct estimation of their flux at the focal plane. If the SPs flux is too high, further expedients become necessary, such as magnetic shielding, as well as an adequate scheduling of the observational plan. 

For a correct evaluation of the expected flux at the focal plane, efforts must be done on both the theoretical and experimental side. A few experimental measurements of scattering of low energy protons at grazing incidence from X-ray mirrors already exist. They have been done on mirror samples of \xmm{} by \cite{Rasmussen} and of \erosita{} \citep[\textit{extended ROentgen Survey with an Imaging Telescope Array}, hosted on board of the \textit{Spektr-RG} mission,][]{Predehl2016} by \cite{Diebold,SPIE2017}. Amongst the physical models, \cite{Remizovich} proposed an analytic formulation, under the assumption of non-elastic scattering, while the same phenomenon in elastic approximation was treated by \cite{Firsov1972} \citep[see also][]{mashkova1985}.

\cite{Fioretti} implemented the aforementioned model in its elastic approximation into \textit{Geant4} \citep[\textit{GEometry ANd Tracking},][]{Geant2003,Geant2006,Geant2016} and then compared it with the scattering measurements performed by \cite{Diebold}, together with the \textit{Geant4} Single Scattering model based on the Coulomb scattering cross section. The Remizovich elastic approximation resulted in a scattering probability about 4--5 times higher than the Single Scattering model at the peak of the distribution of the scattering efficiency as a function of the polar scattering angle. However, the lack of fine data coverage at small ($<1^\circ$) scattering angles in the 2015 data set did not allow for a validation of any of the proposed models.

In the present work, we use the formula of \cite{Remizovich} under non-elastic approximation to model all the currently available experimental data sets of scattering of protons on X-ray mirrors. We propose a semi-empirical model that can be implemented in any ray-tracing code or particle transport simulator for the optics of present and future X-ray missions, provided that experimental measurements on the respective type of optics are performed.

\section{The Remizovich physical model in non-elastic approximation}
\label{section:model}

\begin{figure}[ht]
\centering
\includegraphics[scale=0.25]{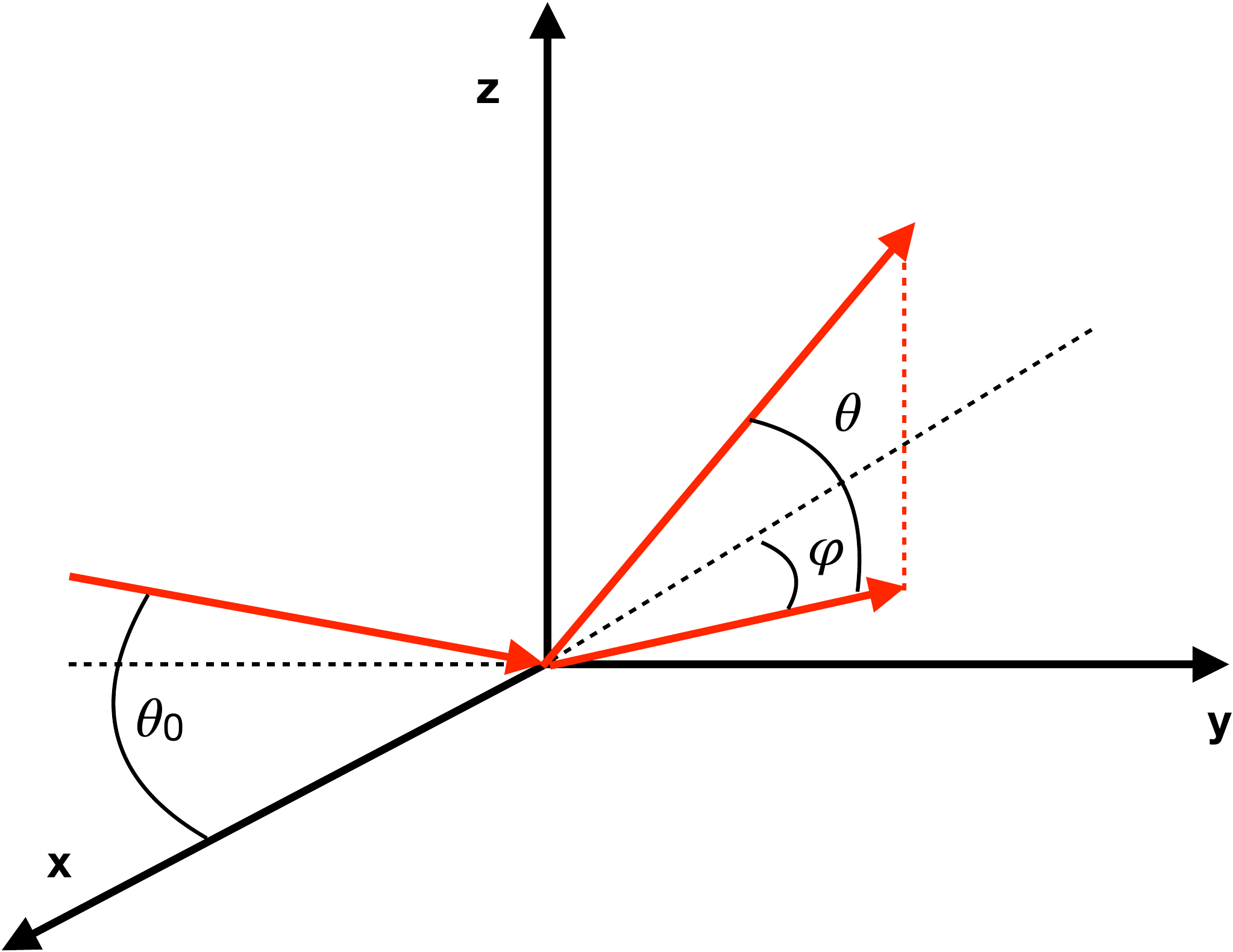}
\caption{Geometric scheme of the system: the incident beam hits the surface (in the $xy$ plane) with an angle $\theta_0$ and it is scattered with a polar angle $\theta$ and an azimuthal angle $\varphi$ \citep{Fioretti}.}
\label{schema}
\end{figure}

Following the schematisation of \cite{Remizovich}, let us suppose that a particle hits a reflecting surface with a grazing angle $\theta_0$ and it is  scattered with a polar angle $\theta$ and an azimuthal angle $\varphi$ (see the geometric scheme of the system in Fig.\,\ref{schema}). For the sake of convenience, we define the dimensionless polar and azimuthal angles $\psi$ and $\chi$ as:
\begin{equation}
\psi=\frac{\theta}{\theta_0} \quad \text{and}\quad \chi=\frac{\varphi}{\theta_0}.
\label{angles_definition}
\end{equation}
and the dimensionless energy of the scattered particle as:
\begin{equation}
u=\frac{T}{T_0},
\label{energy_definition}
\end{equation}
where $T_0$ and $T$ are its initial and final kinetic energy.

\cite{Remizovich} treated the interaction under the small-angle approximation, that assumes that the product of the mean-squared value of the scattering angle per unit path $\braket{\theta_s^2(T)}$ that the particle covers through consecutive collisions with the ions inside the medium and the whole range $R_0$, at the given incident energy $T_0$, is much smaller than one, i.e.
\begin{equation}
\braket{\theta_s^2(T)}R_0\ll1.
\end{equation}
Under the condition of small incidence angles (${\theta_0\ll1}$ rad), the thickness of the layer crossed by a single particle before emerging from the target is proportional to  $\theta_0^3/\braket{\theta_S^2}$. If the energy $T_0$ of the incident particles is small enough ($T_0\ll\nobreak1$ GeV for protons), the process of deceleration of particles in the medium can be modelled as a continuous energy loss (continuous slowing down approximation, CSDA). The process is not conservative, i.e. the incident particle loses part of its energy when interacting with the atomic lattice of the mirror. However, if the spectrum of the reflected particles has a sharp maximum close to the input energy $T_0$, it is possible to assume \citep{Firsov1972}:
\begin{equation}
    \braket{\theta_s^2(T)}\approx\braket{\theta_s^2(T_0)}=\text{const.}
\end{equation} 

\subsection{The differential scattering function}
Under all the assumptions stated above, the scattering probability computed as the ratio of the number of reflected particles in a given direction from a unit surface area per unit time to the number of incident particles on the same unit area per unit time expressed as function of the dimensionless variables $\psi$, $\chi$ and $u$ is \citep{Remizovich}:
\begin{equation}
\begin{split}
W(\psi, \chi, u)=&\frac{3^{1/2}}{2\pi^2}\frac{T_0\psi}{R_0\varepsilon(u)}\frac{\exp\{-[4(\psi^2-\psi+1)+\chi^2]/4\sigma s(u)\}}{\sigma^{3/2}[s(u)]^{5/2}}\\
&\times\text{Erf}\left(\left(\frac{3\psi}{\sigma s(u)}\right)^{1/2}\right)
\end{split}
\label{eq:non_el}
\end{equation}
where: $\varepsilon(u)=-\braket{du/dl}$ is the average energy loss per unit path, i.e the stopping power, which varies with the energy of the beam and with the chemical composition of the reflecting material; $R(T)=\int dT/\varepsilon(T)$ is the resulting average particle range, which is a function of the energy; $R_0$ is the range at the specific incident energy; $s(u)$ is defined as $s(u)=L(T)/R_0=1-R(T)/R_0$, being $L(T)=R_0-R(T)$ the path travelled by a particle with energy $T$;  $\sigma$ is a dimensionless parameter defined as:
\begin{equation}
    \sigma=\braket{\theta^2_s(T_0)}R_0/4\theta^2_0
    \label{eq:sigma}
\end{equation} 

The integration of Eq.\,\ref{eq:non_el} over the energy and angle coordinates gives the total scattering efficiency:
\begin{equation}
    \eta_{tot}=\int_E \int_\Omega W(\psi, \chi, u) d\psi d\chi du
    \label{eq:eff_integrale}
\end{equation}
so that $1-\eta_{tot}$ is the probability that the particle is not reflected\footnote{The scattering probability can be expressed also as a function of the energy alone \citep[see equation 41 of ][]{Remizovich}, when integrating over the solid scattering angle.}. 

\begin{figure}[ht]
\centering
\includegraphics[height=0.3\textheight]{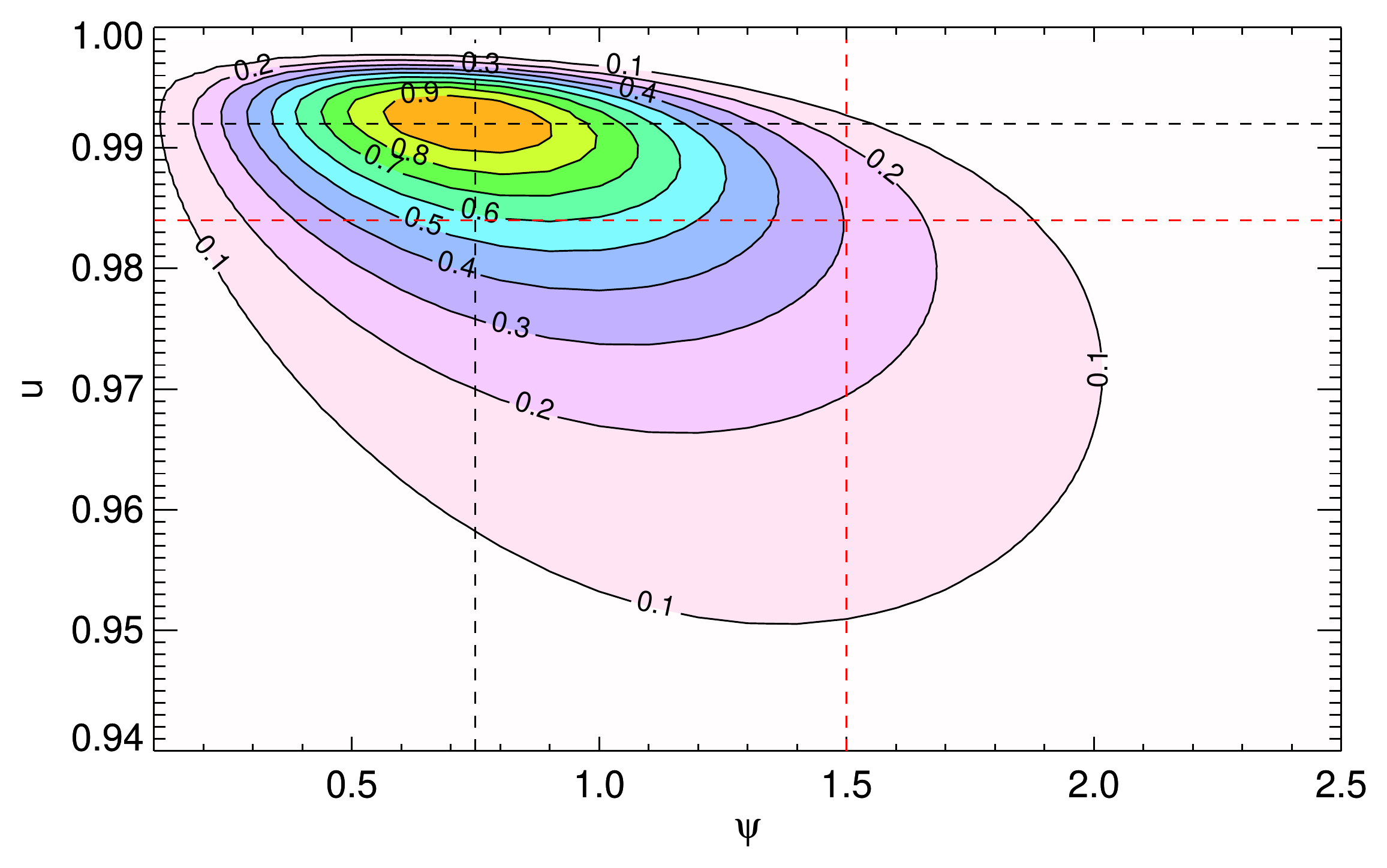} \quad
\caption{Contour plot of the scattering probability  $W(\psi,\chi,u)$ as a function of the polar scattering angle $\psi$ and of the energy $u$, for $\theta_0=0.36^\circ$, $T_0=250$ keV, $\chi$=0 and $\sigma=50$. The plot is normalised to the maximum of the distribution.}
\label{contour_u_psi_50}
\end{figure}

\begin{figure}[ht]
\centering
{\includegraphics[width=0.46\textwidth]{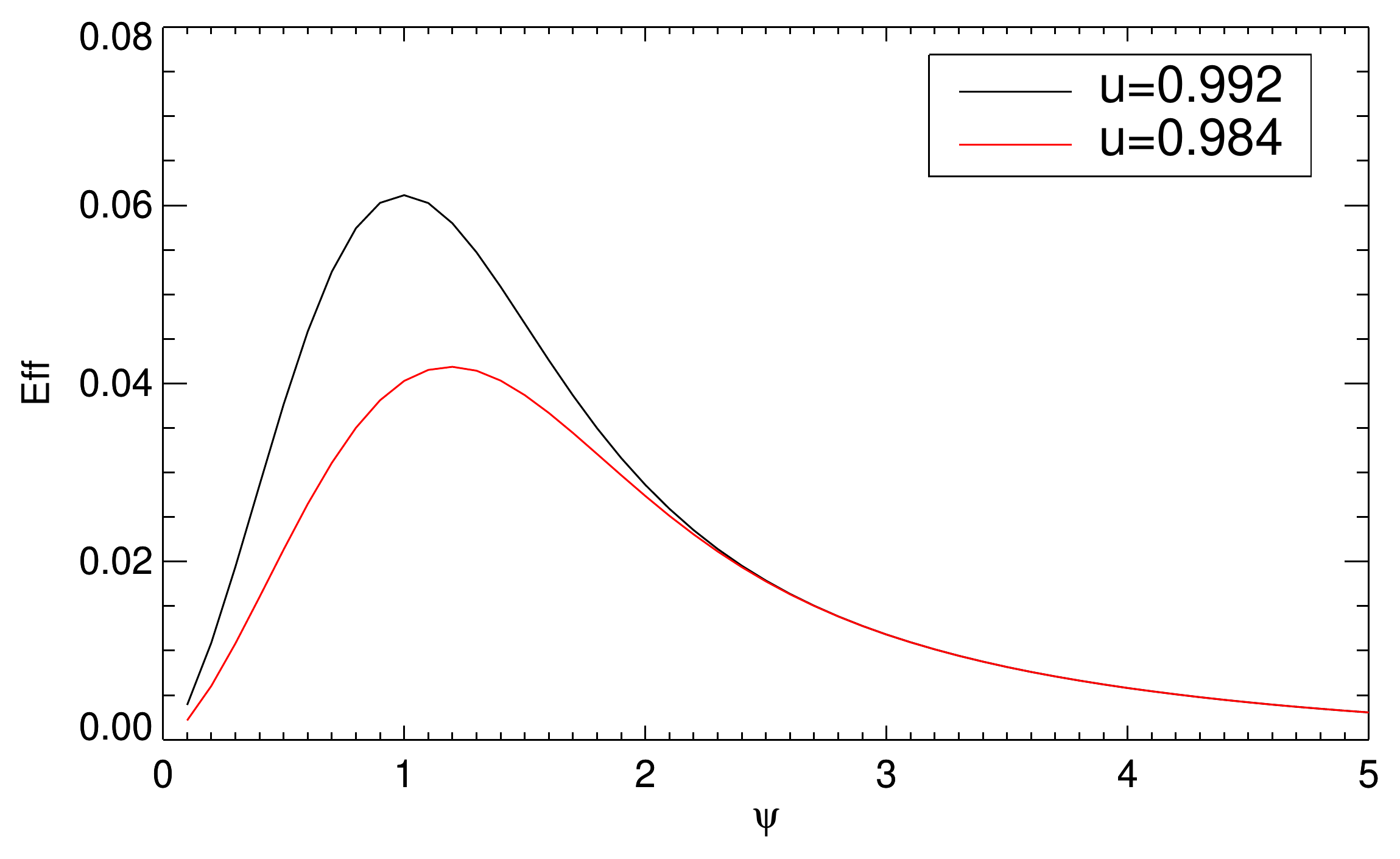}} \quad
{\includegraphics[width=0.48\textwidth]{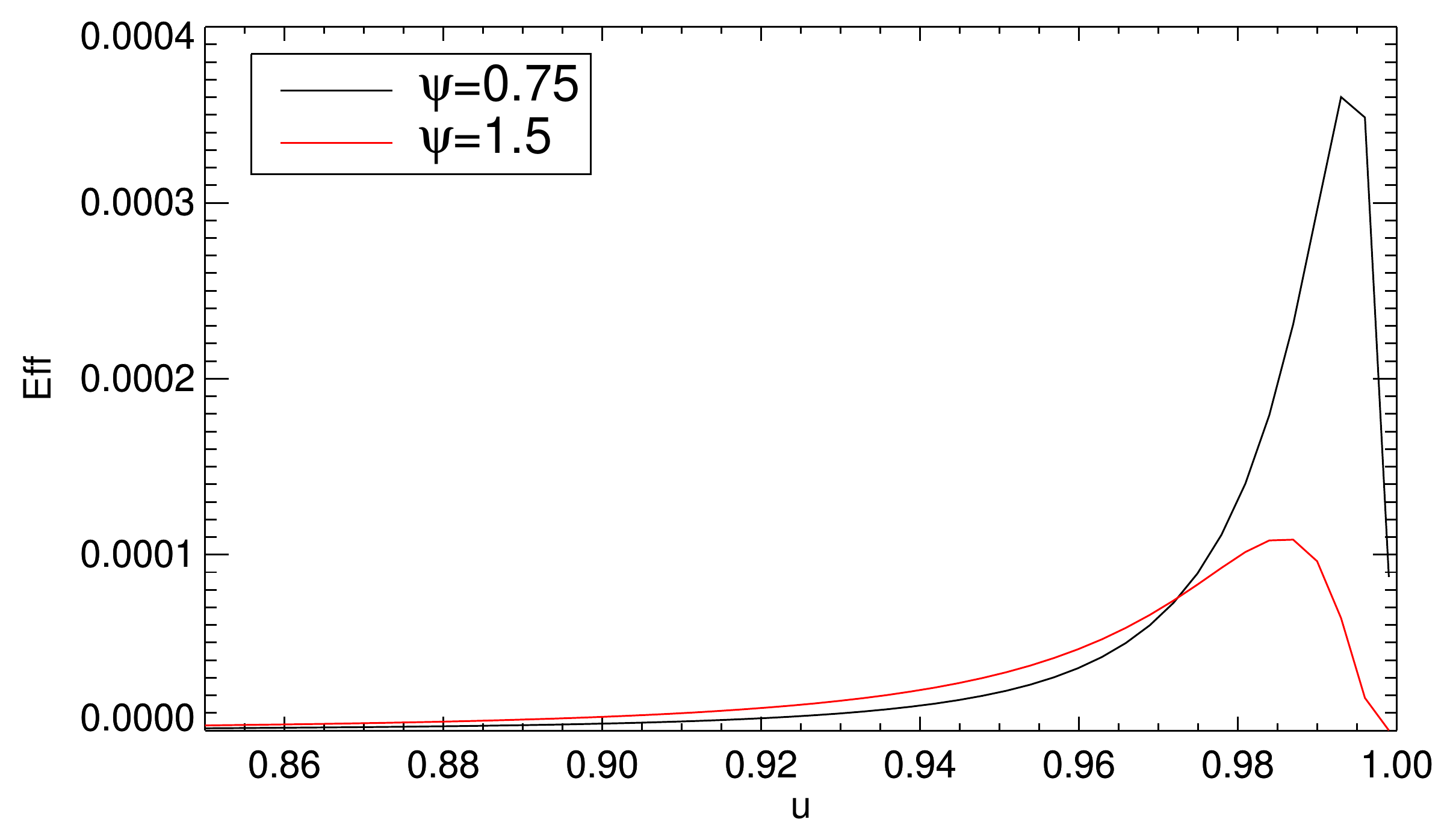}}
\caption{Scattering efficiencies along the red and black dashed lines shown in Fig.\,\ref{contour_u_psi_50}, for $\theta_0=0.36^\circ$, $T_0=250$ keV, $\chi$=0 and $\sigma=50$. The left panel shows the curve as a function of $\psi$ at $u$=0.992 (\textit{black line}) and $u$=0.984 (\textit{red line}); the right panel is relative to the efficiency distribution vs. $u$ at $\psi$=0.75 (\textit{black line}) and $\psi$=1.5 (\textit{red line}). Efficiency values are not normalized.}
\label{scattering_energy_50}
\end{figure}

The main characteristics of the scattering distribution can be summarised as follows:
\begin{itemize}
\item the maximum of the distribution in the plane $\chi=0$ peaks at $\psi\sim0.85$, while it peaks at $\psi\sim1$ when integrated over the azimuthal angle $\chi$ and the energy $u$;
\item the distribution is  symmetric with respect to the scattering azimuthal angle $\chi$, with its maximum at $\chi=0$;
\item smaller values of $\sigma$ produce lower and broader peaks of the distribution; 
\item the value of $\psi$ relative to the maximum of the distribution changes also with $\sigma$;
\item the scattering distribution depends on the final energy $u$, but the same scattering probability can be obtained with different values of $\sigma$ at different $u$;
\end{itemize}
Fig.\,\ref{contour_u_psi_50} shows an example of contour plot of the scattering function (Eq. \ref{eq:non_el}) for a target of Au, with $\theta_0=0.36^\circ$, $T_0$=250 keV and $\sigma$=50, at $\chi$=0, in the space $u$--$\psi$, normalised to its maximum, while Fig.\,\ref{scattering_energy_50} shows the 1-D distributions as a function of $\psi$ and of $u$ corresponding to the values highlighted in the contour plot with black and red dashed lines. 

Eq.\,\ref{eq:non_el} includes several  parameters (e.g. $\varepsilon(u)$, $R(T)$, etc.) that can be found in literature.  In our work, $\varepsilon(u)$ and of $R(T)$ were computed interpolating the values retrieved from the NIST PSTAR Database\footnote{\url{https://physics.nist.gov/PhysRefData/Star/Text/PSTAR.html}}. The Au density was set to 19.3 g/cm$^3$. 

\subsection{The parameter $\sigma$}
The parameter $\sigma$ in the Remizovich formula (Eq.\,\ref{eq:non_el}) determines the total number of particles reflected from the surface: the larger this value, the larger the number of reflected particles, and the narrower the peak of the distribution \citep{mashkova1985}. According to Eq.\,\ref{eq:sigma}, $\sigma$ can be computed knowing the mean-square scattering angle per unit path and the range, which depends on the scattering properties of the medium. Different approximations have been adopted to evaluate $\braket{\theta_s^2(T)}$, depending on the energy and on the angle of the incident particle. In the energy range of the experimental data used in this paper, it can be obtained with the following formula  \citep{Remizovich,Firsov1958}:
\begin{equation}
    \braket{\theta_s^2(T)}=2 \pi n_0 \frac{Z_1^2 Z_2^2 r_e^2}{T^2}L_k
    \label{eq:theta2}
\end{equation} 
where $n_0$ is the density of the atoms in the target, $Z_1$ and $Z_2$ are the nuclear charge of the incident particle and of the material of the target, respectively, $r_e$ is the classical electron radius, $T$ the particle energy in units of $mc^2$ and $L_k$ the Coulomb logarithm, which, in this specific case, can be approximated as:
\begin{equation}
    L_k=\ln\left(1+0.7\frac{T_{ev}}{30.5 eV}\frac{Z_1Z_2}{\sqrt{Z_1^{2/3}+Z_2^{2/3}}}\right)
    \label{eq:coulomb_log}
\end{equation} 
where $T_{ev}$ is the energy of the incident charge in unit of electronvolt.

Eq.\,\ref{eq:theta2} is a good approximation of values derived from a theoretical computation based on the assumptions that the inelastic process occurring during the collision can be obtained using the potential for elastic interactions and that the energy of the incident particle is significantly greater than the ionization potential of the atoms \citep{Firsov1958}.

\section{Experimental data sets}
\label{section:data}

In our analysis, we used all the data sets involving the scattering of protons at grazing incidence on X-ray mirrors available so far. In all the cases, samples were made of nickel and coated with gold, with a coating thickness $>$50 nm for \erosita{} \citep{Friedrich2008} and 0.2 $\mu$m for \xmm\ \citep{Stockman2001}.
Incidence angles and energies for each data set are listed in Tab.\,\ref{table_erosita}.

The first measurements were performed by \cite{Rasmussen} on \xmm\ optics, at the Harvard University, Cambridge Accelerator for Materials Science. The facility included a tandem Van de Graaff accelerator, which produced a monoenergetic proton beam with energy tunable from 0.1 to 3 MeV. The beam divergence was reduced to 3 arcmin level, with consecutive collimating apertures. The mirror sample was mounted on a holder, so that the plane of the sample exactly bisected the beam. 
The position of the detector was fixed at three different scattering angles (0.75$^{\circ}$, 1.40$^{\circ}$ and 2.38$^{\circ}$), while the incidence angles varied between 0$^{\circ}$ and 1.75$^{\circ}$ in steps of 0.25$^{\circ}$. The proton beam had the following energies: 300 keV, 500 keV and 1.3 MeV (see Tab.\,\ref{table_erosita}). For each configuration, the scattering efficiencies and the output spectra are reported. However, the authors published only uncalibrated spectra from which no useful information on the energy loss could be extracted. In our analysis, we use only data that are not affected by the occlusion of the mirror face by the mirror bulk, which caused a drop in the scattering efficiency. Errors on the scattering distribution are derived from the uncertainties on the beam flux and correspond approximately to 30\% of the scattering efficiency.

More recent data were obtained by \citet{Diebold,SPIE2017}, using a piece of a spare mirror shell of the \erosita\ telescope, at the ion accelerator facility at the University of T\"ubingen, a 3 MV single-ended Van de Graaff accelerator, working in the energy range 400 keV--2.5 MeV. The beam line consisted of a pair of entrance slits, a pinhole aperture of 0.1--1 mm diameter, a $\sim$80 cm long collimator, with apertures of 1.0 mm at the entrance and of 0.3 mm at the exit, which limited the maximum opening angle to 0.1$^\circ$. To achieve low proton energies, a metal degrader foil was put after the pinhole aperture. It widened the beam and reduced the energy down to 250 keV, 500 keV and 1 MeV in the first campaign \citep{Diebold} and to 300 keV in the second one \citep{SPIE2017}. 
The mirror target was located on a shiftable plane. The detector, a silicon surface barrier with a low energy threshold of 100 keV and an energy resolution of 10--20 keV, was mounted at a distance of $\sim$1 m along the beam line, shiftable to a maximum distance of 75 mm, corresponding to a maximum angle $\theta$ of about 4.5$^\circ$. The beam reached the detector through a 1.2 mm aperture, corresponding to a solid angle of $\sim$1.3 $\mu$sr. 
Furthermore, only the data from \cite{Diebold} reported explicitly both the scattering efficiency and the energy loss measurements. 

\begin{table}[ht]
\centering
\caption{incidence angles for each incident energy for the \xmm\ and \erosita{} mirror targets used in this work. }
\begin{threeparttable}
\begin{tabular}{lll}
\hline\hline\noalign{\smallskip}
 $E$ (keV)& \multicolumn{1}{c}{$\theta_0$ (deg) }  & Reference \\
\noalign{\smallskip}\hline\noalign{\smallskip}
250  & 0.36, 0.51, 0.67, 0.89, 1.06, 1.23 & \cite{Diebold}\tnote{1} \\
300 & 0.50, 0.75, 1.00, 1.25, 1.50, 1.75 & \cite{Rasmussen}\\
& 0.50, 0.64, 0.81& \cite{SPIE2017}\tnote{2} \\
500 & 0.50, 1.00 & \cite{Rasmussen}\\
& 0.33, 0.48, 0.64, 0.85, 1.02, 1.19& \cite{Diebold}\tnote{1} \\
1000 & 0.30, 0.46, 0.61, 0.83, 1.00, 1.17& \cite{Diebold}\tnote{1} \\
1300 & 0.25, 0.50, 0.75, 1.00, 1.25, 1.50, 1.75 & \cite{Rasmussen}\\
\noalign{\smallskip}\hline\hline
\end{tabular}
\begin{tablenotes}
        \footnotesize
        \item[1] Dataset with energy losses explicitly reported.
        \item[2] Dataset with off-axis measurements at azimuthal angles of about $\pm2^\circ$.
    \end{tablenotes}
\end{threeparttable}
\label{table_erosita}
\end{table}

\begin{figure}[ht]
\centering
\includegraphics[width=.85\textwidth]{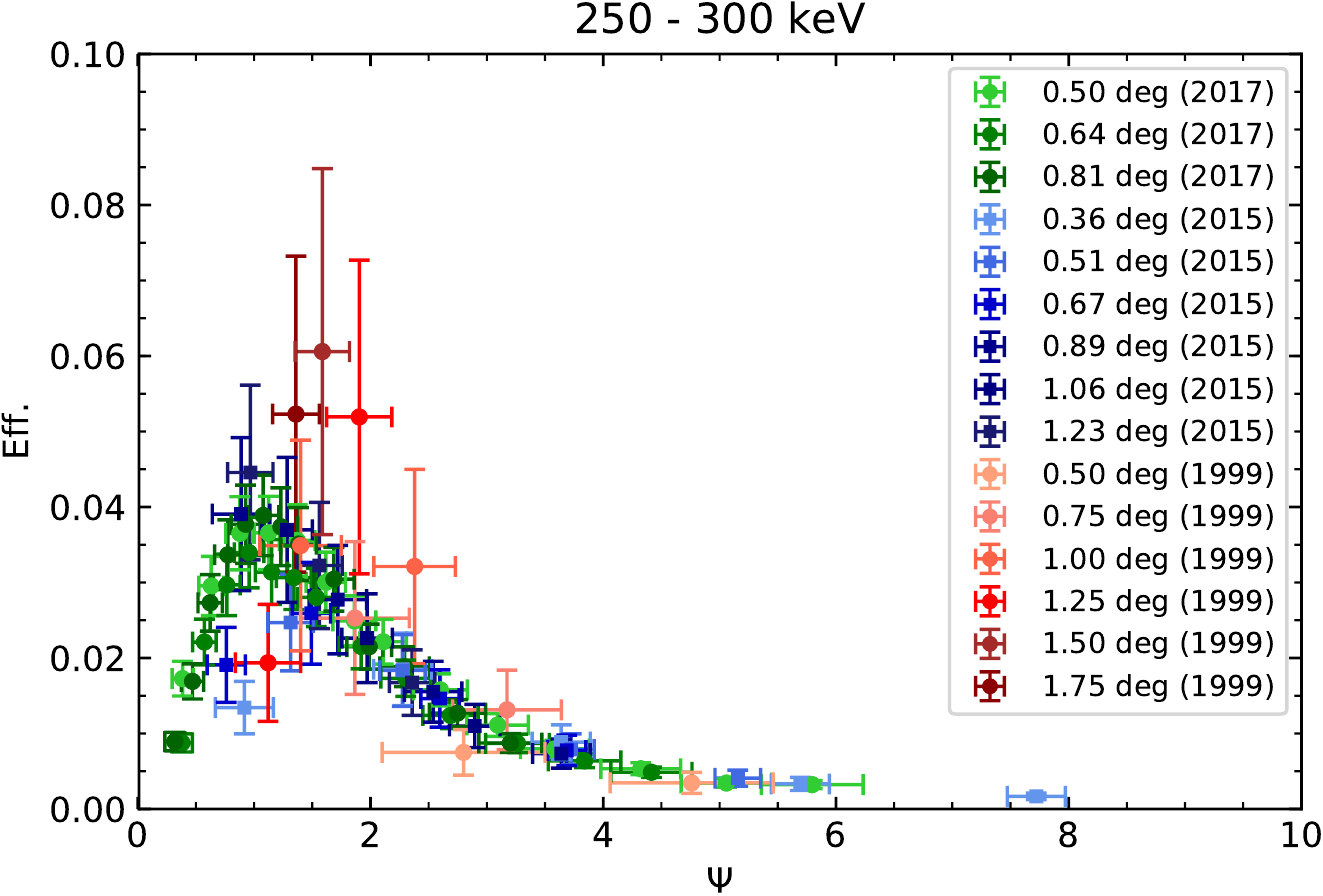} \quad
\includegraphics[width=.85\textwidth]{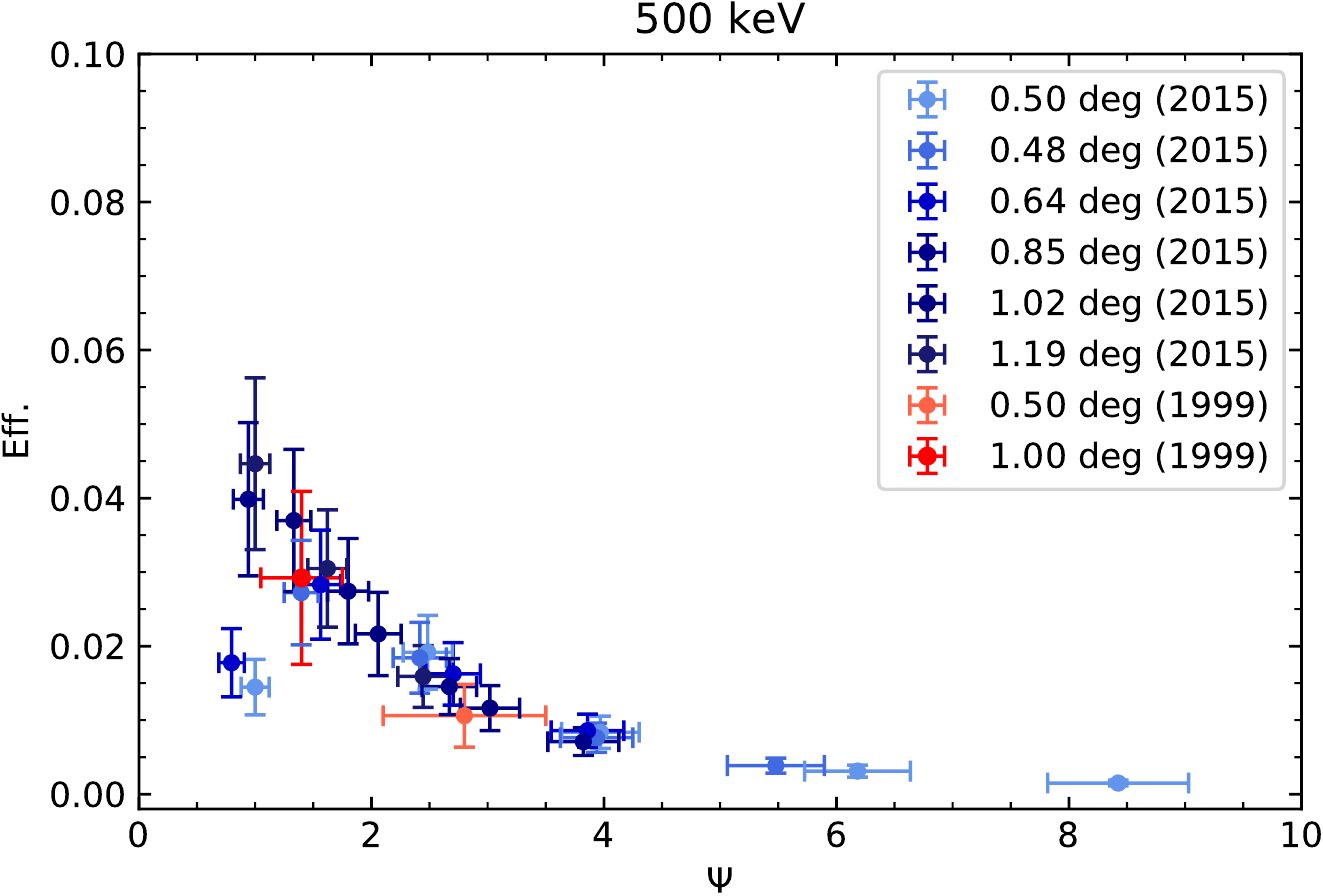}
\caption{Scattering efficiencies as a function of the scattering angle $\psi=\theta/\theta_0$, for two representative energies of the incident proton beam: 250-300 keV (\textit{upper panel}) and 500 keV (\textit{lower panel}). The blue and green dots stands for measurements on \erosita\ optics \citep{Diebold,SPIE2017}, the red ones for \xmm\ optics \citep{Rasmussen}. Incidence angles are shown in the legends; errors on \xmm{} scattering angles are at the nominal value of 21 arcmin.} 
\label{fig:all_data}
\end{figure}

If we express all the experimental data in the normalised coordinate space of Eq.\,\ref{angles_definition}-\ref{energy_definition} (i.e. $\psi=\theta/\theta_0$, $\chi=\phi/\theta_0$, $u=T/T_0$) and coherently normalise the scattering efficiency as: 
\begin{equation}
    \eta=\eta_{exp}\,\theta_0^2
\end{equation}
where $\eta_{exp}$ is the measured efficiency (in units of sr$^{-1}$), we can make a direct comparison of all the data. Fig.\,\ref{fig:all_data} shows two representative examples, for the incident energies of 250--300 keV and 500 keV. All data points from \erosita{} optics are well in agreement at large scattering angles ($\psi>$1.5), while a modest spread in data relative to the first campaign \citep{Diebold} is observed at angles close to the incident one ($\psi \, \simeq$1). This spread is not present in \cite{SPIE2017} data. \xmm\ measurements seem not to follow the same  trend of \erosita{} data (Fig.\ref{fig:all_data}, lower panel): the peaks appear to be shifted towards higher scattering angles and the efficiencies are slightly higher and more spread-out. Moreover, the low number of available data points (e.g. only two data points are available for the incident energy of 500 keV) prevents us to state more on the comparison.

\section{Analysis of the experimental data with the Remizovich model}
\label{section:fitting}

The analytic expression of Eq.\,\ref{eq:non_el} depends
on the parameter
$\sigma$ (Eq.\,\ref{eq:sigma}, with $\braket{\theta^2_s(T_0)}$  given by Eq.\,\ref{eq:theta2}-\ref{eq:coulomb_log}). However, after calculating the value of 
this parameter, the theoretical curves
never led to consistent results with the experimental data, as the
theoretical scattering functions were much higher and the energy losses much lower than the experimental points. Assuming the target surface made of nickel instead of gold (nickel being the material of the substrate of the optics of both \xmm{} and \erosita{})
also did not significantly change the mismatch. 
Hence, we decided to adopt a semi-empirical approach and to determine the parameter $\sigma$ directly from the data. We fit the data with the Remizovich formula given in Eq.\,\ref{eq:eff_integrale}, treating $\sigma$ as a free parameter of the fit.  
Since the total scattering efficiency is a function of the scattering angle and of the energy at the same time, we could use only the data sets that included both these variables (\cite{Diebold}). It must be stressed that the model we propose is an empirical best fit model based on the Remizovich solution and hence it depends on the accuracy of the experimental data. 

\subsection{Empirical evaluation of $\sigma$}

We performed a simultaneous fit of the scattering efficiency and  energy loss data points with the Remizovich formula, leaving $\sigma$ as the only free parameter. 

The fit model was computed taking into account the experimental set-up. More in detail, the scattering efficiencies were obtained by the integration of the scattering function (Eq.\,\ref{eq:non_el}) over the solid angle subtended by the detector ($\sim$1.3$\mu$sr) and over the energy interval between the energy of the incoming proton beam and 100 keV, this being the nominal low energy threshold of the detector. Because the energy of the protons from the laboratory beam is not perfectly monochromatic, but has a Gaussian profile around a nominal value, we considered several input energies with a Gaussian distribution whose center and sigma are given in \cite{Diebold}. The scattered spectra relative to each input energy were then added in a single output spectrum. The energy losses were obtained
using the same method as \cite{Diebold}, by fitting with a Gaussian the output spectrum and taking the differences with respect to the input one.

The goodness of the fit was established using a least-squares minimization without taking into account uncertainties, because points at large scattering angles, which have smaller errors, would have strongly biased the fit, while we are mainly interested in modelling the data around the peak (see Sect.~\ref{section:discussion}).

We define a total RMS as the sum of the RMS of the scattering efficiencies (RMS$_S$) and of the energy losses (RMS$_E$), normalised to the total efficiency and to the incident energy, respectively:
\begin{equation}
\text{RMS}=\frac{\text{RMS$_S$}}{\eta_{tot}}+\frac{\text{RMS$_E$}}{T_0}=\frac{\sqrt{\sum_{i=1}^n (S_i-\eta_i)^2}}{\eta_{tot}}+\frac{\sqrt{\sum_{i=1}^n(E_i-\epsilon_i)^2}}{T_0}
\end{equation}
where $S_i$ is the measured scattering efficiency for each $i$-th scattering angle, $\eta_i$ is the corresponding efficiency given by the model, $\eta_{tot}$ is the total scattering efficiency (Eq.~\ref{eq:eff_integrale}), $E_i$ is the experimental energy loss, $\epsilon_i$ is the energy loss given by the model and $T_0$ is the energy of the incident beam. To compute the errors on the parameter $\sigma$, we produced 1000 Monte Carlo simulations of the scattering and energy loss distributions per each data set, sorting the values from Gaussian distributions whose means and widths were equal to the data and their relative errors, respectively. We fit every simulated data set with Eq.~\ref{eq:eff_integrale}, assuming errors on $\sigma$ at the 90\% confidence interval. 

\begin{figure}[ht]
\centering
\includegraphics[width=.85\textwidth]{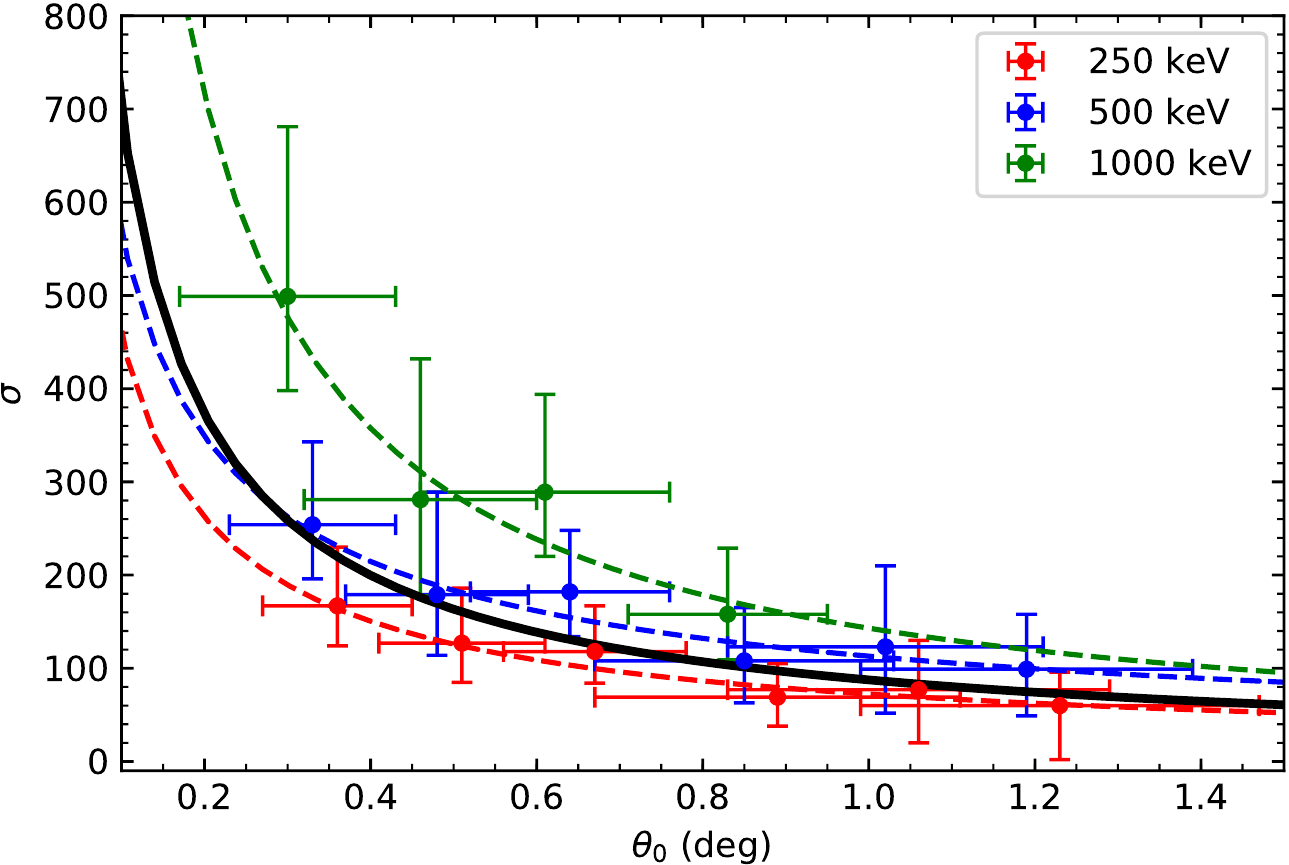}
\caption{Values of $\sigma$ derived from the fit as a function of the incidence angle $\theta_0$, for 250 keV (\emph{red}), 500 keV (\emph{blue}) and 1 MeV (\emph{green}), error bars on the values of $\sigma$ at 95\%. The dashed lines represent the best fit curve for each energy, while the solid \emph{black} line stands for the best fit curve of all the values of $\sigma$.}
\label{fig:sigma}
\end{figure}

The best fit values of $\sigma$ as a function of input angle and energy are reported in Tab.\,\ref{table_sigma}, together with the RMSs, and shown in Fig.\,\ref{fig:sigma}.

\begin{figure}[ht]
\centering
\includegraphics[width=.48\textwidth]{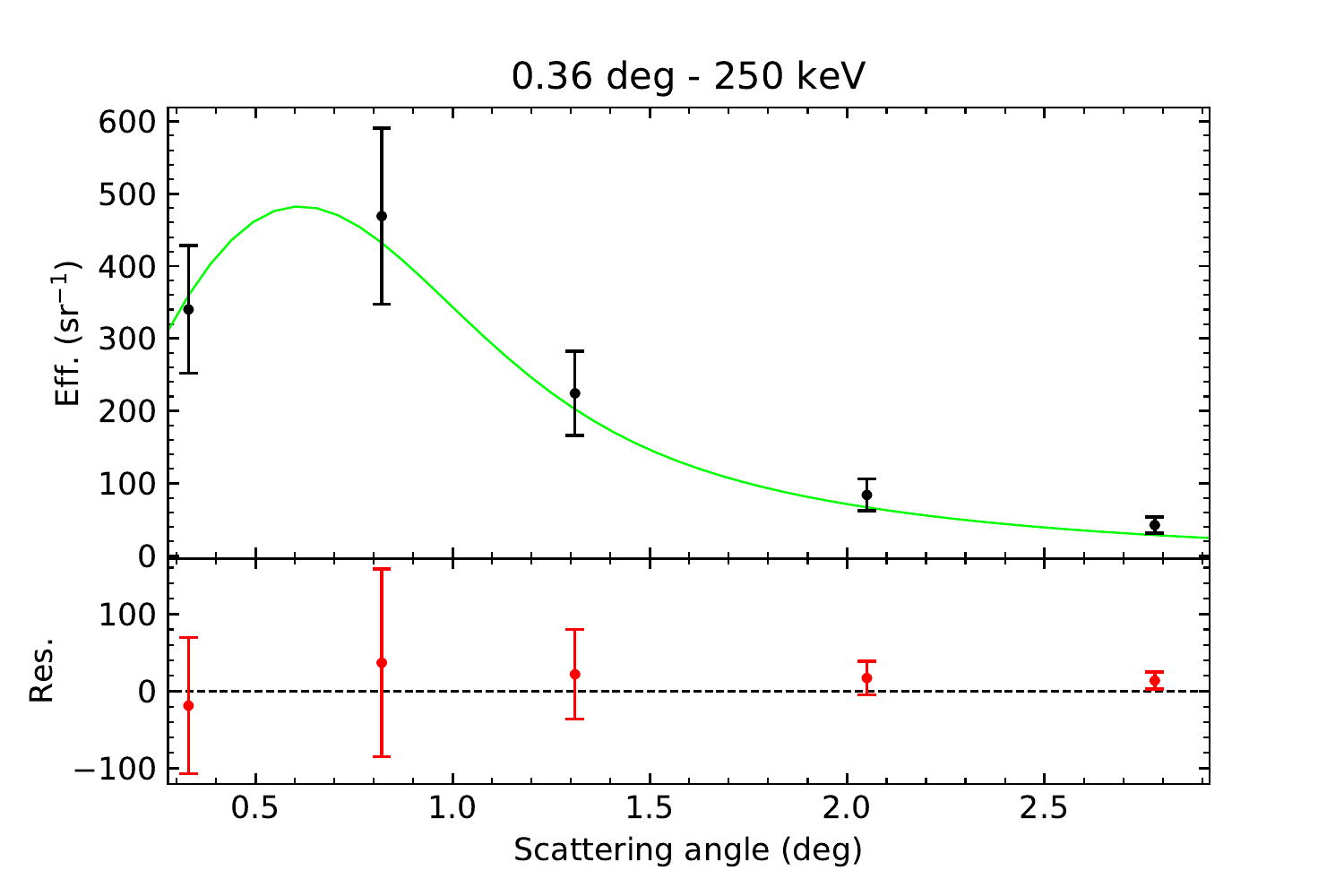}
\includegraphics[width=.48\textwidth]{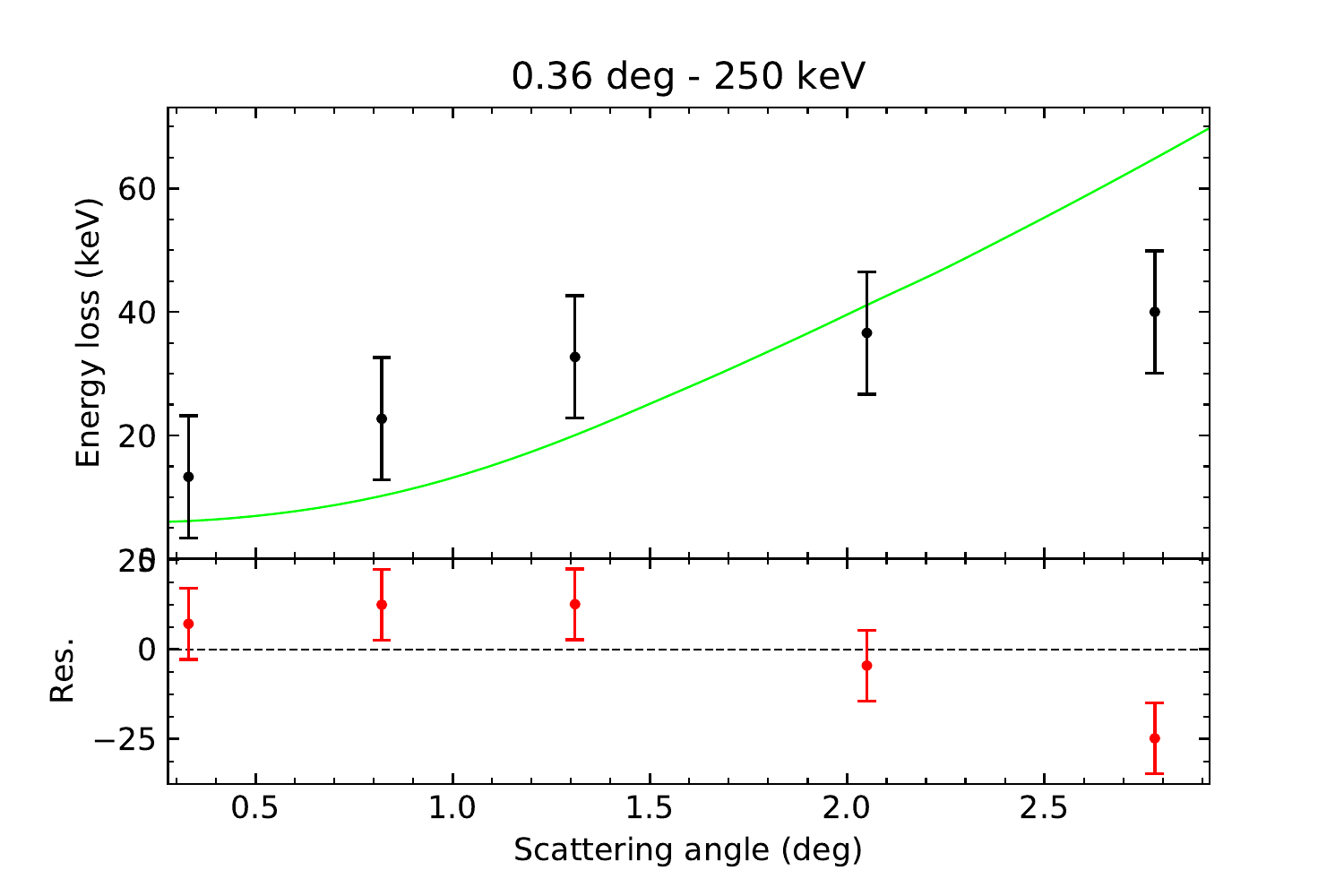}
\caption{Scattering efficiency (\emph{left panel}) and energy loss (\emph{right panel}) distribution as a function of the scattering angle, for the incidence angle of 0.36$^{\circ}$ and the incident energy at 250 keV, fit with the Remizovich model in non-elastic approximation. \emph{Bottom panels}: residuals of the fit.}
\label{fig:scat_250kev_0p36}
\end{figure}

The model is always in good agreement with the experimental scattering efficiencies, but it is not with the energy losses, which show a small consistency only for the lowest incident energy (250 keV). Fig.\,\ref{fig:scat_250kev_0p36} shows one representative example, at 250 keV, for an incidence angle of 0.36$^\circ$, while the whole sample can be view in Appendix \ref{sec:appendixA}. 

\begin{table}[ht]
\centering
\caption{Best fit values of the parameter $\sigma$ and corresponding values of RMS of the scattering (RMS$_S$) and of the energy loss (RMS$_E$), with the number of data points ($n$).}
\begin{tabular}{llccc}
\hline\hline\noalign{\smallskip}
 & $\theta_0$ ($^{\circ}$) & $\sigma$ & RMS$_S$($n$) & RMS$_E$($n$)\\
\noalign{\smallskip}\hline\noalign{\smallskip}
250 keV & 0.36 & $167^{+63}_{-43}$ & 23(5) & 14(5) \\
& 0.51 & $127^{+59}_{-42}$ & 11(4) & 14(4) \\
& 0.67 & $118^{+49}_{-34}$ & 7(4) & 17(4) \\
& 0.89 & $69^{+36}_{-31}$ & 16(4) & 24(4) \\
& 1.06 & $77^{+53}_{-57}$ & 10(3) & 26(3) \\
& 1.23 & $60^{+36}_{-58}$ & 12(3) & 28(3) \\
\noalign{\smallskip}\hline\noalign{\smallskip}
500 keV & 0.33 & $254^{+89}_{-58}$ & 52(5) & 18(5) \\
& 0.48 & $179^{+110}_{-65}$ & 10(4) & 17(4) \\
& 0.64 & $182^{+66}_{-48}$ & 12(4) & 21(4) \\
& 0.85 & $108^{+57}_{-45}$ & 15(4) & 19(4) \\
& 1.02 & $123^{+87}_{-71}$ & 10(3) & 22(3) \\
& 1.19 & $99^{+59}_{-50}$ & 13(3) & 23(3) \\
\noalign{\smallskip}\hline\noalign{\smallskip}
1 MeV & 0.30 & $499^{+182}_{-101}$ & 71(4) & 19(4) \\
& 0.46 & $281^{+151}_{-103}$ & 18(4) & 20(4) \\
& 0.61 & $289^{+105}_{-69}$ & 14(4) & 25(4) \\
& 0.83 & $158^{+71}_{-49}$ & 7(3) & 25(3) \\
\noalign{\smallskip}\hline\hline
\end{tabular}
\label{table_sigma}
\end{table}

The values of $\sigma$ show a clear trend with respect to the incidence angle $\theta_0$ (Fig.\,\ref{fig:sigma}), that we tried to describe analytically using a power law $\sigma \propto A \theta_0^{-\alpha}$.  Results of the fits are reported in Tab.\,\ref{table_gamma}. We also noted that, even if a systematic trend with the energy is visible in the results, the $\sigma$ relative to the same incidence angle are generally consistent with each other (apart from a few points), as well as the best fit parameters shown in Tab.\,\ref{table_gamma}.
We then fit the $\sigma$ all together with the same power law, obtaining an index value of $\alpha=-(0.9\pm0.3)$. This is different from what is stated by \cite{Remizovich}, for which $\sigma\propto\theta_0^{-2}$ (see Eq.\,\ref{eq:sigma}).

\begin{table}[ht]
\centering
\caption{Best fit values of the $\sigma$ parameters, fit with a power law of the type $f(x)=Ax^{-\alpha}$, and $\chi^2$ values at 2.7$\sigma$ level.}
\begin{tabular}{lccc}
\hline\hline\noalign{\smallskip}
$T_0$ (keV) & $A$ & $\alpha$ & $\chi^2$(d.o.f.)\\
\noalign{\smallskip}\hline\noalign{\smallskip}
250 keV & 73$\pm$34 & 0.8$\pm$0.4 & 0.3(5) \\
500 keV & 113$\pm$55 & 0.7$\pm$0.4 & 0.4(5) \\
1 MeV & 143$\pm$60 & 1.0$\pm$0.5 & 0.4(3)\\
\noalign{\smallskip}\hline\noalign{\smallskip}
All & 88$\pm$28 & 0.9$\pm$0.3 & 9(15)\\
\noalign{\smallskip}\hline\hline
\end{tabular}
\label{table_gamma}
\end{table}

\section{Discussion}
\label{section:discussion}

In this work we presented a comprehensive application of the model proposed by \cite{Remizovich} to describe the scattering of low energy protons at grazing incidence from the optics of astronomical X-ray observatories. We used all the  experimental data available so far \citep{Rasmussen,Diebold,SPIE2017} to verify the limitations
of this model in predicting the proton scattering distributions. 
The model under examination is based on the non-elastic approximation and expresses the scattering efficiency as a function of the angular distribution and of the energy loss of
the incident particles. 
The complex micro-physics of the interaction between
 the incident particle and the target lattice is condensed
into one parameter, $\sigma$, which depends on the material density and incidence angle. In other words, this parameter tips the scale of the scattering: the higher its value, the larger is the fraction of reflected particles and the narrower is their energy spectrum.
As stated by \cite{Remizovich}, the parameter $\sigma$ is proportional to the ratio between the mean-squared value of the scattering angle over the whole path to the squared incidence angle (${\sigma=\braket{\theta_s^2(T_0)}R_0/4\theta_0^2}$, Eq.~\ref{eq:sigma}). \cite{Remizovich} reported also an analytic expression to compute the value of the parameter $\braket{\theta_s^2(T_0)}$, which depends upon well known quantities (see Eq.\,\ref{eq:theta2}-\ref{eq:coulomb_log}). However, if we compute the parameter $\sigma$ in this way and successively use it to estimate the scattering efficiency, we obtain values almost two orders of magnitudes higher than those derived from the experimental data. There are also alternative derivations of the mean-squared scattering angle $\braket{\theta_s^2(T_0)}$, based on  different initial assumptions \citep[see][]{mashkova1985}. We also tested them, with no convincing results.

Hence, we chose to determine the value of the parameter $\sigma$ directly by fitting the data. The resulting values of $\sigma$ are shown in Tab.\,\ref{table_sigma}, where we indicated separately the RMS of the scattering efficiency (RMS$_S$) and of the energy loss (RMS$_E$) distributions, and plotted as a function of the incidence angle in Fig.~\ref{fig:sigma}. According to Eq.\,\ref{eq:sigma}, $\sigma$ should be proportional to $\theta_0^{-2}$, but the best fit model of all the obtained $\sigma$ resulted instead in a power-law index of 0.9\,$\pm$\,0.3 (error at 2.7$\sigma$), therefore more consistent with a $\sigma\propto\theta_0^{-1}$ law. Since this index is not in agreement with what is stated in literature, we argue that some of the initial assumptions in treating this problem  analytically might not fully hold, though 
we cannot still claim a complete rule-out of the model as more data are necessary to significantly diminish the uncertainty on this parameter.

The gold coating of the \erosita\ mirrors is tens of nm thick \citep{Merloni2012}. For the energies under consideration, the mean penetration length of protons is of the order of $\sim10^1$--$10^{-2}$ nm, depending on the energy of the incident beam. It is possible, then, that some of the incident protons pass through the gold layer and are scattered by the underlying nickel lattice. This led us to repeat the calculations by substituting density, range, stopping power and atomic number of gold with the ones specific for nickel. Nevertheless, the range of the values of $\sigma$ found for the nickel, between 500 and 40, are perfectly consistent with the ones found using gold and no significant improvements in the fits were obtained. Our conclusion is either the model is weakly dependent on the choice between the two metals or there is a more complex cumulative effect due the presence of the double layer. We also considered a potential deposit of water on the reflecting surface. It may happen that water molecules are trapped within the superficial layers of the lattice, altering the scattering properties of the medium. So, we computed the expected $\sigma$ for the water and found much smaller values than the best fit ones. Clearly the presence of water cannot be entirely excluded, but, in any case, the comprehensive analysis of multiple layers or materials is far beyond the goals of this work.

We also attempted to fit separately the scattering efficiencies and the energy loss distributions, but the two sets of fits returned different values of $\sigma$, not always consistent with each other. Moreover,
the $\sigma$ obtained from the scattering efficiency were systematically lower and flatter than those in Tab.\,\ref{table_sigma}, when plotted as a function of the incidence angle, while those from the energy were systematically higher and steeper. 
Therefore, we conclude that the two distribution should be fit simultaneously. 

Overall, the fit is mainly driven by the scattering efficiencies, while the energy loss distributions seem to contribute very weakly. The angular scattering distributions appear always well modelled by the Remizovich function and have lower RMS values in most of the cases.

\subsection{Comprehensive analysis of all the data sets}

To fully test the validity of the model, we applied it to the other data sets \citep{SPIE2017,Rasmussen} that could not be fit due to their lack of any energy loss information, in two different ways. 

\begin{figure}[ht]
\centering
{\includegraphics[width=0.48\textwidth]{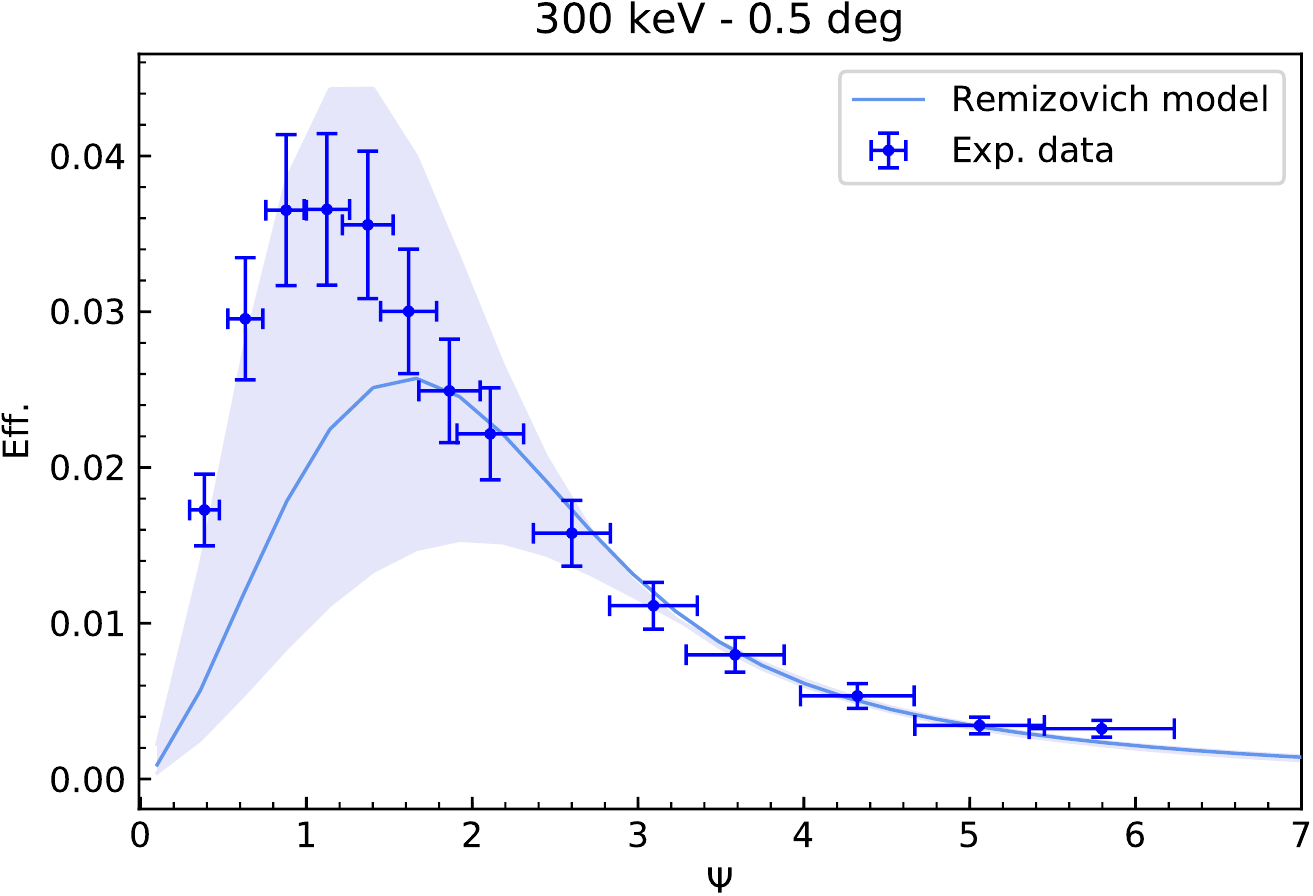}} \quad
{\includegraphics[width=0.48\textwidth]{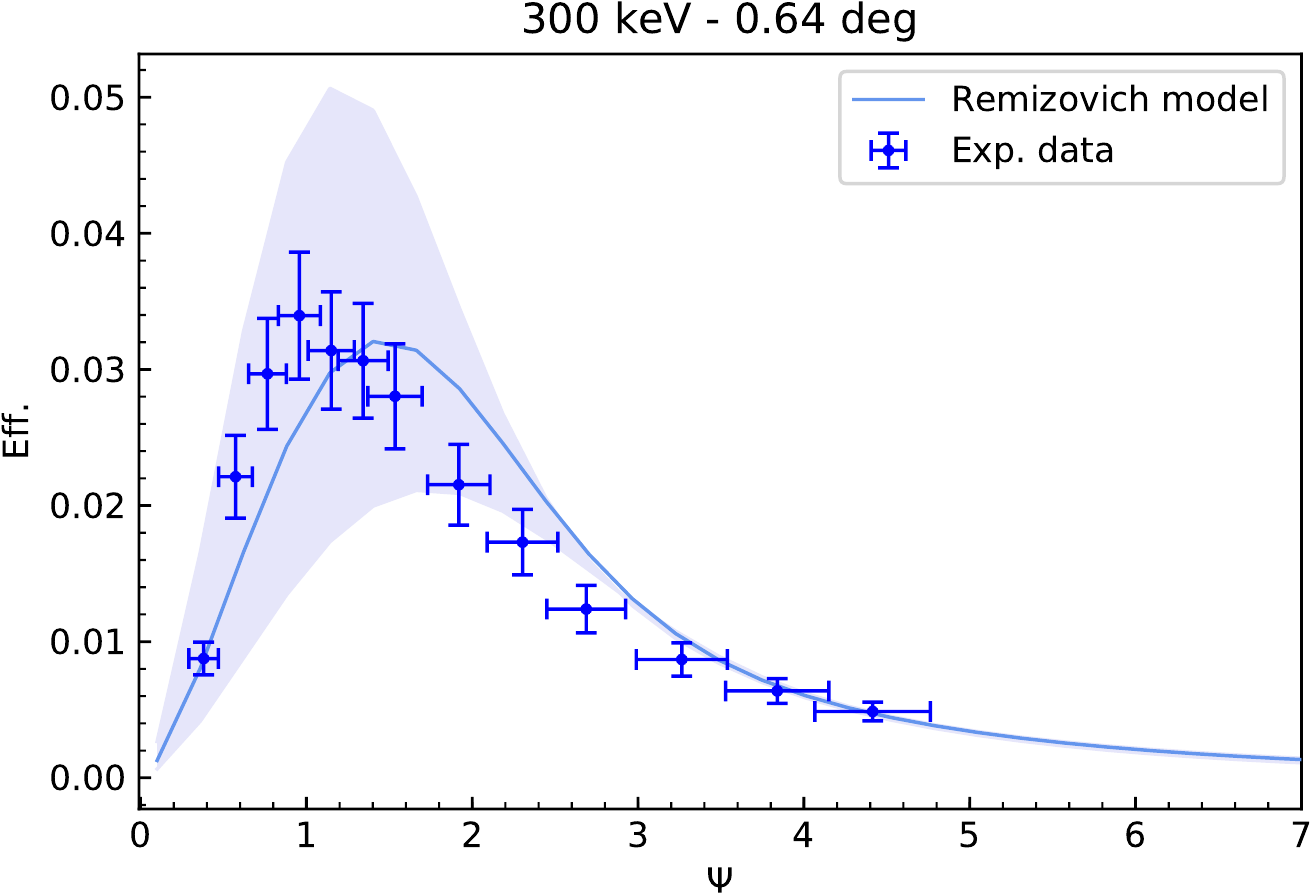}} \quad
{\includegraphics[width=0.48\textwidth]{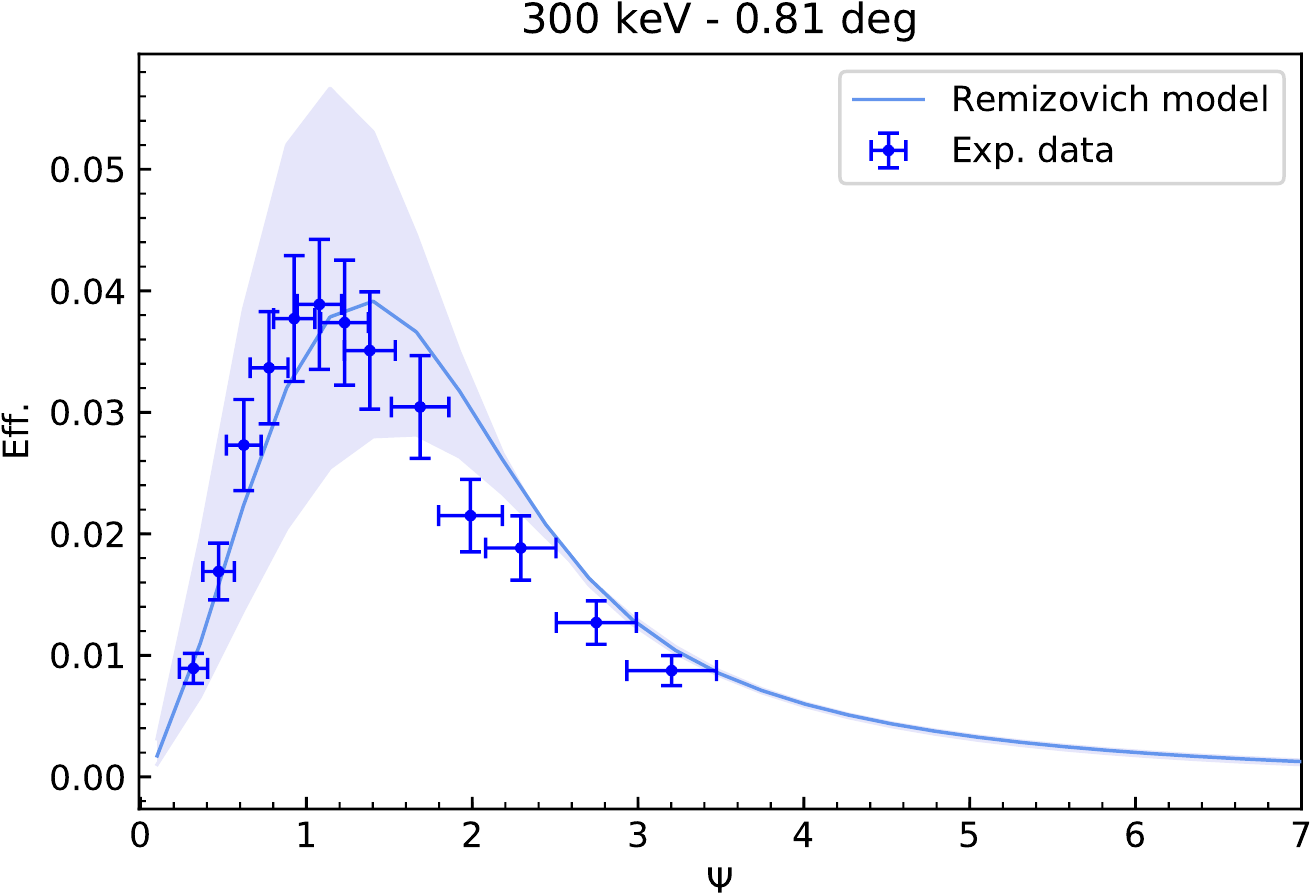}} \quad
{\includegraphics[width=0.48\textwidth]{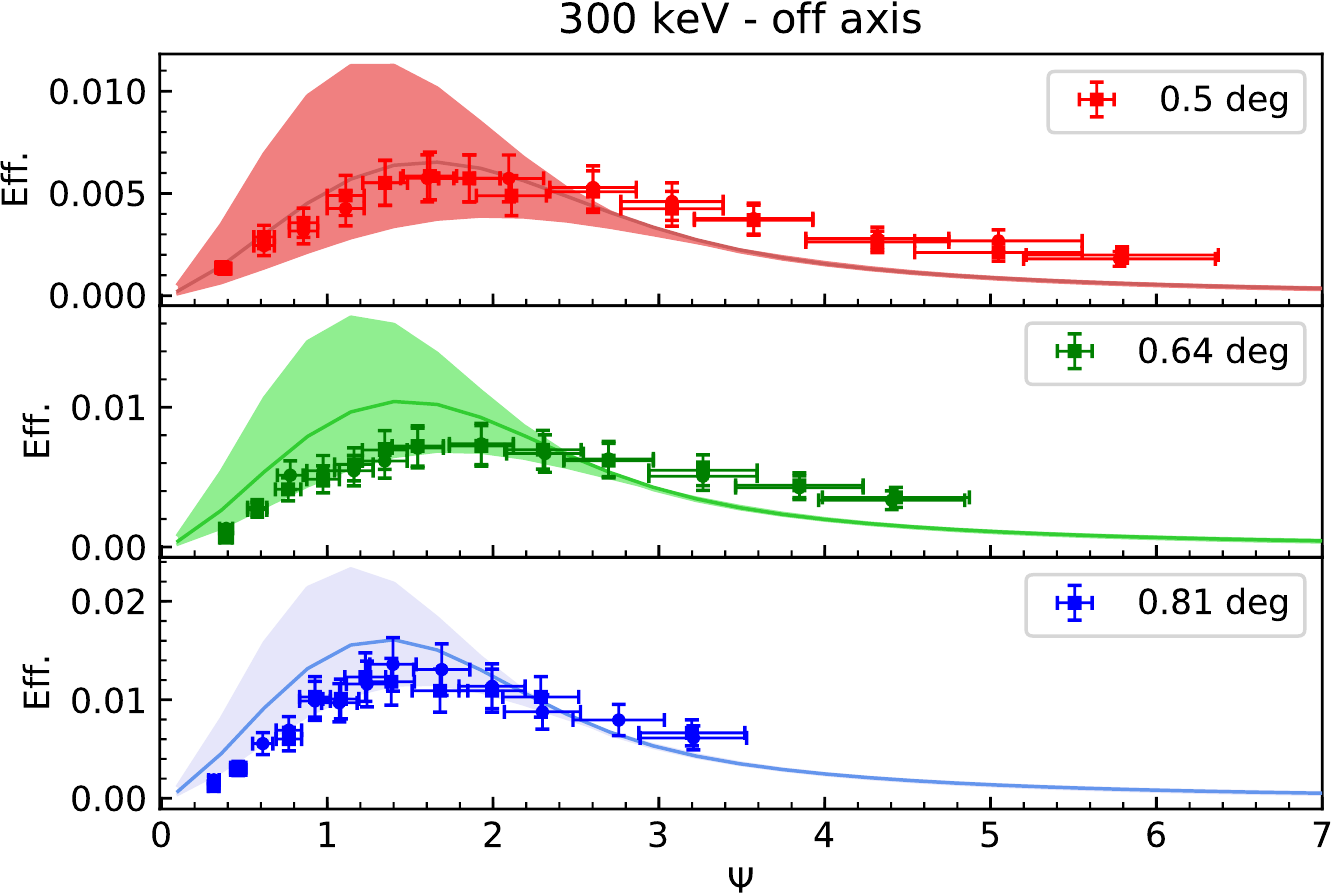}}
\caption{Data and model for the scattering efficiencies at 300 keV \citep{SPIE2017}. The solid line corresponds to the model obtained from the best fit value of the parameter $\sigma$, the coloured area to the maximum and minimum of the distribution, according to the error on $\sigma$ (errors at 2.7$\sigma$). The \emph{bottom right} panel shows the same comparison for the off-axis data.}
\label{fig:model_2017data}
\end{figure}

First, we computed the expected scattering probability distributions for the experimental measurements of \cite{SPIE2017} and \cite{Rasmussen}, using the results of Table \ref{table_sigma}, and compared it to the data. Fig.~\ref{fig:model_2017data} shows the over plot of the experimental measurements on \erosita{} sample \citep{SPIE2017} with the model computed with the best fit power law value of $\sigma$. In the case of the on-axis configuration, the scattering efficiency curve for the smallest incidence angle of 0.5$^\circ$ is noticeably underestimated in the peak, while the curves for the other two incidence angles of 0.64$^\circ$ and 0.81$^\circ$ are closer to the data, though they do not perfectly reproduce the experimental trend. However, if we consider the maximum and the minimum of the expected scattering efficiency distributions (coloured area in Fig.~\ref{fig:model_2017data}), resulting by the maximum and minimum error on the parameter $\sigma$, then the data can be considered acceptably well modelled, especially at the peaks, even though the spread in efficiency is so high that it prevents any more precise evaluation. For the off-axis configuration, instead, the expected scattering efficiencies are slightly overestimated in the peak, while the tails are underestimated (Fig.\,\ref{fig:model_2017data}, bottom right panel). A correct modelling of the peak, rather than of the tail of the distribution, is essential to predict the expected flux of proton funnelled through the X-ray optics. For the first time, this semi-empirical approach is the closest to the experimental data in giving a correct modelling of the peak. We remark here that having a larger extent of experimental data, i.e. more data points per set, covering wider angular and energetic ranges, remains necessary for better assessing the experimental value of $\sigma$.

\begin{figure}[ht]
\centering
{\includegraphics[width=0.48\textwidth]{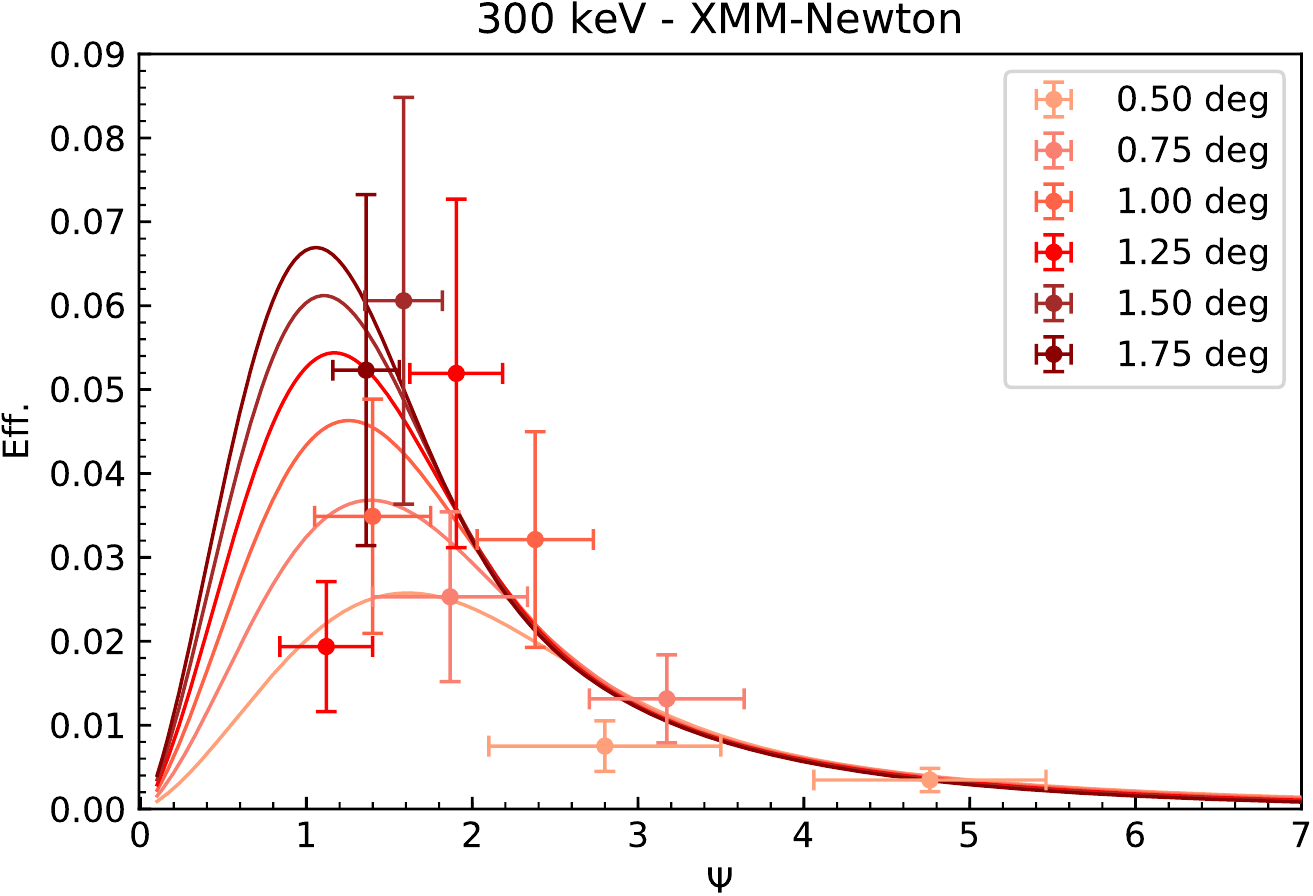}} \quad
{\includegraphics[width=0.48\textwidth]{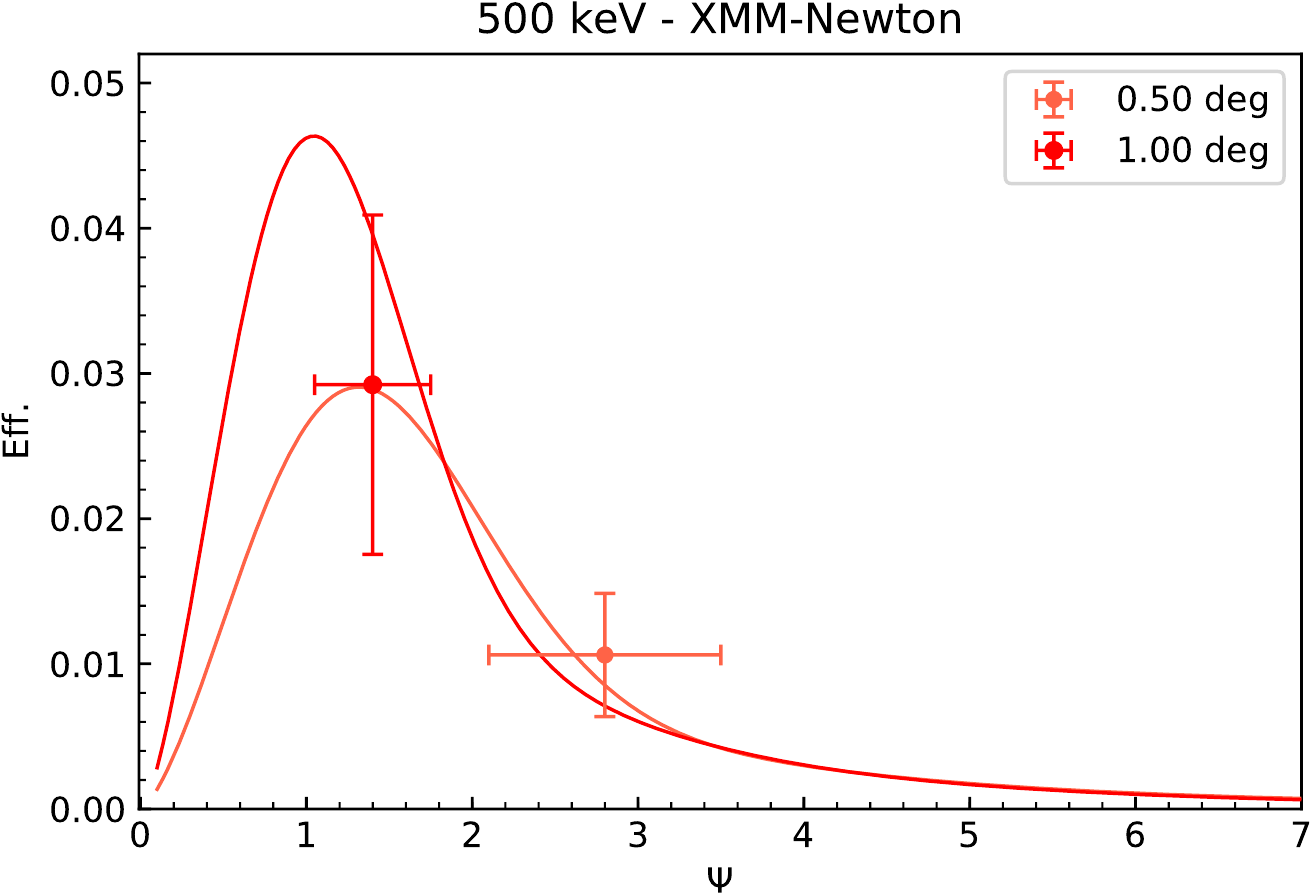}} \quad
{\includegraphics[width=0.48\textwidth]{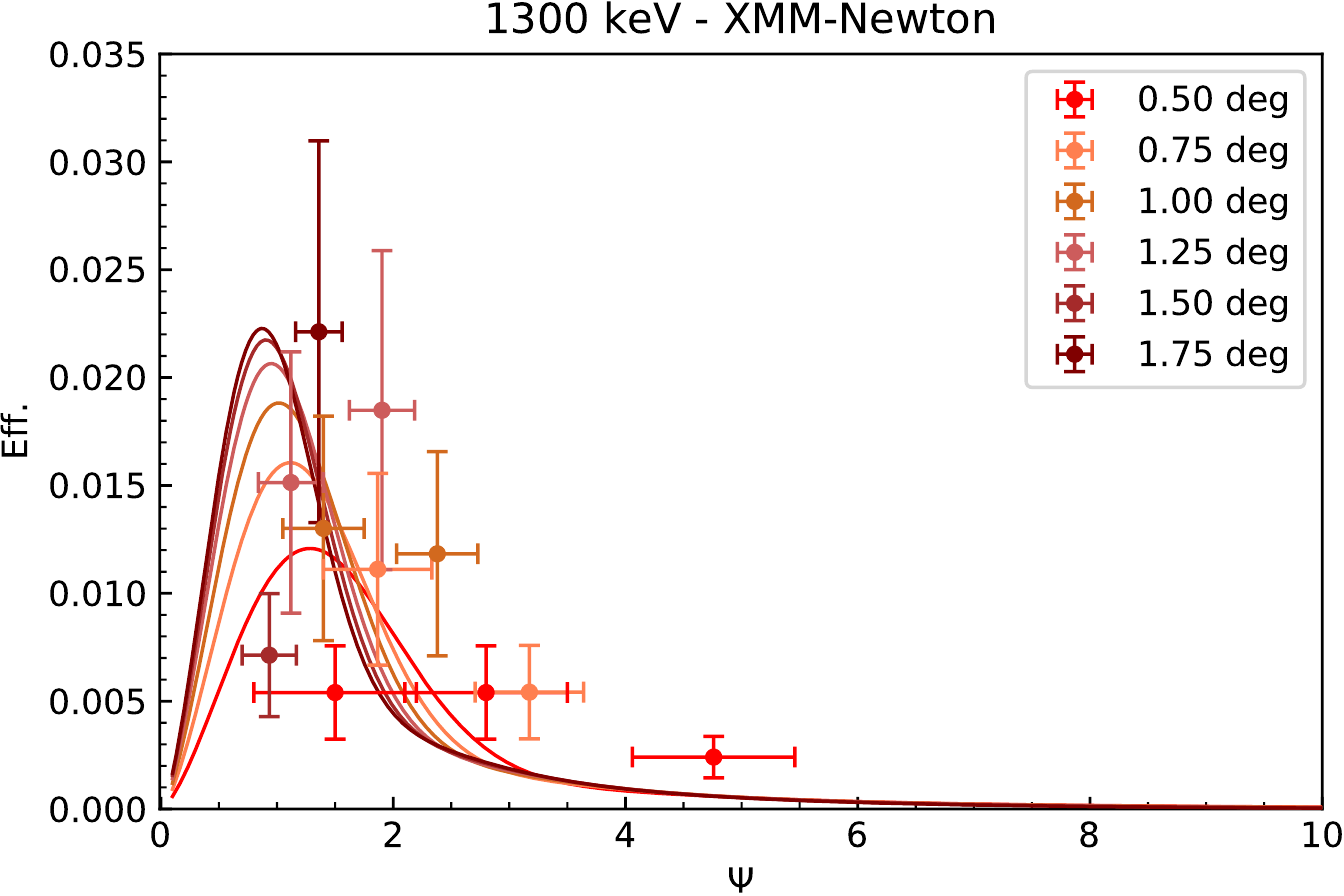}} \caption{Data and scattering efficiency distribution predicted by the model with the best fit value of $\sigma$ for the \xmm{} mirror sample \citep{Rasmussen}.}
\label{fig:model_xmmdata}
\end{figure}

For completeness, we took into account also the measurements on \xmm{} mirrors \citep[][Fig.\,\ref{fig:model_xmmdata}]{Rasmussen}, though the paucity of data does not really allow us to put tighter constraints. In this case, the model is not consistent with the data, since the peaks of the distributions are always shifted towards lower scattering angles, as we already noticed when comparing these data with the \erosita{} sets (Fig.~\ref{fig:all_data}, Section \ref{section:data}).

\begin{figure}[ht]
\centering
\includegraphics[width=0.7\textwidth]{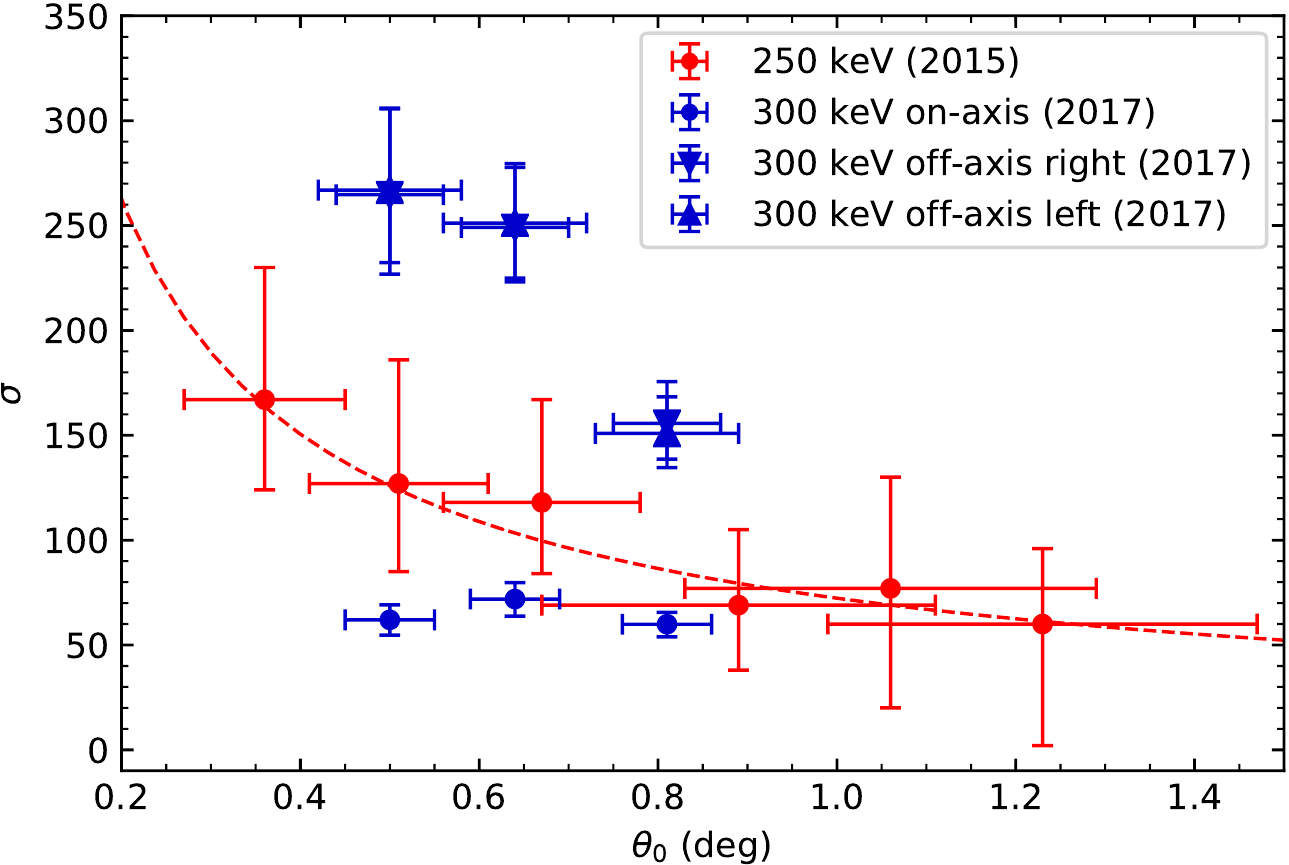}
\caption{Best fit values of $\sigma$ of the 2017 data sets, compared with the previous values for the incident energy of 250 keV (see Fig.~\ref{fig:sigma}). Error bars on the values of $\sigma$ at 95\%.}
\label{fig:sigma2017}
\end{figure}

\begin{figure}[ht]
\centering
{\includegraphics[width=0.48\textwidth]{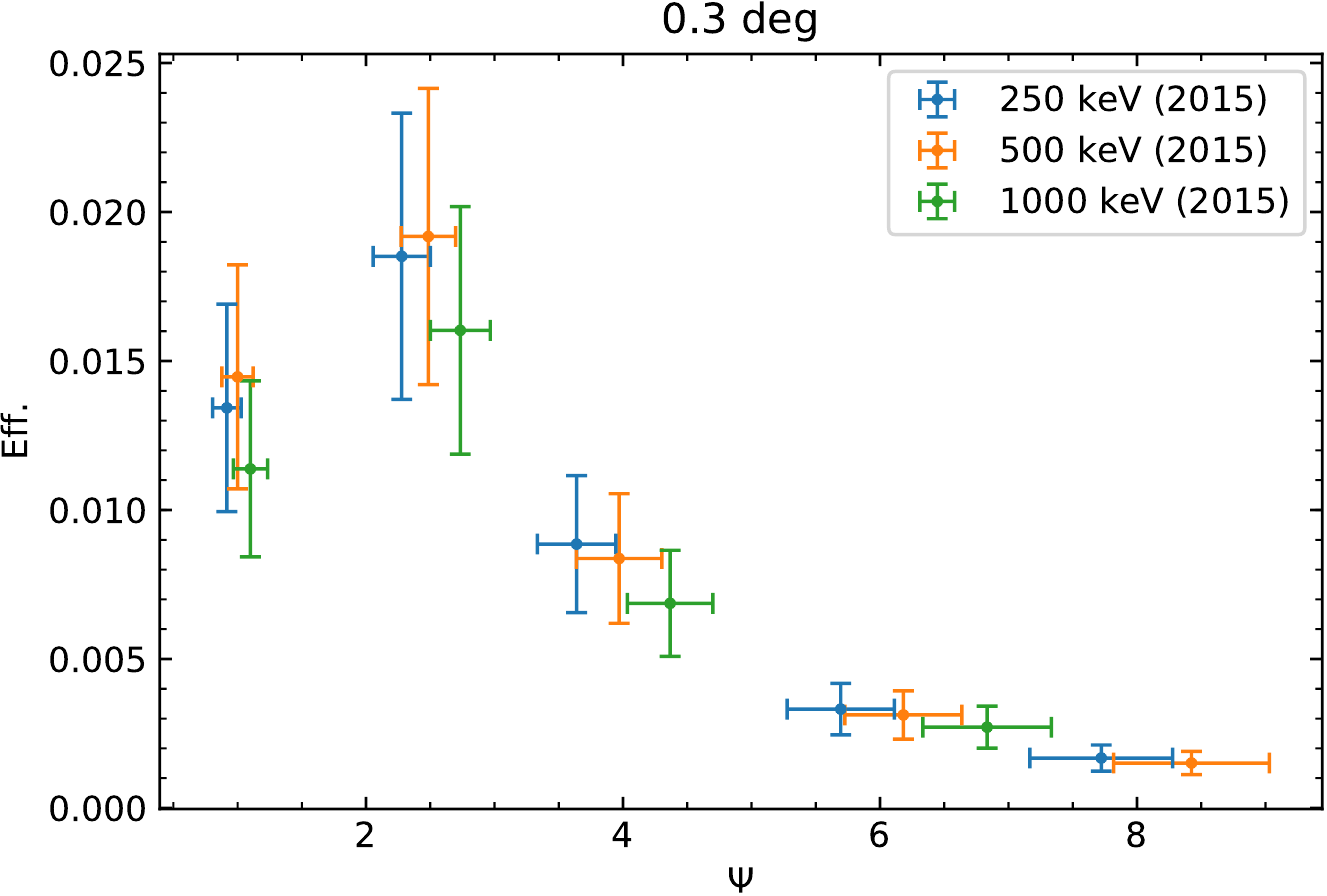}} \quad
{\includegraphics[width=0.48\textwidth]{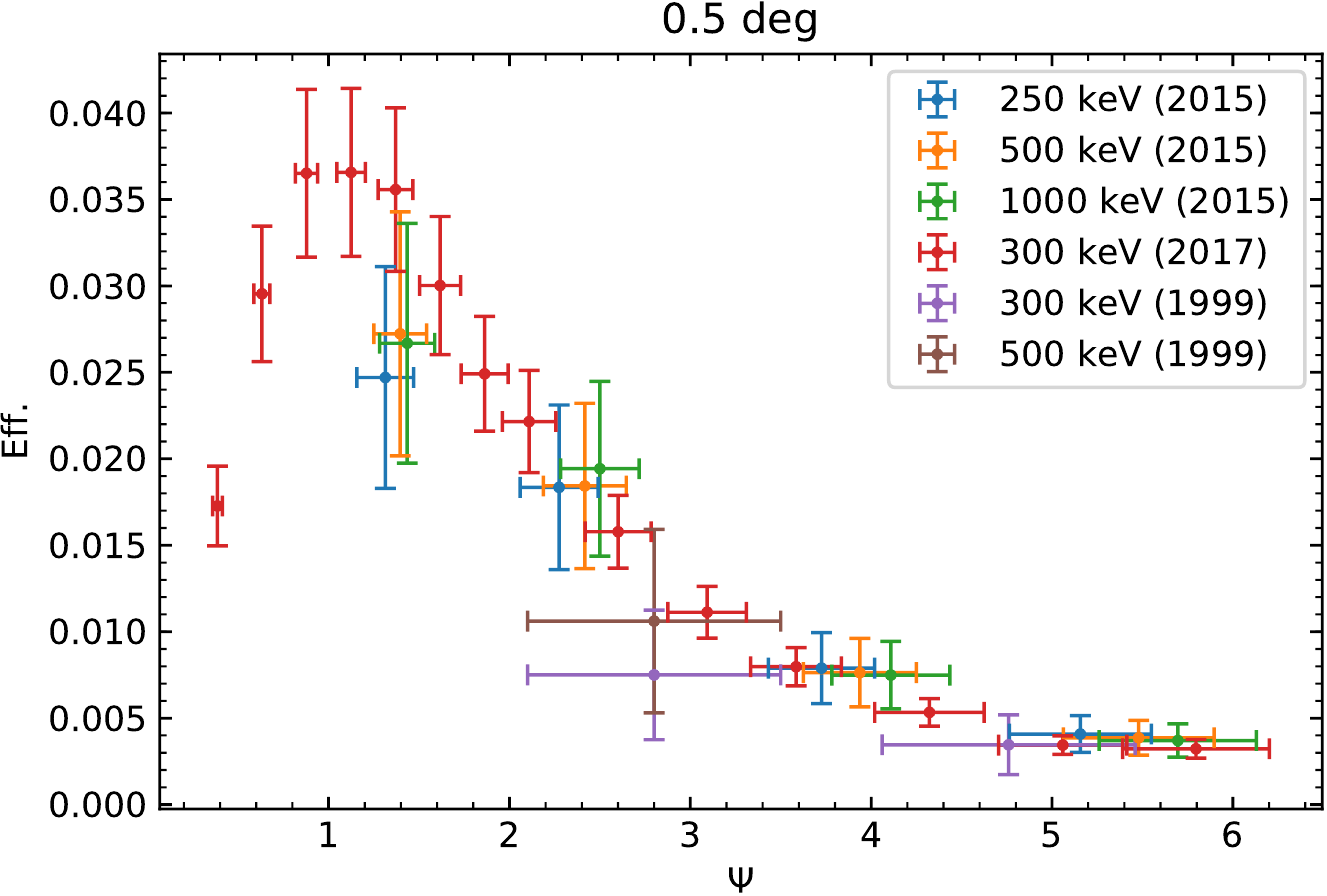}} \quad
{\includegraphics[width=0.48\textwidth]{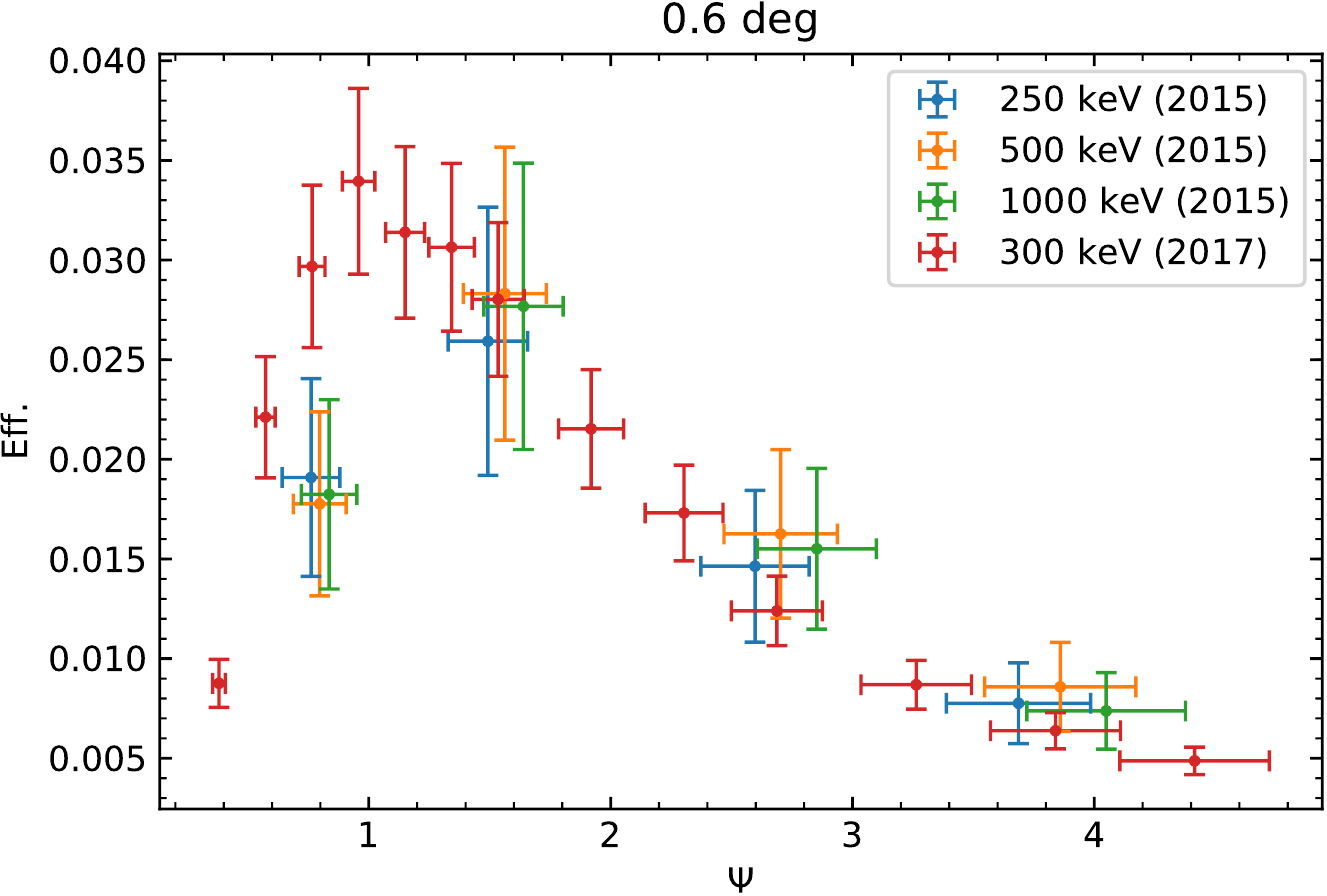}} \quad
{\includegraphics[width=0.48\textwidth]{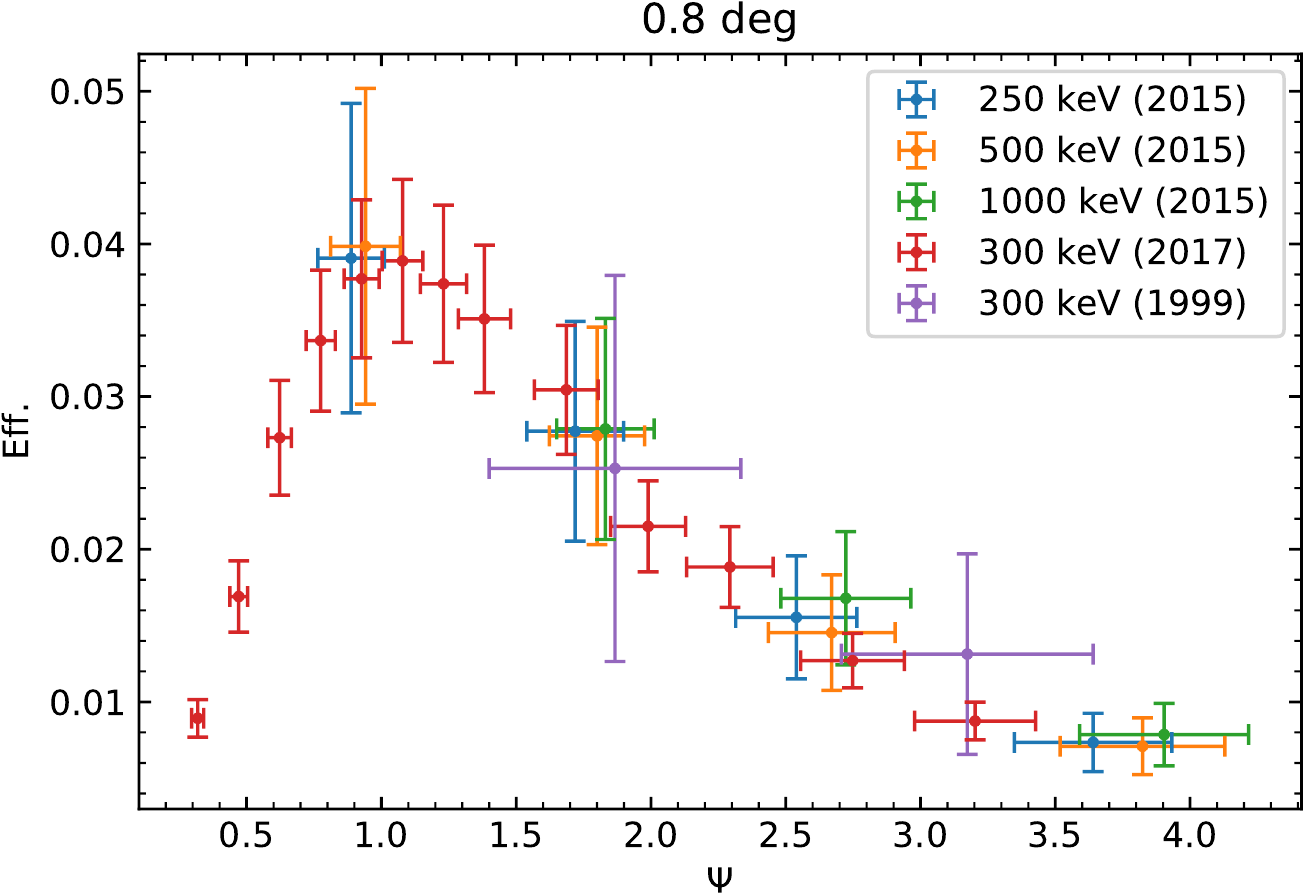}}\quad
{\includegraphics[width=0.48\textwidth]{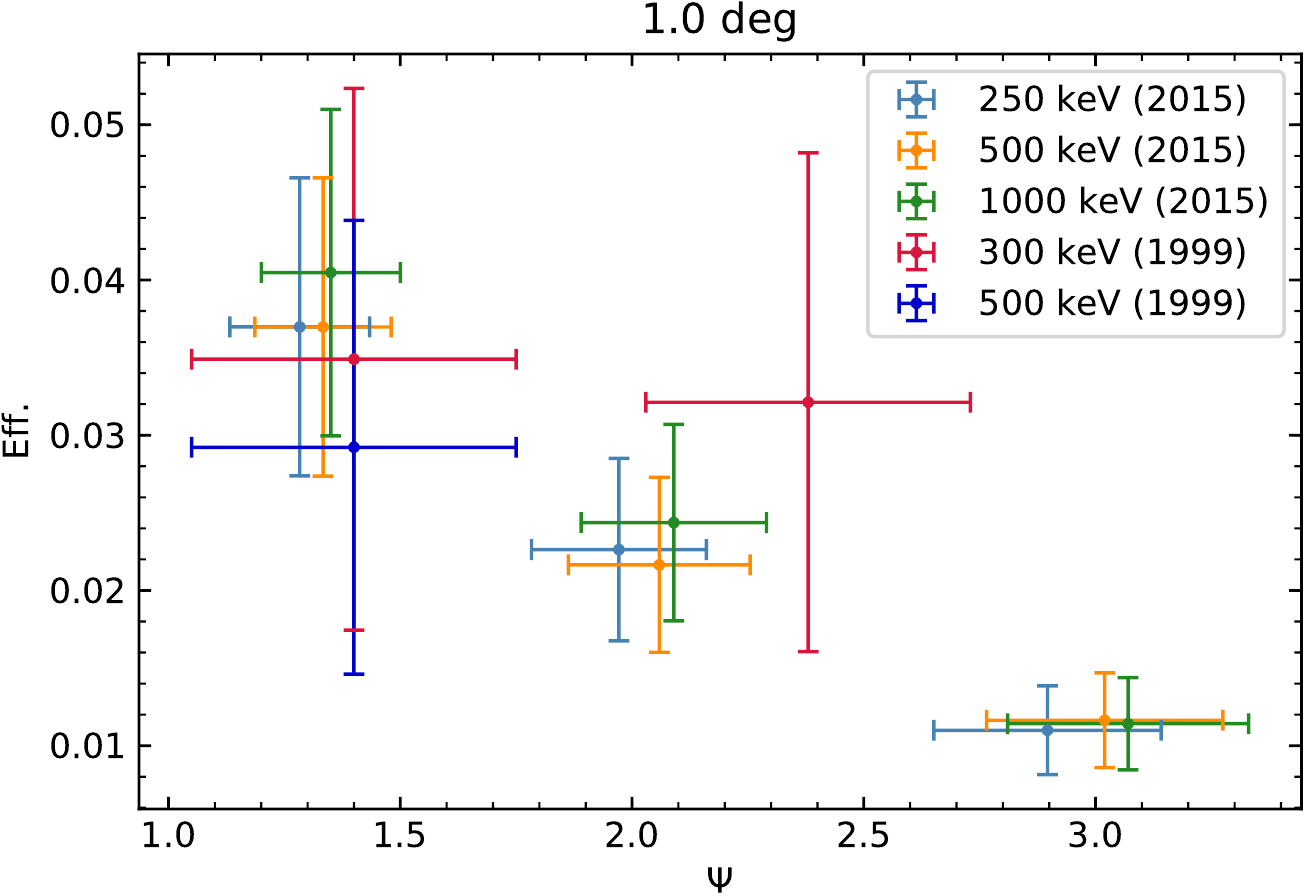}} \quad
{\includegraphics[width=0.48\textwidth]{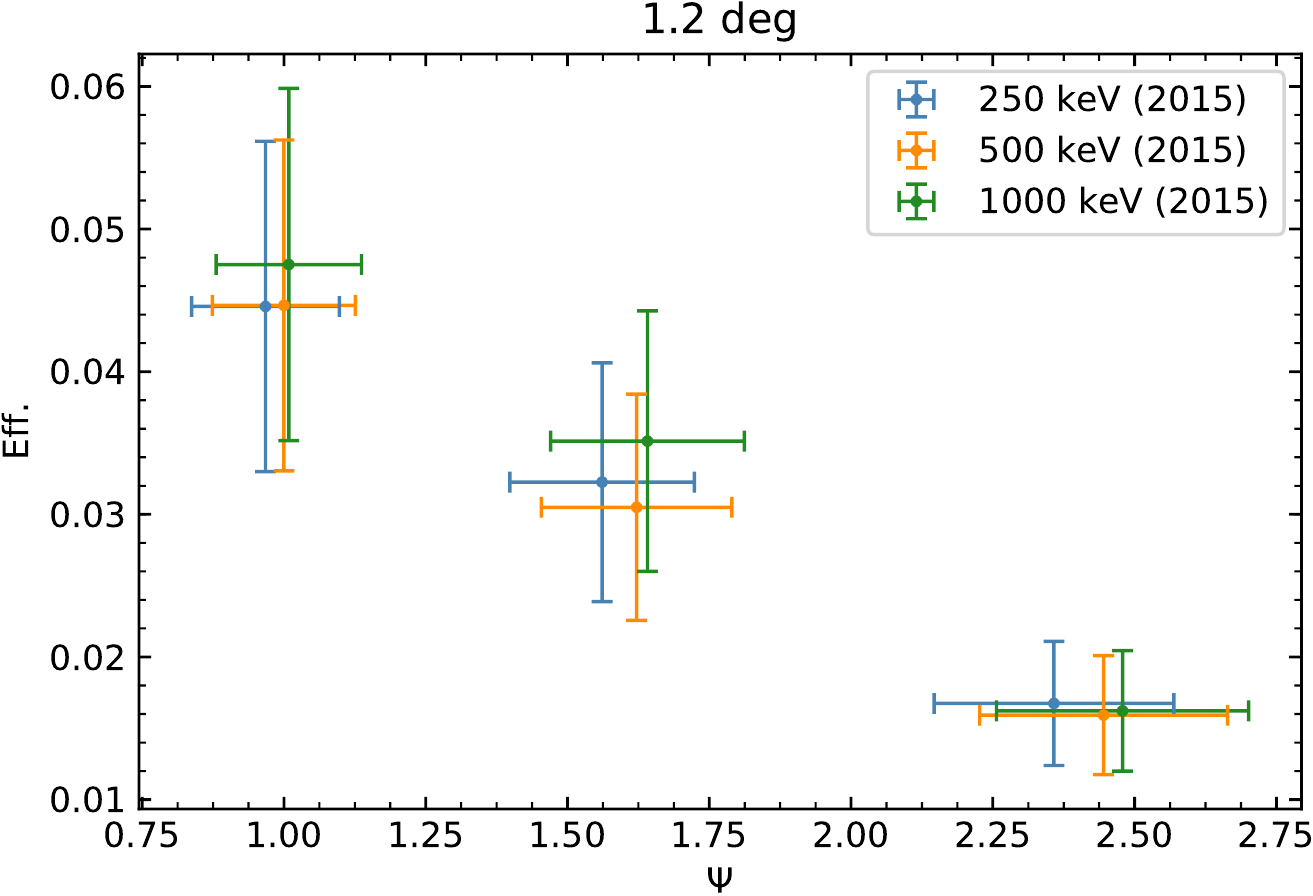}}
\caption{Scattering efficiencies of all the available data sets as a function of the scattering angles $\Psi$ for different incidence angles. Errors on \xmm{} efficiencies are at 40\%.}
\label{fig:data_sort_by_angle}
\end{figure}

Secondly, since the fit is weakly dependent on the energy losses, we directly fit the data of \cite{SPIE2017}, without accounting for them. However, the on-axis measurements result on the whole in smaller values than the previous ones and the values of $\sigma$ for the on-axis and off-axis configurations are not consistent with each other (Fig.~\ref{fig:sigma2017}). This stresses once again that the energy losses are necessary to constrain the fit.

Overall, the consistency of almost all the $\sigma$ of Tab.\,\ref{table_sigma}, regardless of the initial energy, leads to the hypothesis that the 
scattering efficiency is not dependent upon the energy of the impinging proton beam. To verify this assumption, we sort all the data simply by the incidence angle, irrespective of the energies (Fig.\,\ref{fig:data_sort_by_angle}), and, as a matter of fact, all the scattering efficiencies appear consistent with each other.

Finally, one minor concern regards the microroughness of mirroring surfaces, which is already known to be responsible of reducing the reflection efficiency of X-ray photon, by causing scattering in other directions than the incident one \citep{Spiga2007}. The same effect might apply to protons as well, although the higher mass of protons suggests that almost all the impinging particles penetrate the surface, instead of being scattered in the surrounding directions. The lack of any experimental estimates on the angular distribution of sided or back scattered protons does not allow us to investigate this issue any further.

\section{Conclusions and future perspectives}
\label{section:conclusions}

In this work we tested all the available experimental measurements of proton scattering efficiency at grazing incidence on X-ray mirrors with the analytic model developed by \cite{Remizovich} under the non-elastic approximation. We came up with a semi-empirical model based on the Remizovich formula, where the parameter $\sigma$ is directly determined by fitting the only experimental data set with energy loss measurements. The main results can be summarized as follows:
\begin{itemize}
    \item all the \erosita{} data sets can be modelled with the same value of the parameter $\sigma$, which can be considered independent from the energy of the incident protons, even if a systematic trend with energy is observed;
    \item there is a clear dependence of the parameter $\sigma$ over the incidence angle $\theta_0$, well reproduce by a power law with $\sigma\propto\theta_0^{-1}$. This is in contrast to what is stated by \cite{Remizovich};
    \item the peaks and the tails of the scattering efficiency are acceptably well modelled. We remind here that a correct evaluation of the scattering efficiency at its peak is crucial for the estimation of the SPs flux expected at the focal plane of every X-ray mission with grazing incidence optics; 
    \item although the energy loss distribution drives marginally the fit, they are necessary in modelling the data and in returning consistent values of the parameter $\sigma$;
\end{itemize}

The semi-empirical model we propose is strictly limited to the actual experimental data sets. For instance, we cannot verify the independence of the angular scattering efficiency distribution from the incident energy also at energies below 250 keV\footnote{Energies below 250 keV are especially relevant for the future X-ray mission \athena{}. Simulations by \cite{Lotti2018} show, indeed, that SPs with energies between 1 and 150 keV produce significant background signals in the working range of the instruments at the focal plane.}. To overcome this weakness of the model and to achieve a better estimation of the parameter $\sigma$,further laboratory activities are necessary. In particular, it would be beneficial to have a higher number of data points on wider ranges of incidence angles and energies, especially in the range 10-100 keV, and on different materials, specific for each X-ray observatory. 

This work is a contribution to recent efforts made by multiple teams for a better comprehension and evaluation of the non X-ray background (NXB) for X-ray satellites operating with grazing incidence optics \citep[see, for instance,][for a thorough study on \xmm\ NXB]{XMMbkg-1,XMMbkg-2,XMMbkg-3,XMMbkg-4}. It improves the methods of estimation of the expected SPs flux at the focal plane of such telescopes. This is especially relevant for X-ray missions that aim to observe faint or extended sources at high redshifts, as \erosita{} itself, successfully launched on July 13, 2019, and \athena{} \citep[\textit{Advanced Telescope for High Energy Astrophysics},][]{Nandra2013}, an ESA mission, planned to fly in the early 2030s. \athena{} will orbit around the Lagrangian point L2, 1.5 million km from Earth, in the opposition to the Sun. As L2 is located in the tail of the Earth magnetosphere, the satellites will experience a strongly variable particle environment, so that a correct evaluation of the expected SPs flux is fundamental in pursuing the scientific goals. At present, the scientific requirement is that the SPs flux at the focal plane must be $<5\times10^{-4}$ cts s$^{-1}$ cm$^{-2}$ keV$^{-1}$ (90\% of the observing time, 10\% of the total NXB) in the 2--7 keV (WFI) and 2--10 keV (X-IFU) energy ranges\footnote{\url{https://www.cosmos.esa.int/documents/400752/507693/Athena_SciRd_iss1v5.pdf}}. A first estimation for both the instruments at the focal plane of \athena{} has been done by \cite{Lotti2018}, by computing a proton response matrix of the telescope through the simultaneous and independent use of a ray-tracing code and \textit{Geant4} simulations, together with a thorough study of the particle environment in L2. \cite{Fioretti2018} studied the SPs induced background for the the ATHENA Wide Field Imager, using the \textit{Geant4} Single Scattering to model the SPs interaction with the  \athena{} Silicon Pore Optics (SPOs).
Their results show the necessity to have a proton diverter on board \textit{ATHENA} to respect the aforementioned scientific requirement.
The physical model adopted by \cite{Lotti2018} to treat the grazing incidence scattering of SPs is elastic, with 100\% scattering efficiency, while the non-elastic semi-empirical model here proposed gives an average scattering efficiency of $\sim$80\%.  

However, to be used directly for \athena{}, laboratory measurements on SPOs samples are necessary\footnote{An experimental campaign of proton scattering at grazing incidence from SPOs is currently ongoing, lead by the EXACRAD group of the University of T\"ubingen, at the facility of the Goethe University of Frankfurt.}. We plan to test in the next future the validity of the model on already available observational X-ray data from \xmm{}. Hence, the next and necessary step will be to implement the model in a ray-tracing code for \xmm{} optics and to build the proper response matrix \citep{Mineo2017}, in order to estimate the SPs fluxes and the spectra detected by the satellite directly from the observed data. Then, the model can be extended to \erosita{} and to all the other X-ray missions equipped with similar optics, provided that experimental measurements of proton scattering on the proper mirror samples are available.

\begin{acknowledgements}
This work was partially supported by the ASI contract no. 2018-11-HH.0, the AREMBES contract no. 4000116655/16/NL/BW, the AHEAD project (grant agreement n. 654215) which is part of the EU-H2020 programme and the \textsl{Bundesministerium f\"{u}r Wirtschaft und Energie} through the \textsl{Deutsches Zentrum f\"{u}r Luft- und Raumfahrt e.V. (DLR)} under the grants number FKZ 50 QR 0702 and 50 QR 1602. AD acknowledges financial contribution from the agreement ASI-INAF n.2017-14-H.O. 
\end{acknowledgements}

\bibliographystyle{spbasic}      

\begin{small}
\bibliography{main.bib}
\end{small}

\clearpage

\appendix
\section{Overall fits}
\label{sec:appendixA}
\label{grafici}
\begin{figure}[ht]
\centering
   {\includegraphics[width=.48\textwidth,height=0.23\textheight]{fit_results_scat_250KeV_0p36.pdf}} 
   {\includegraphics[width=.48\textwidth,height=0.23\textheight]{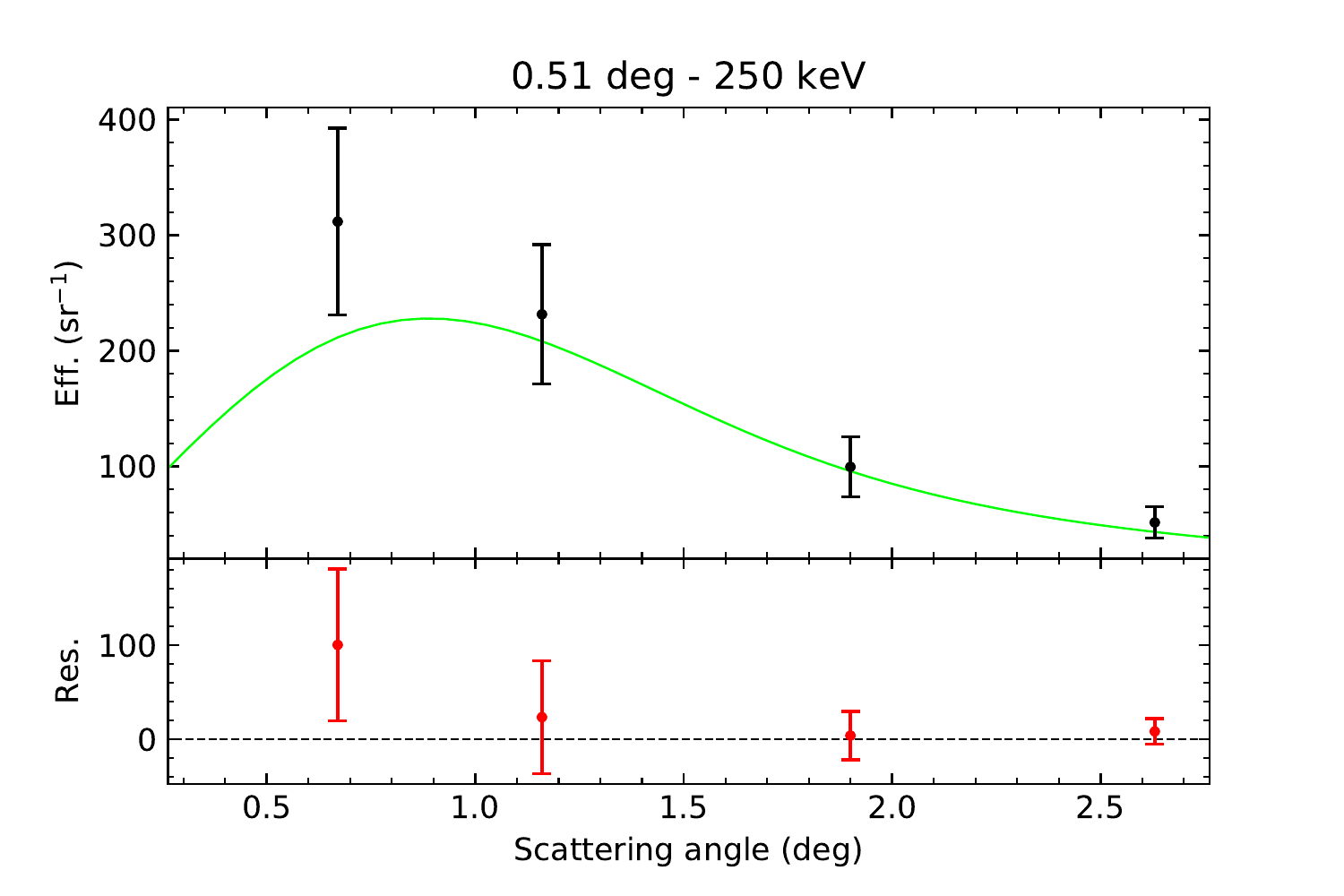}} 
   {\includegraphics[width=.48\textwidth,height=.23\textheight]{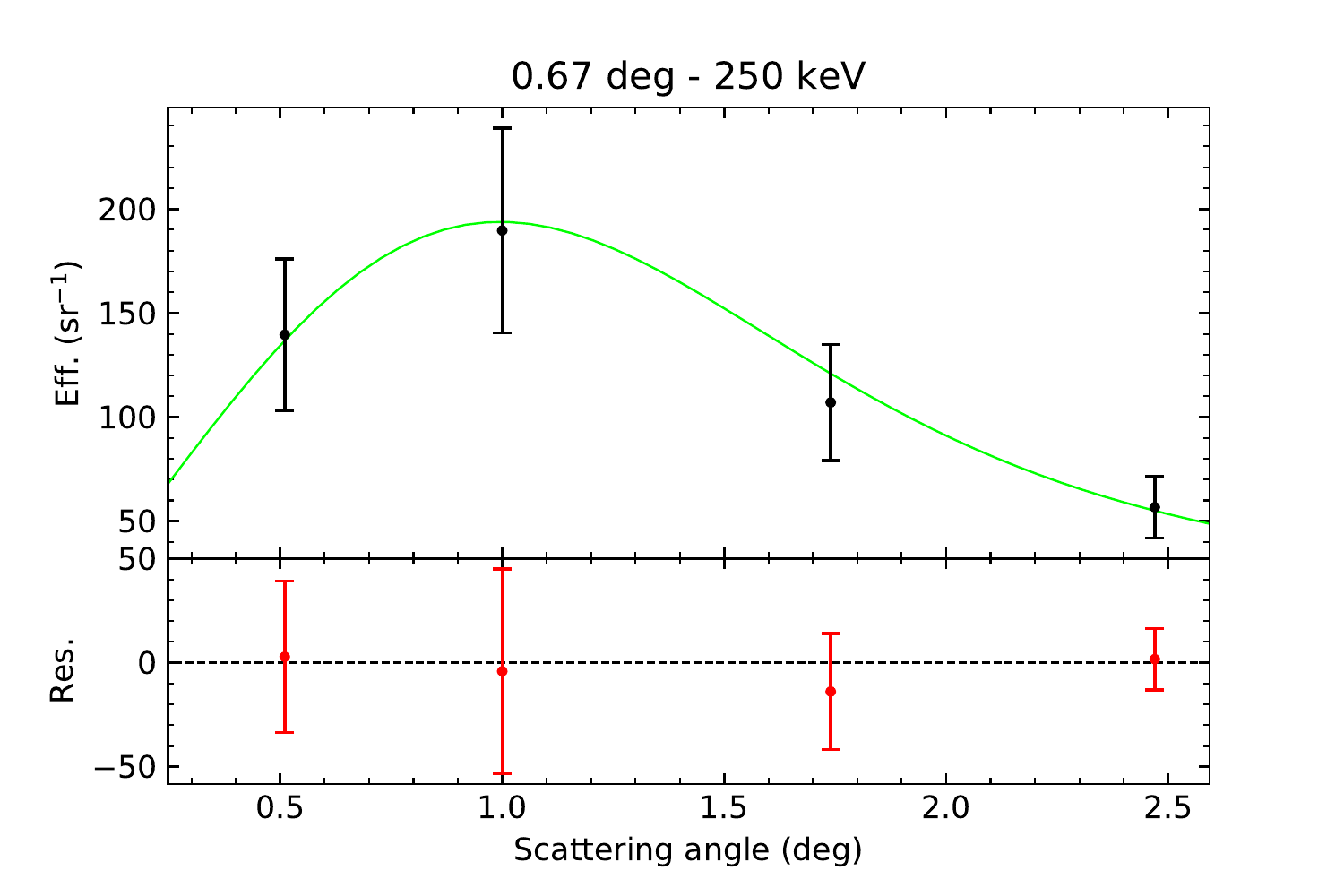}}
   {\includegraphics[width=.48\textwidth,height=.23\textheight]{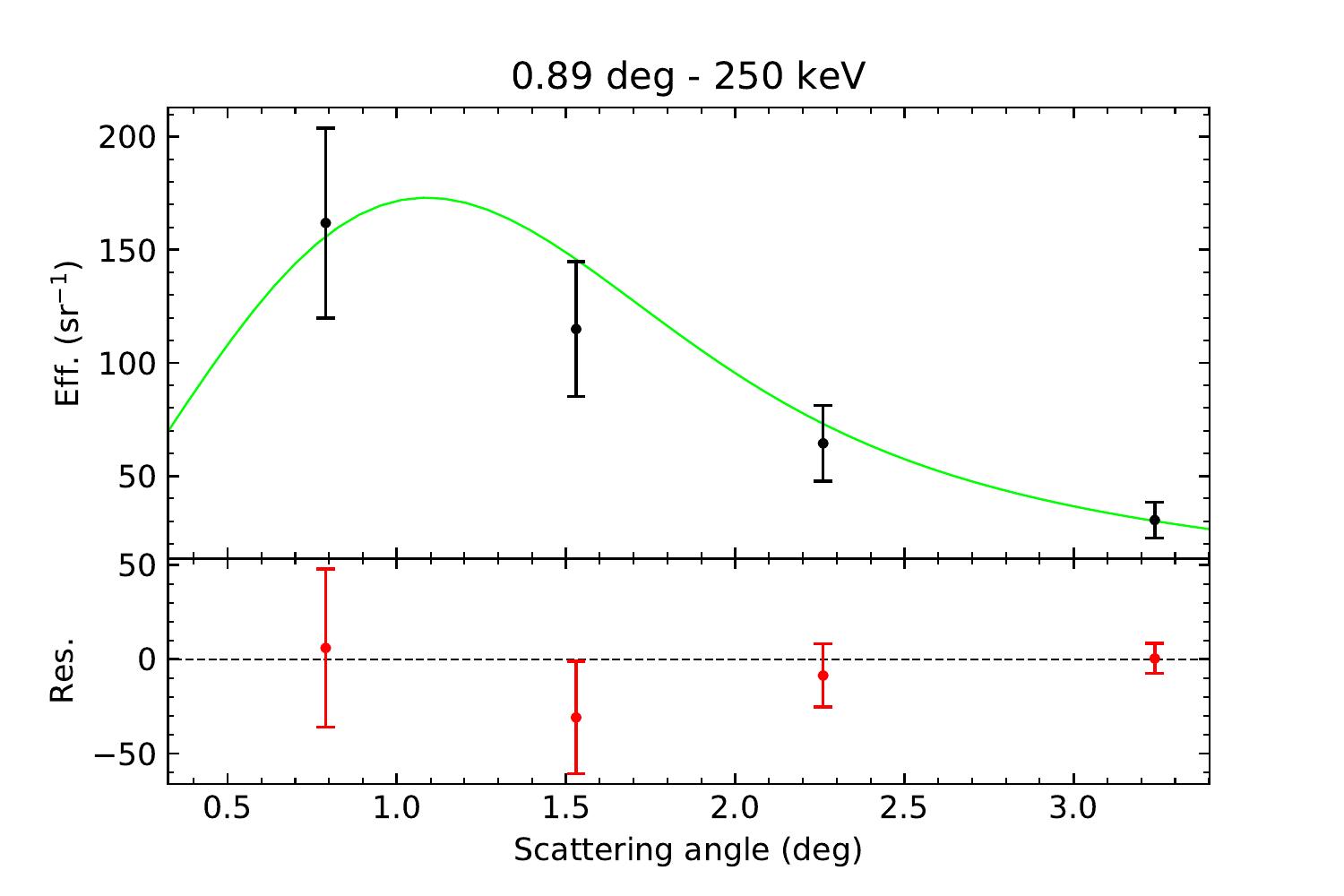}}
   {\includegraphics[width=.48\textwidth,height=.23\textheight]{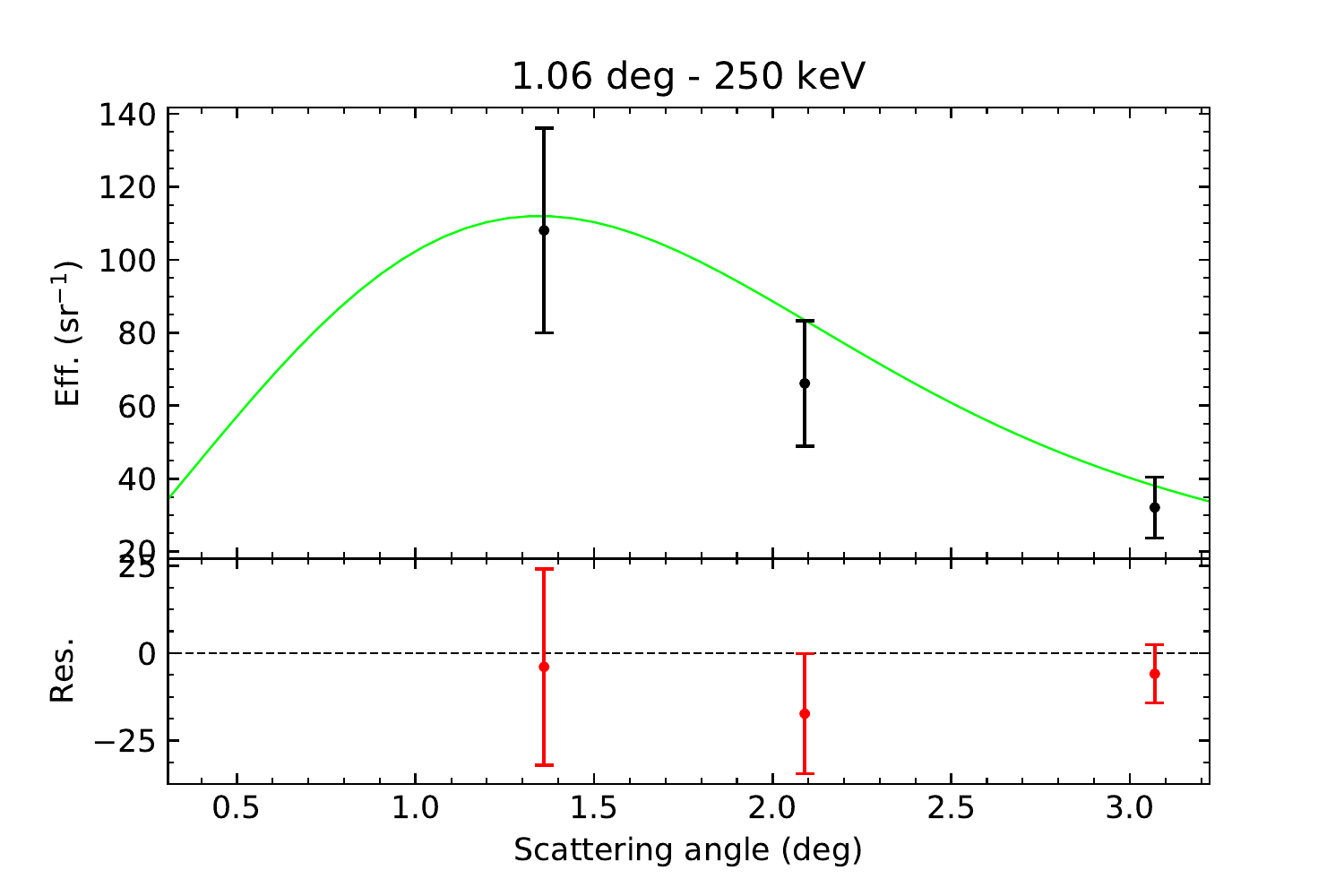}}
   {\includegraphics[width=.48\textwidth,height=.23\textheight]{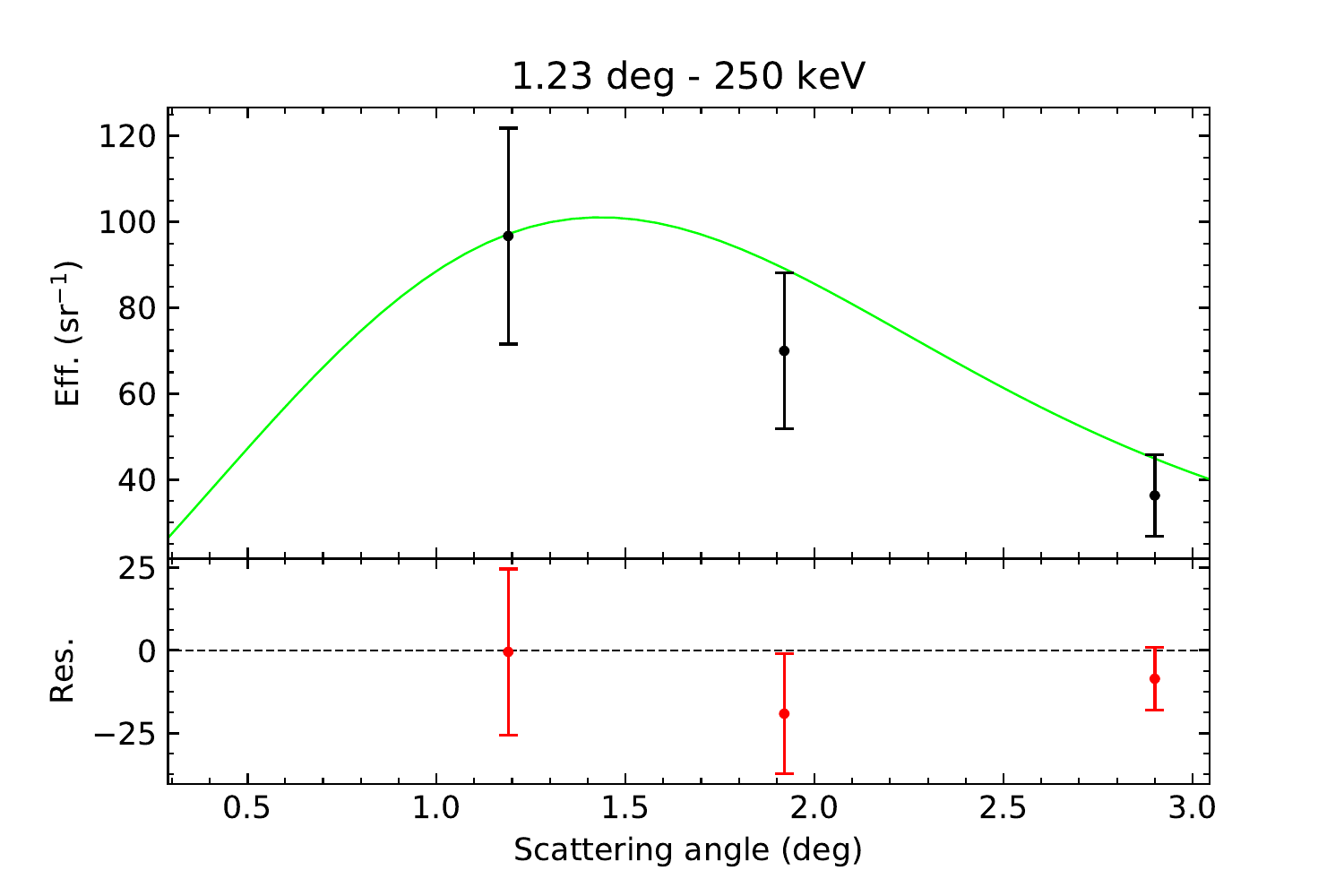}}
\caption{Fitting results with the Remizovich formula (Eq.~\ref{eq:eff_integrale}) in non-elastic approximation. Incident energy of 250 keV.}
\label{fig:fit_all_250}
\end{figure}

\begin{figure}
\centering
   {\includegraphics[width=.48\textwidth,height=.23\textheight]{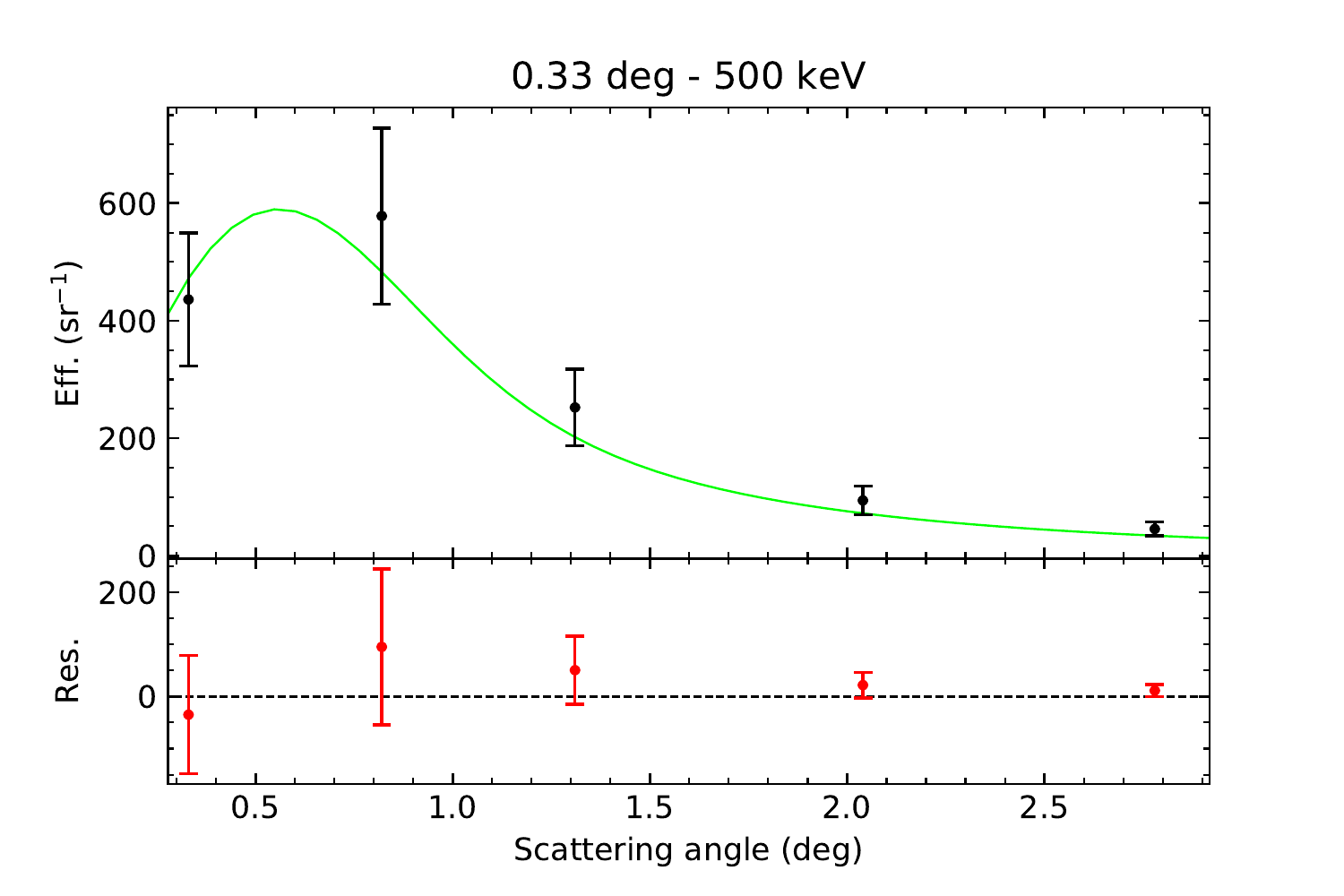}} 
   {\includegraphics[width=.48\textwidth,height=.23\textheight]{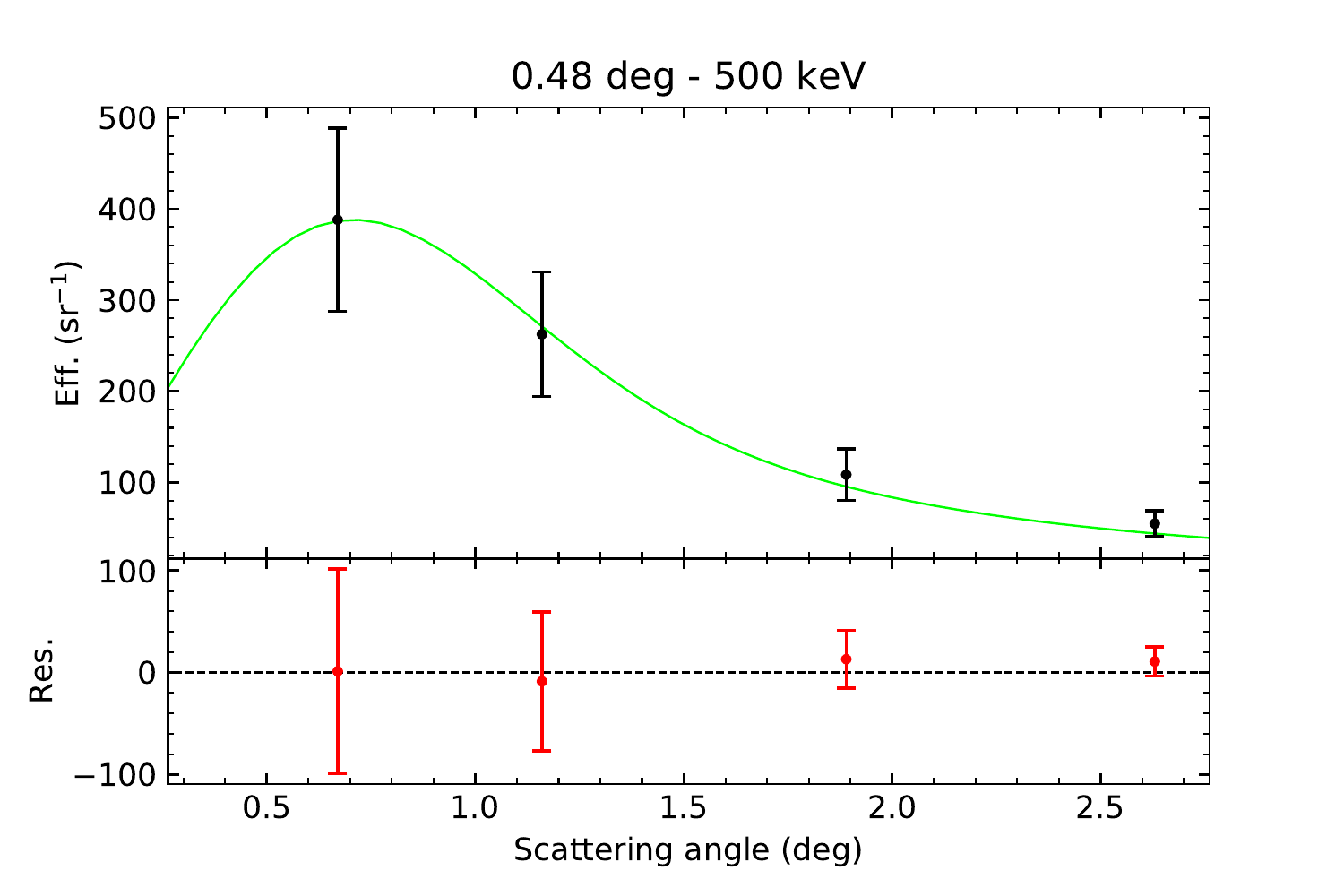}} 
   {\includegraphics[width=.48\textwidth,height=.23\textheight]{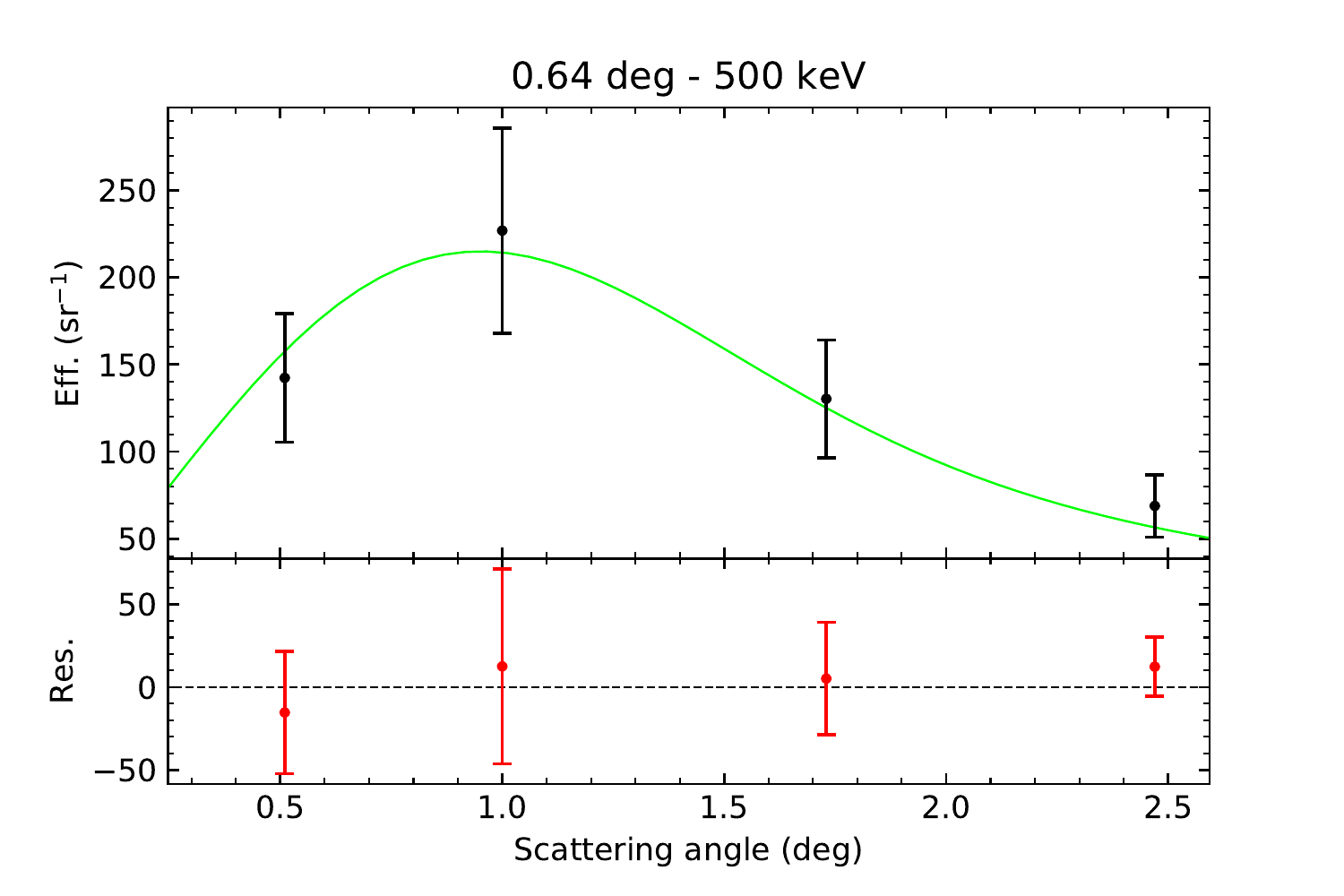}}
   {\includegraphics[width=.48\textwidth,height=.23\textheight]{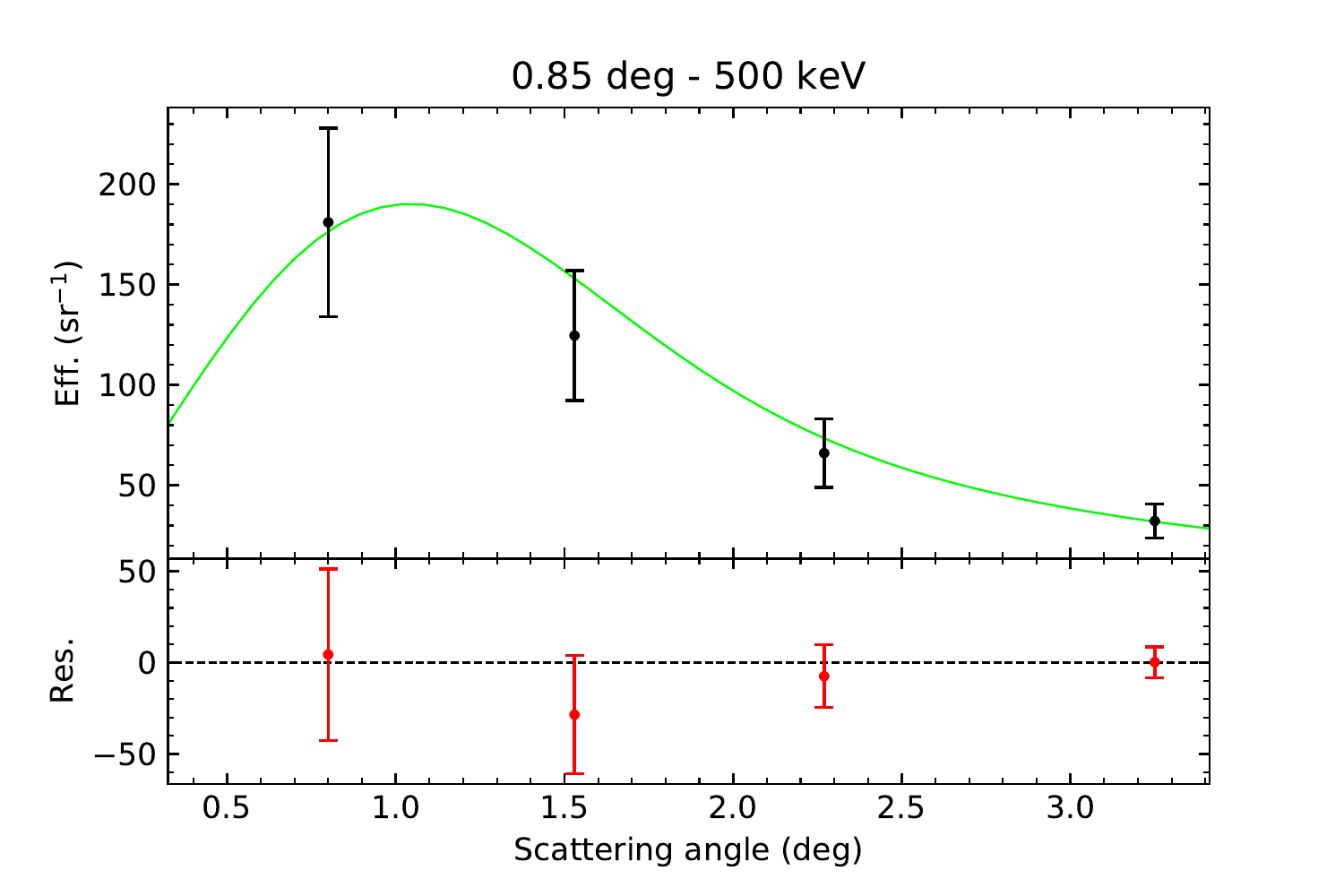}}
   {\includegraphics[width=.48\textwidth,height=.23\textheight]{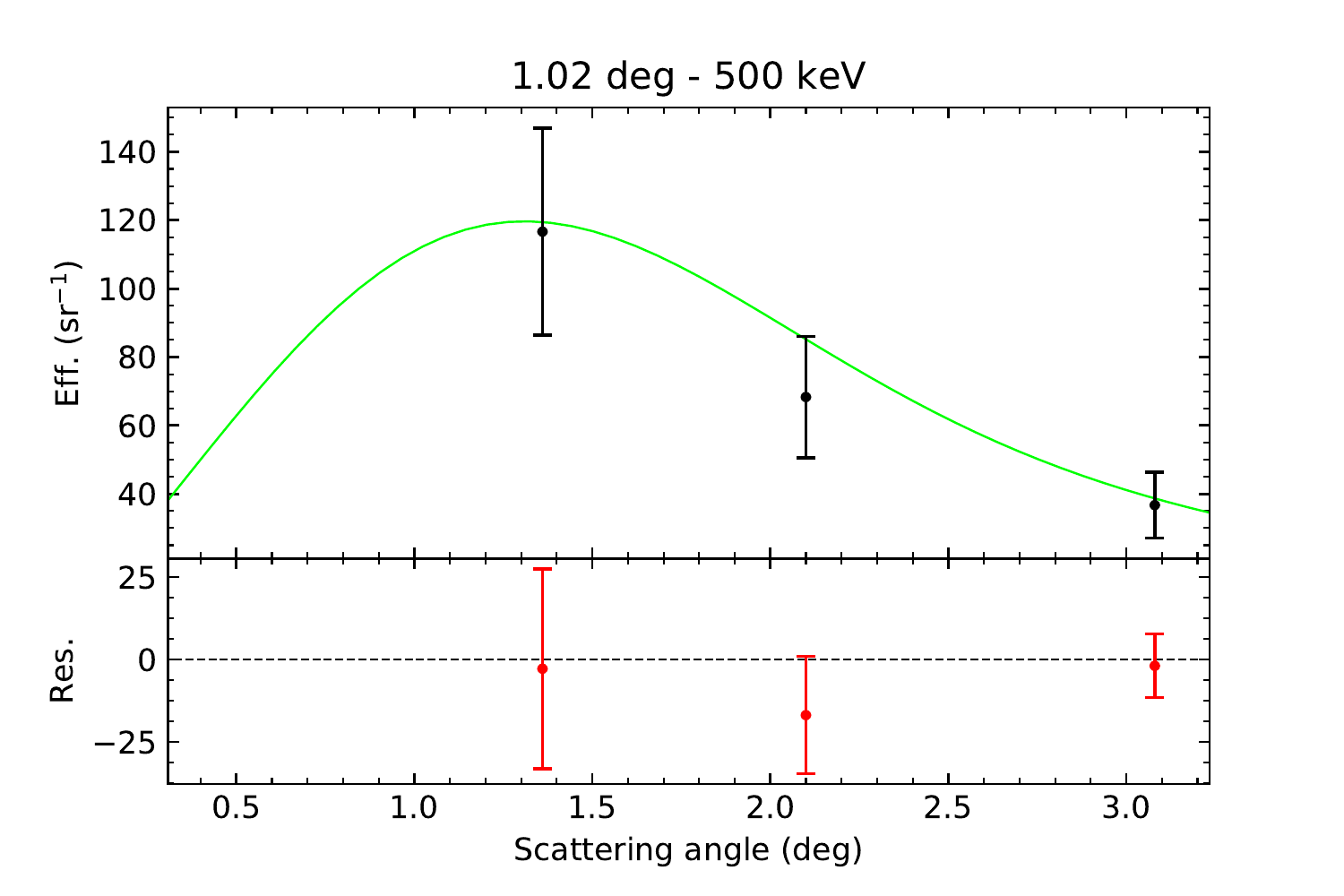}}
   {\includegraphics[width=.48\textwidth,height=.23\textheight]{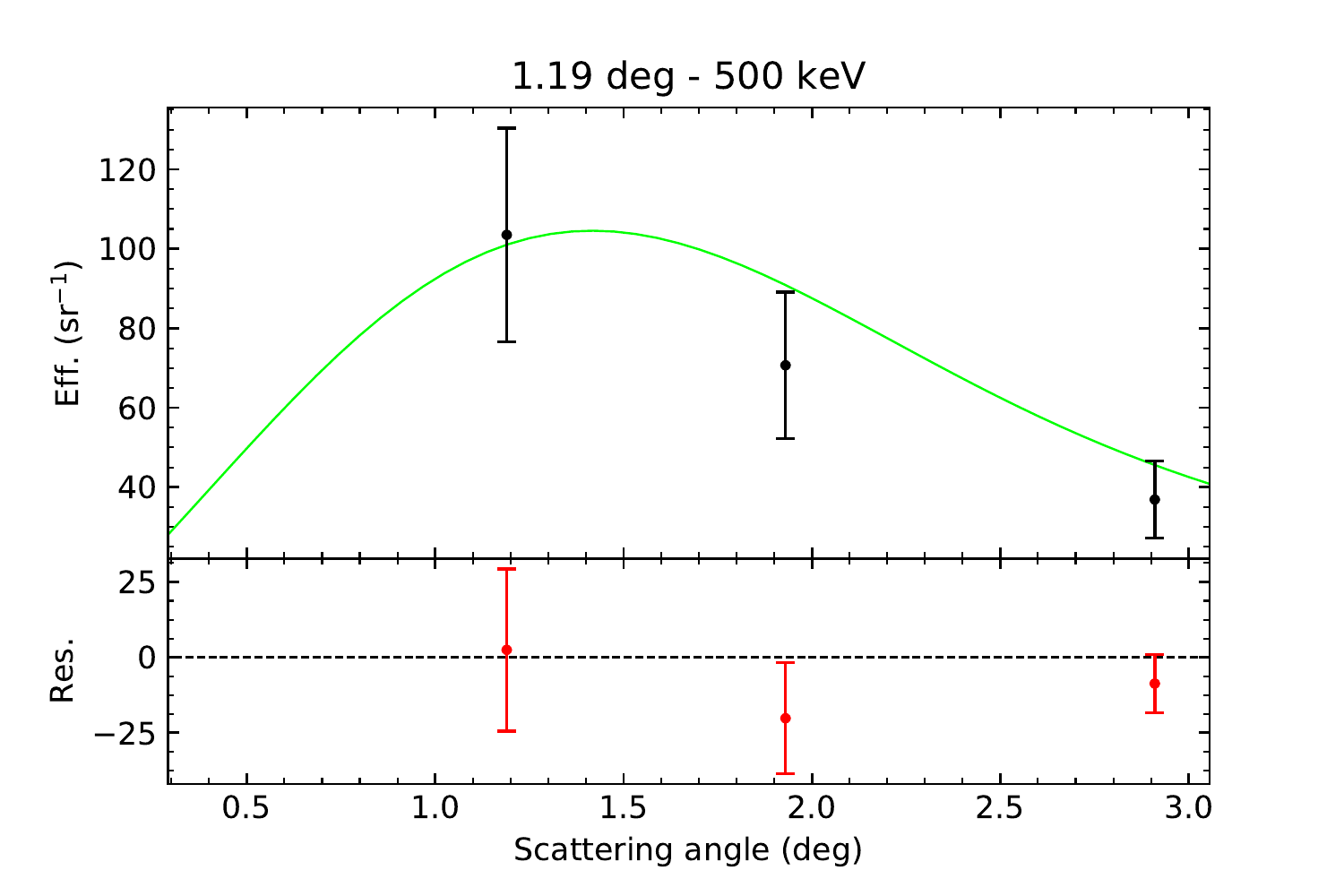}}
\caption{As before, for the incident energy of 500 keV.}
\label{fig:fit_all_500}
\end{figure}
   
\begin{figure}
\centering
   {\includegraphics[width=.48\textwidth,height=.23\textheight]{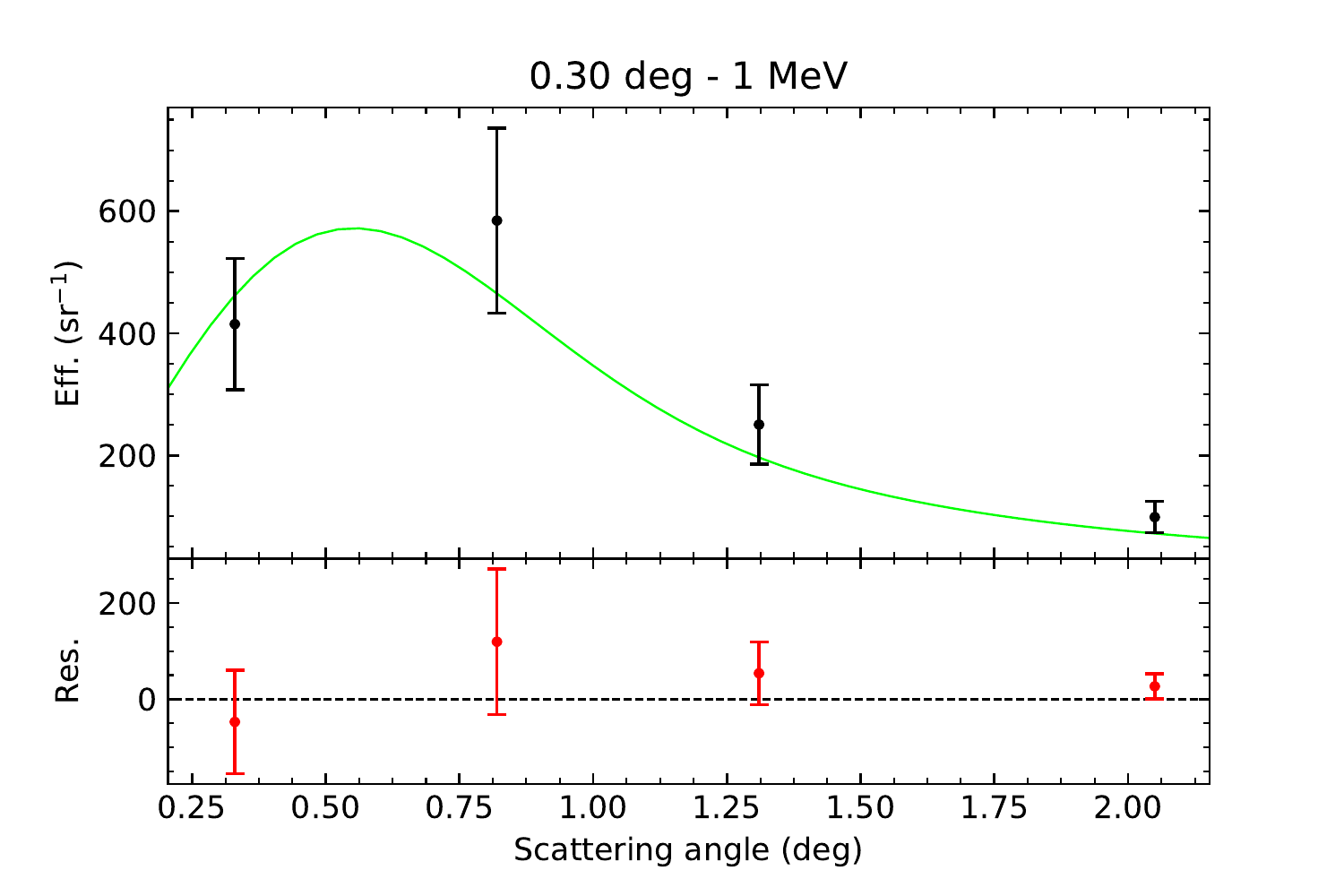}}
   {\includegraphics[width=.48\textwidth,height=.23\textheight]{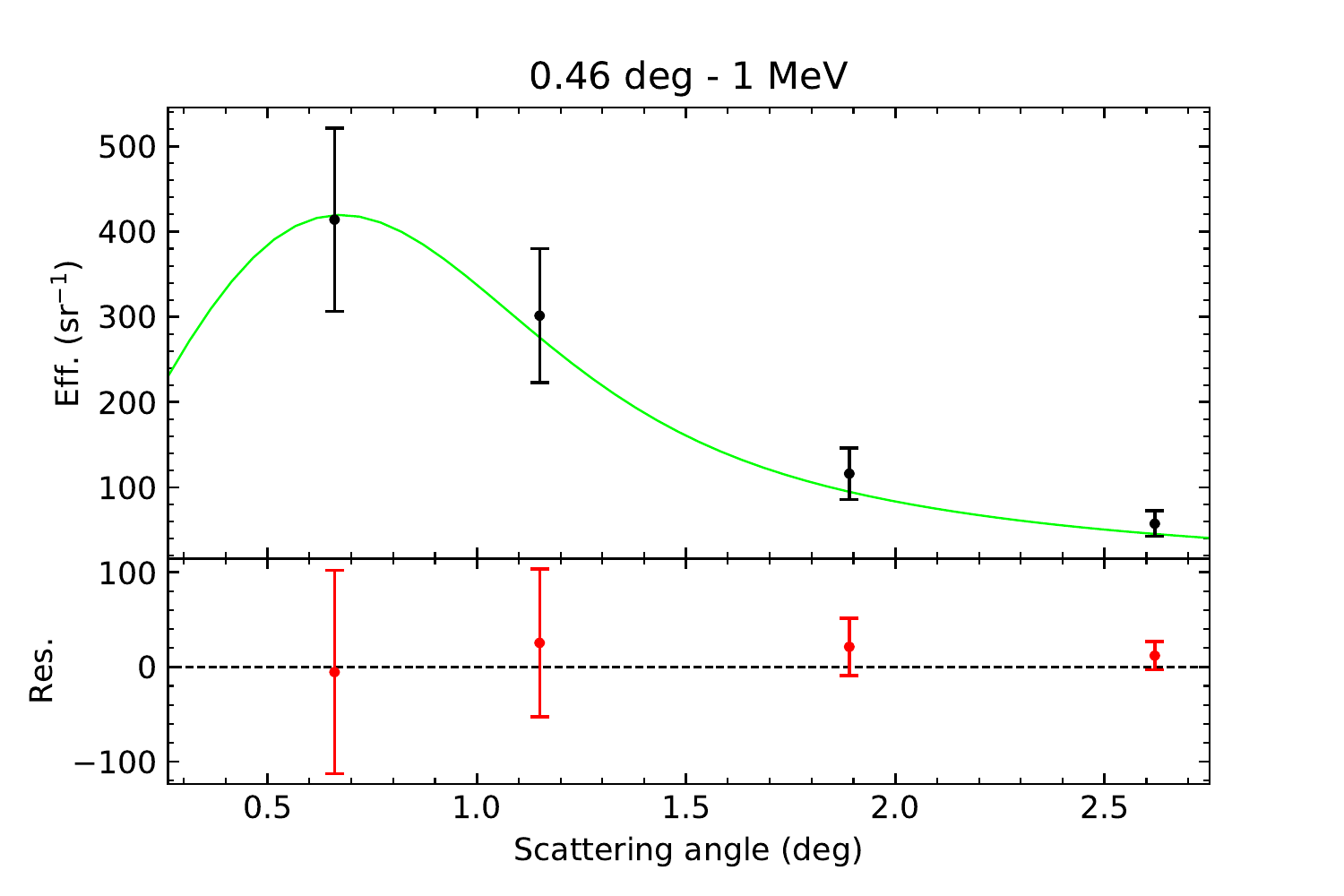}} 
   {\includegraphics[width=.48\textwidth,height=.23\textheight]{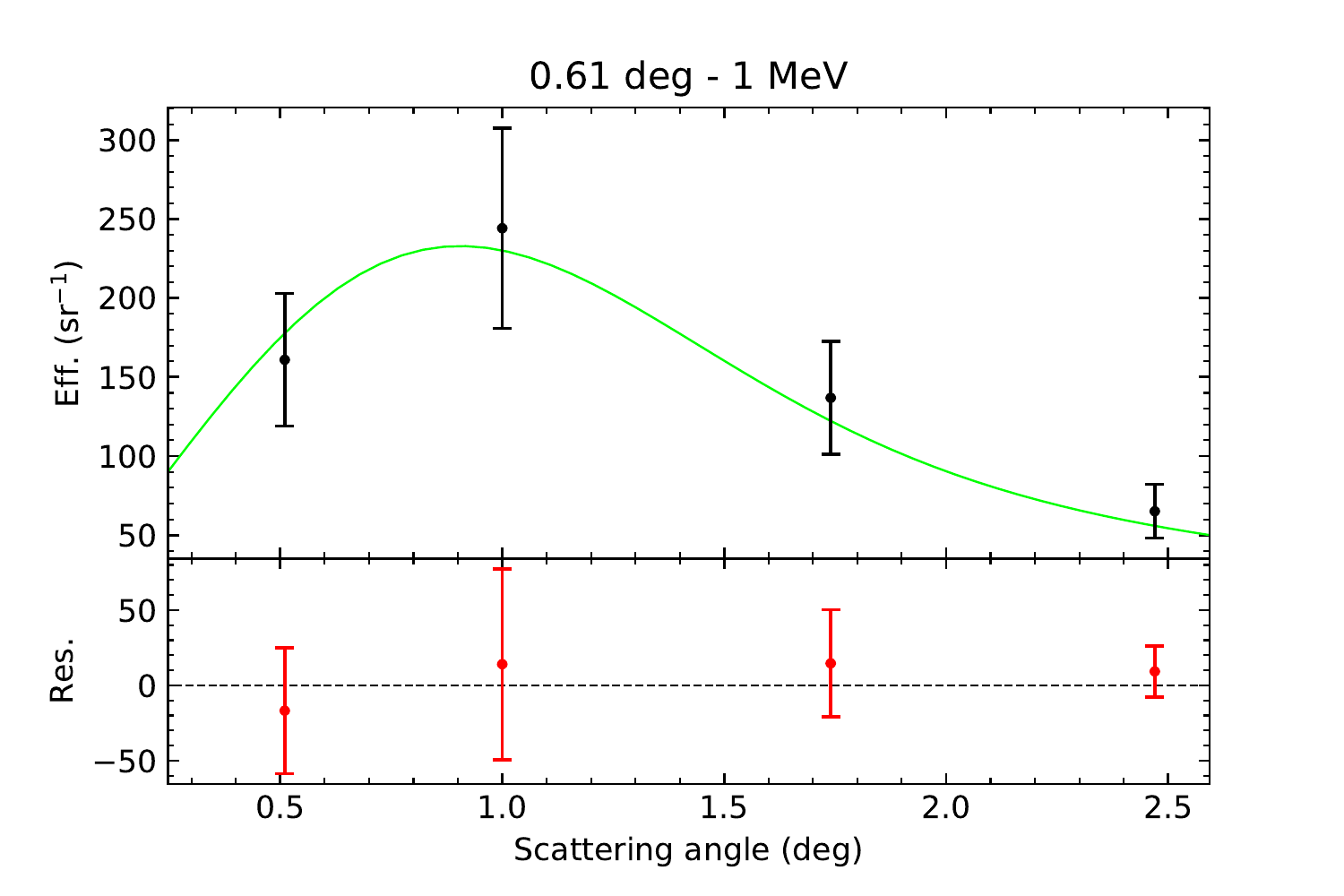}} 
   {\includegraphics[width=.48\textwidth,height=.23\textheight]{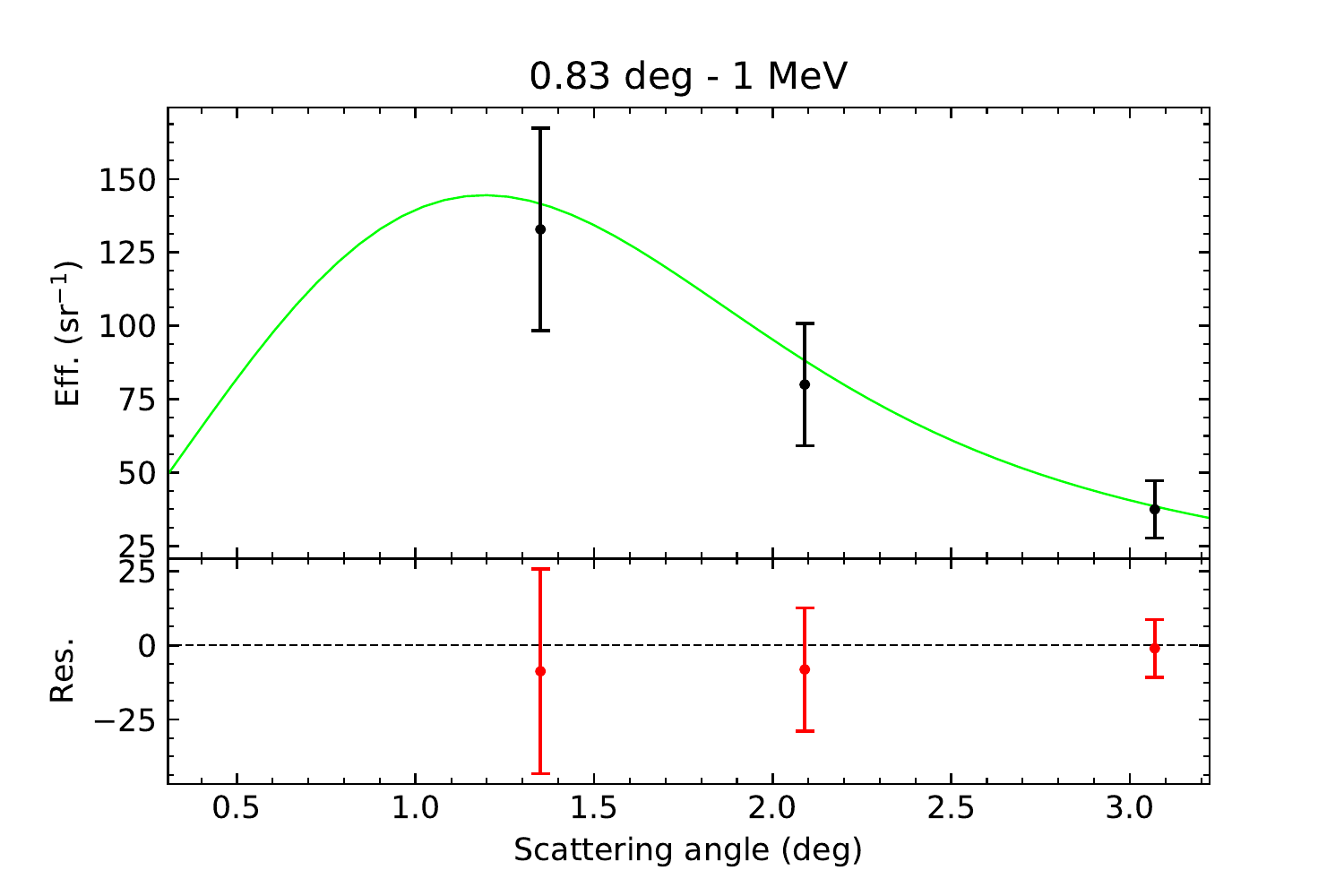}}
\caption{As before, for the incident energy of 1 MeV.}
\label{fig:fit_all_1000}
\end{figure}

\begin{figure}[ht]
\centering
   {\includegraphics[width=.48\textwidth,height=.23\textheight]{fit_results_ene_250KeV_0p36.pdf}} 
   {\includegraphics[width=.48\textwidth,height=.23\textheight]{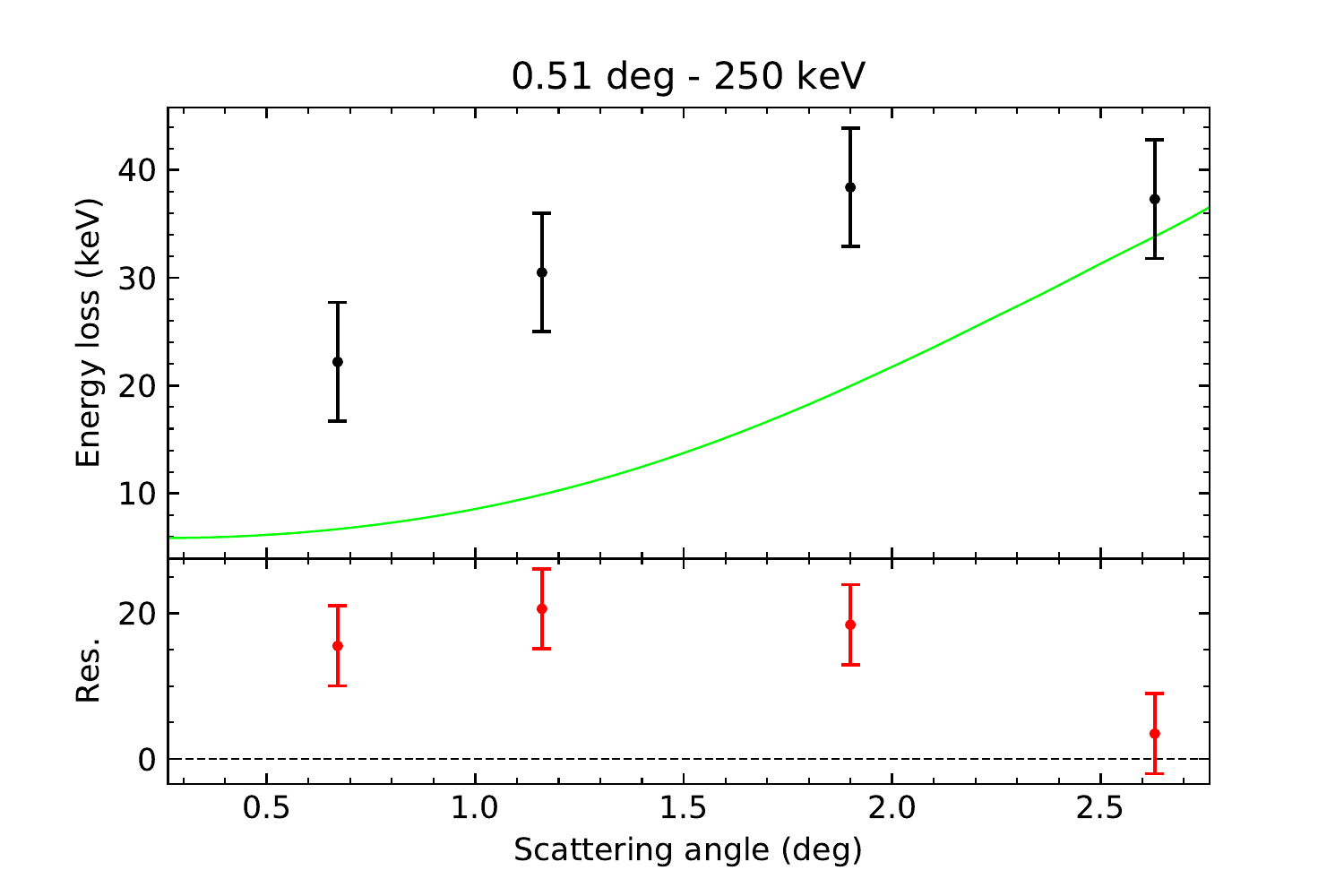}} 
   {\includegraphics[width=.48\textwidth,height=.23\textheight]{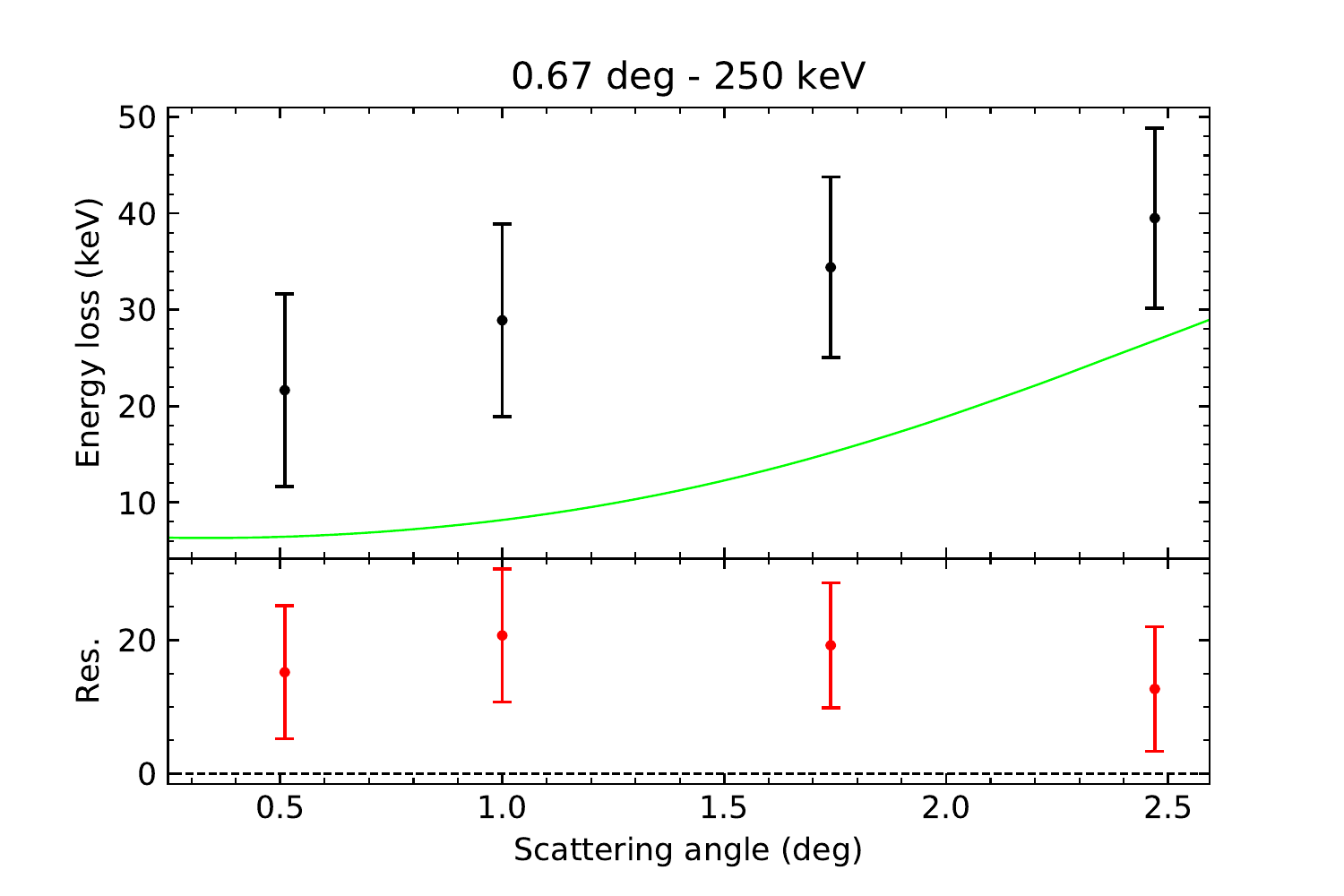}}
   {\includegraphics[width=.48\textwidth,height=.23\textheight]{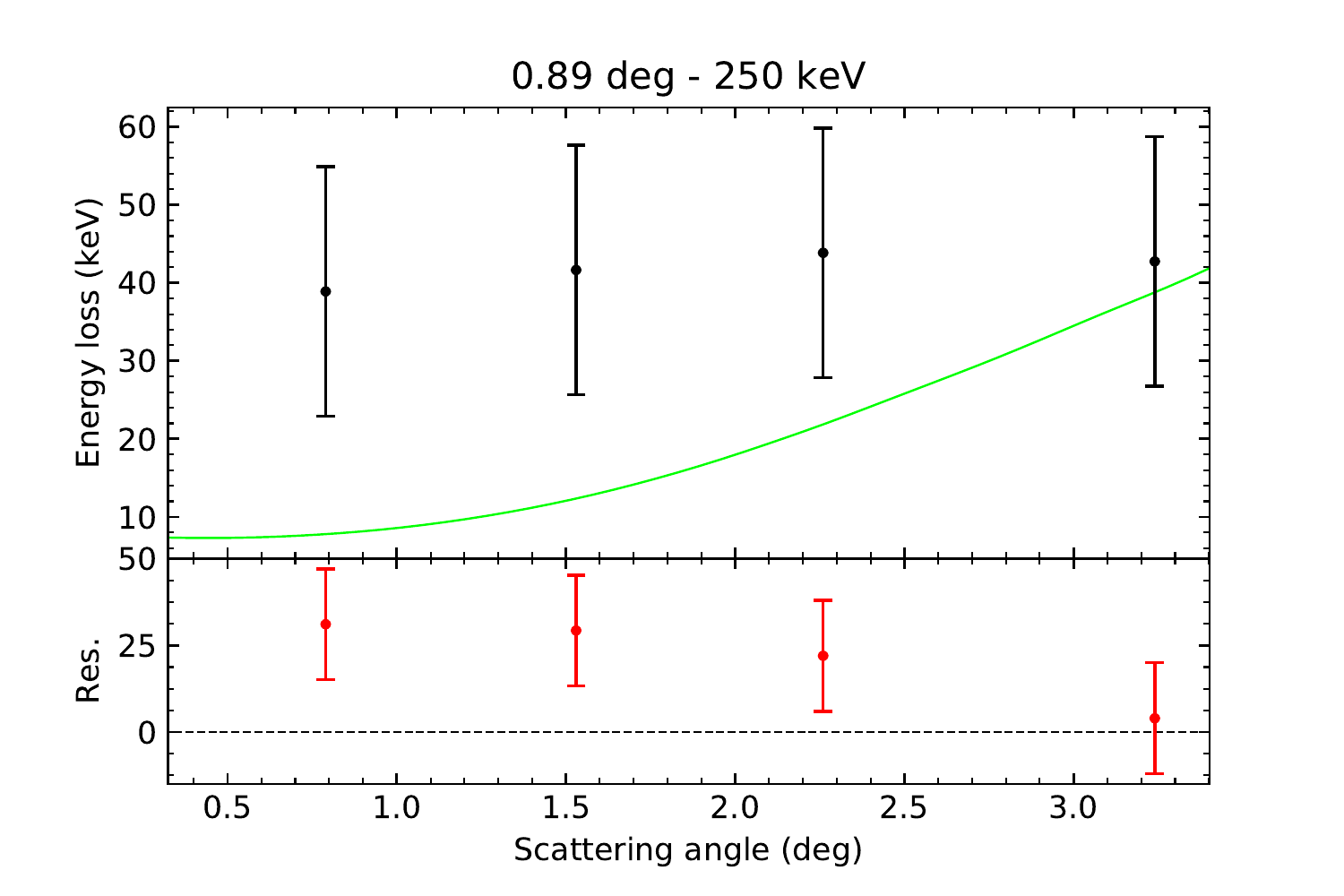}}
   {\includegraphics[width=.48\textwidth,height=.23\textheight]{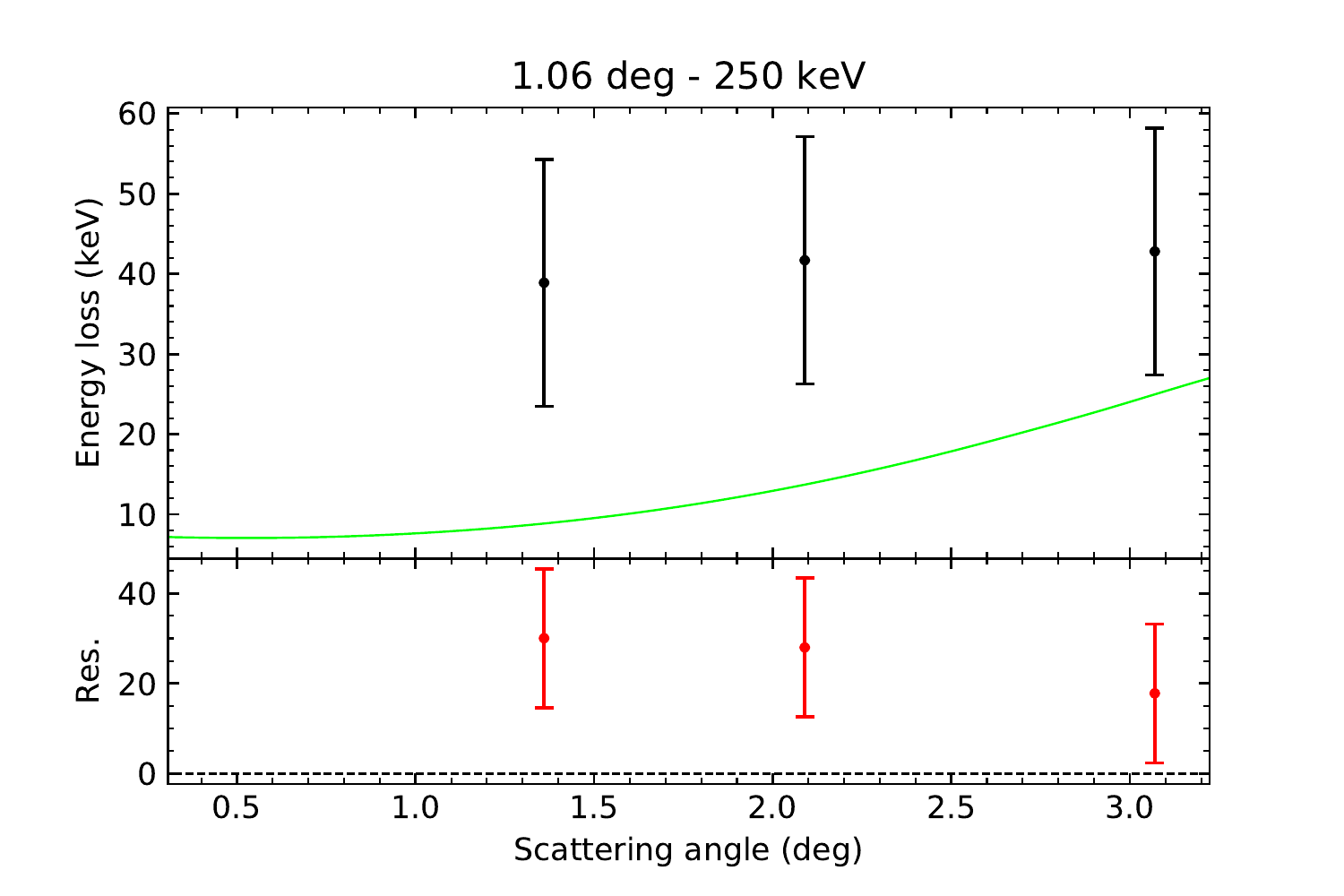}}
   {\includegraphics[width=.48\textwidth,height=.23\textheight]{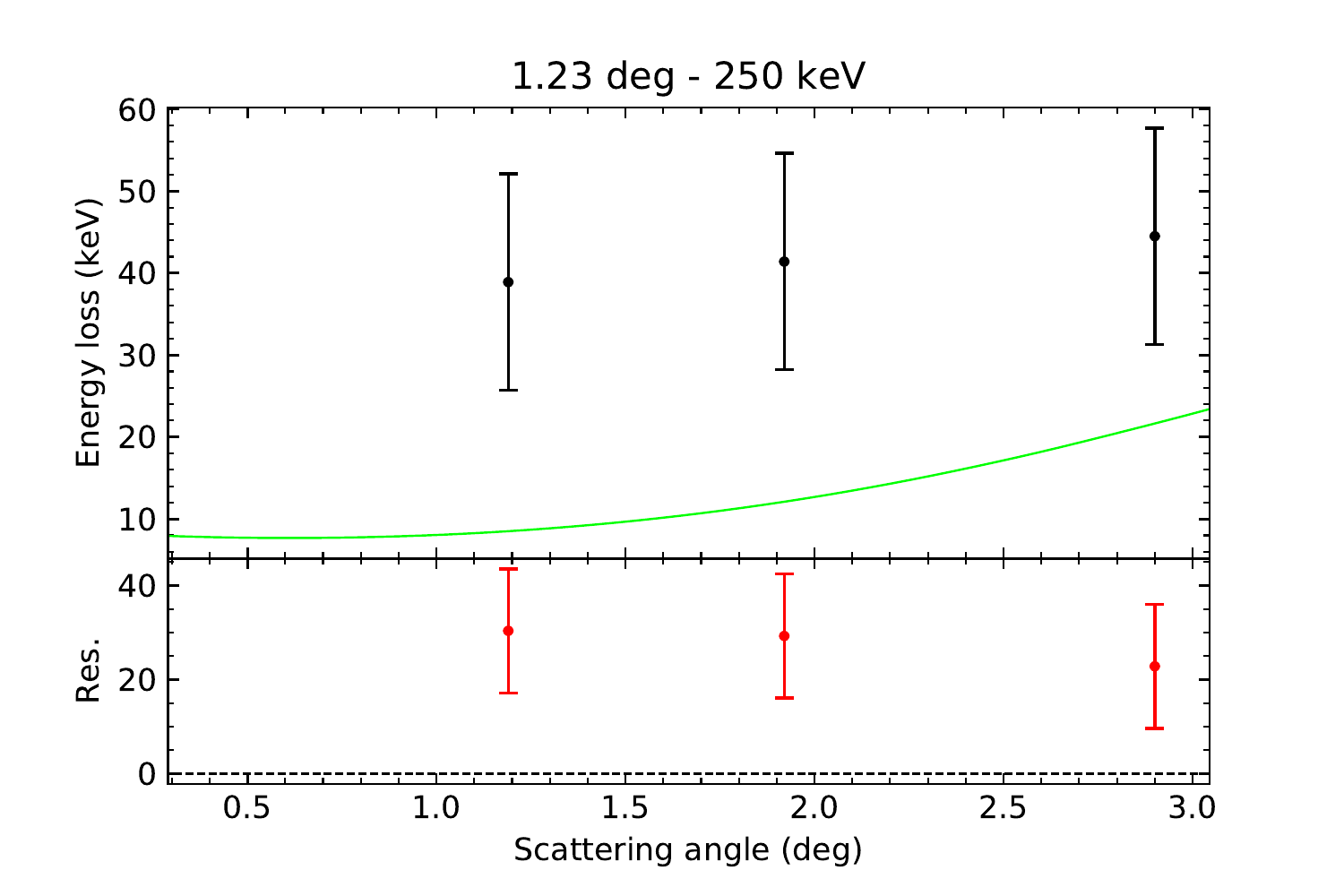}}
\caption{Fitting of the energy losses with the Remizovich formula (Eq.~\ref{eq:eff_integrale}) in non-elastic approximation. Incident energy of 250 keV.}
\label{fig:ene_all_250}
\end{figure}

\begin{figure}[ht]
\centering
   {\includegraphics[width=.48\textwidth,height=.23\textheight]{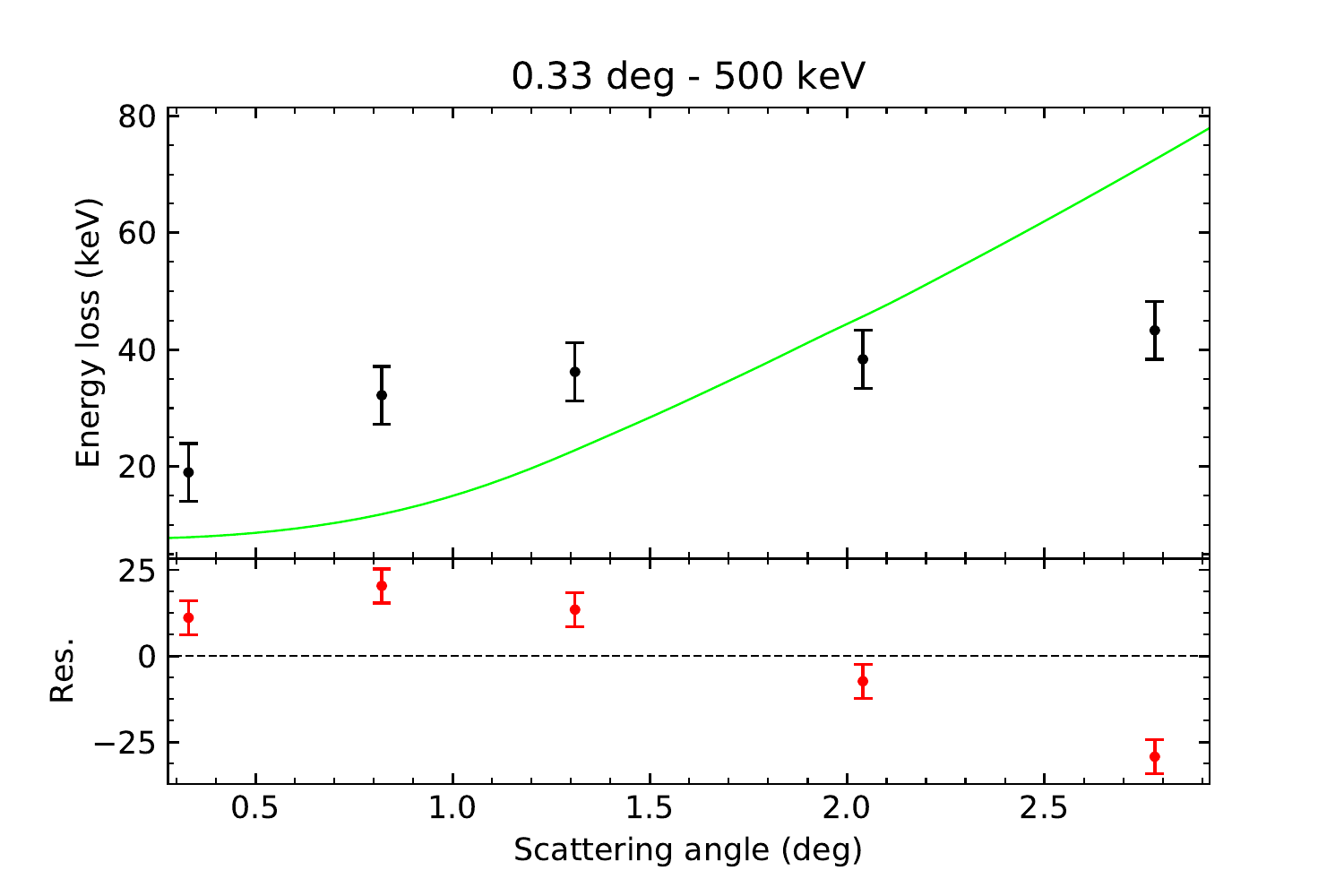}} 
   {\includegraphics[width=.48\textwidth,height=.23\textheight]{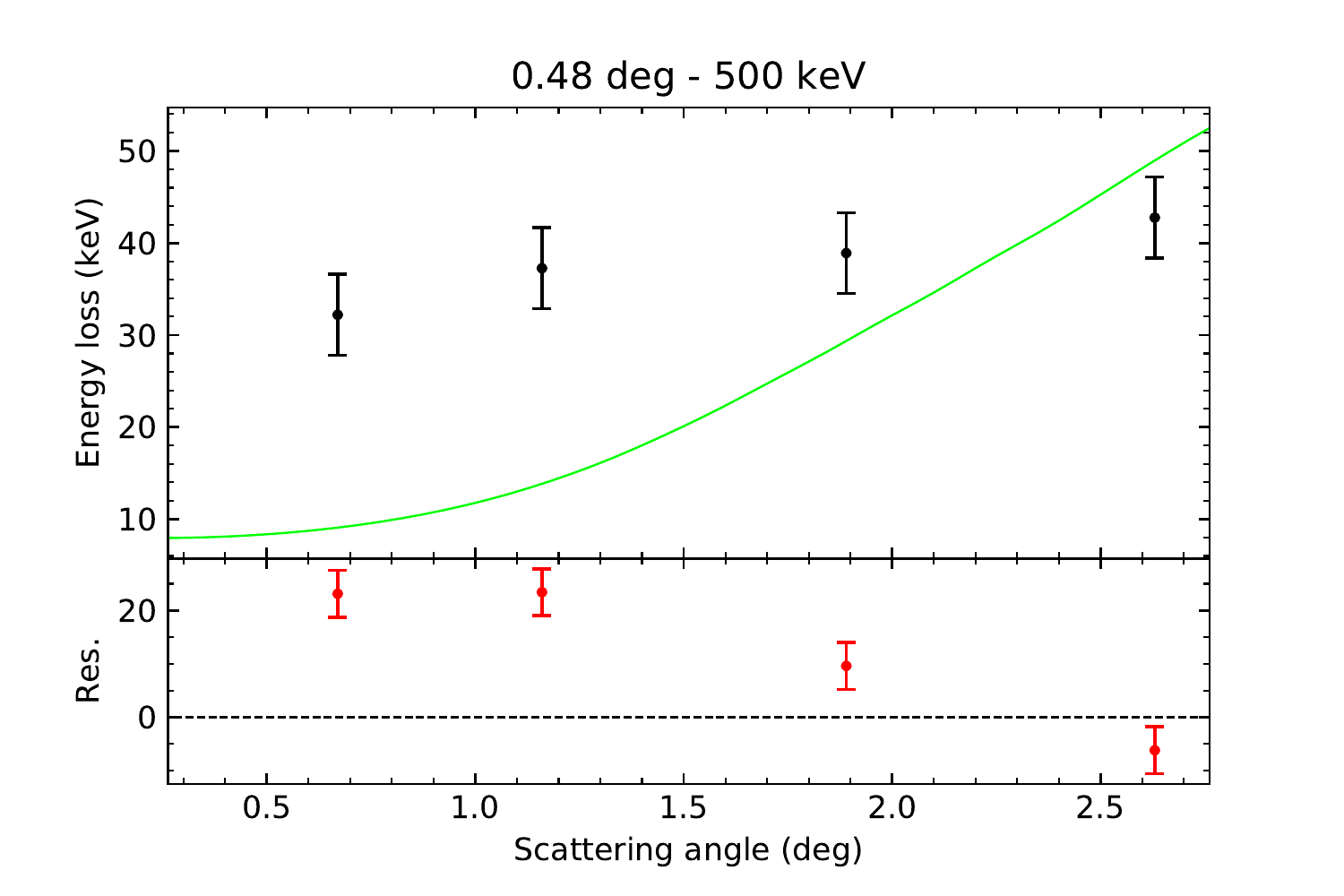}} 
   {\includegraphics[width=.48\textwidth,height=.23\textheight]{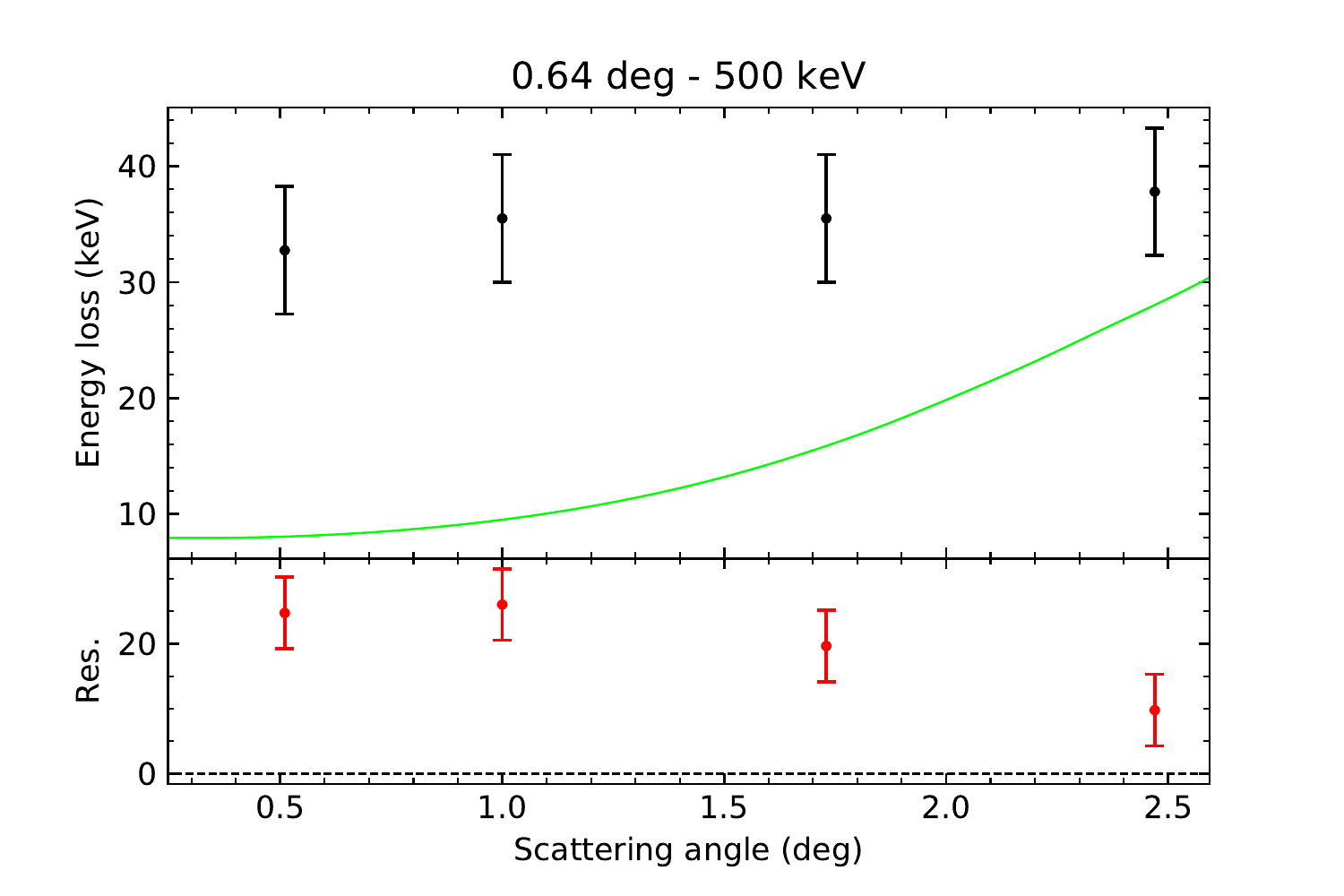}}
   {\includegraphics[width=.48\textwidth,height=.23\textheight]{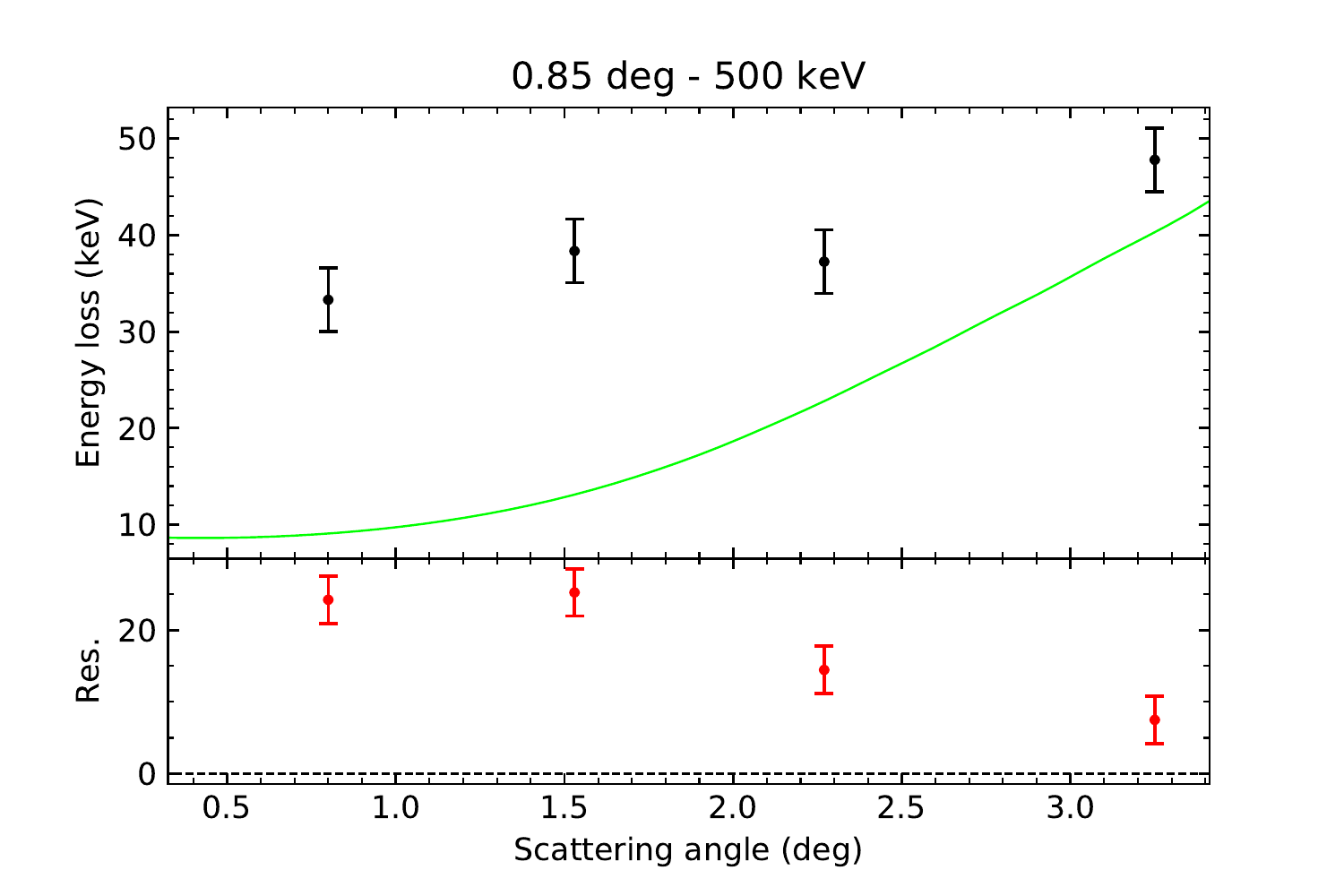}}
   {\includegraphics[width=.48\textwidth,height=.23\textheight]{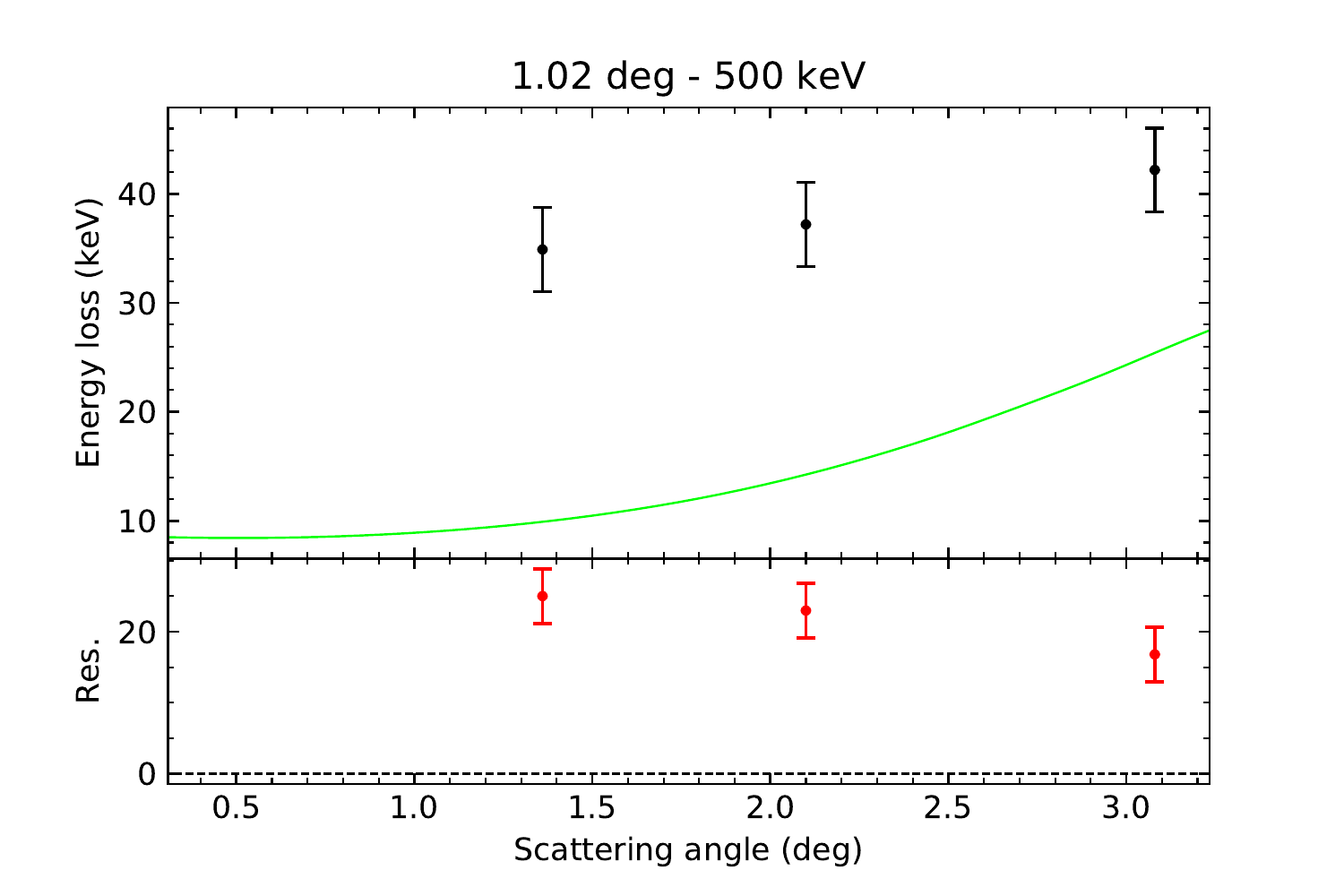}}
   {\includegraphics[width=.48\textwidth,height=.23\textheight]{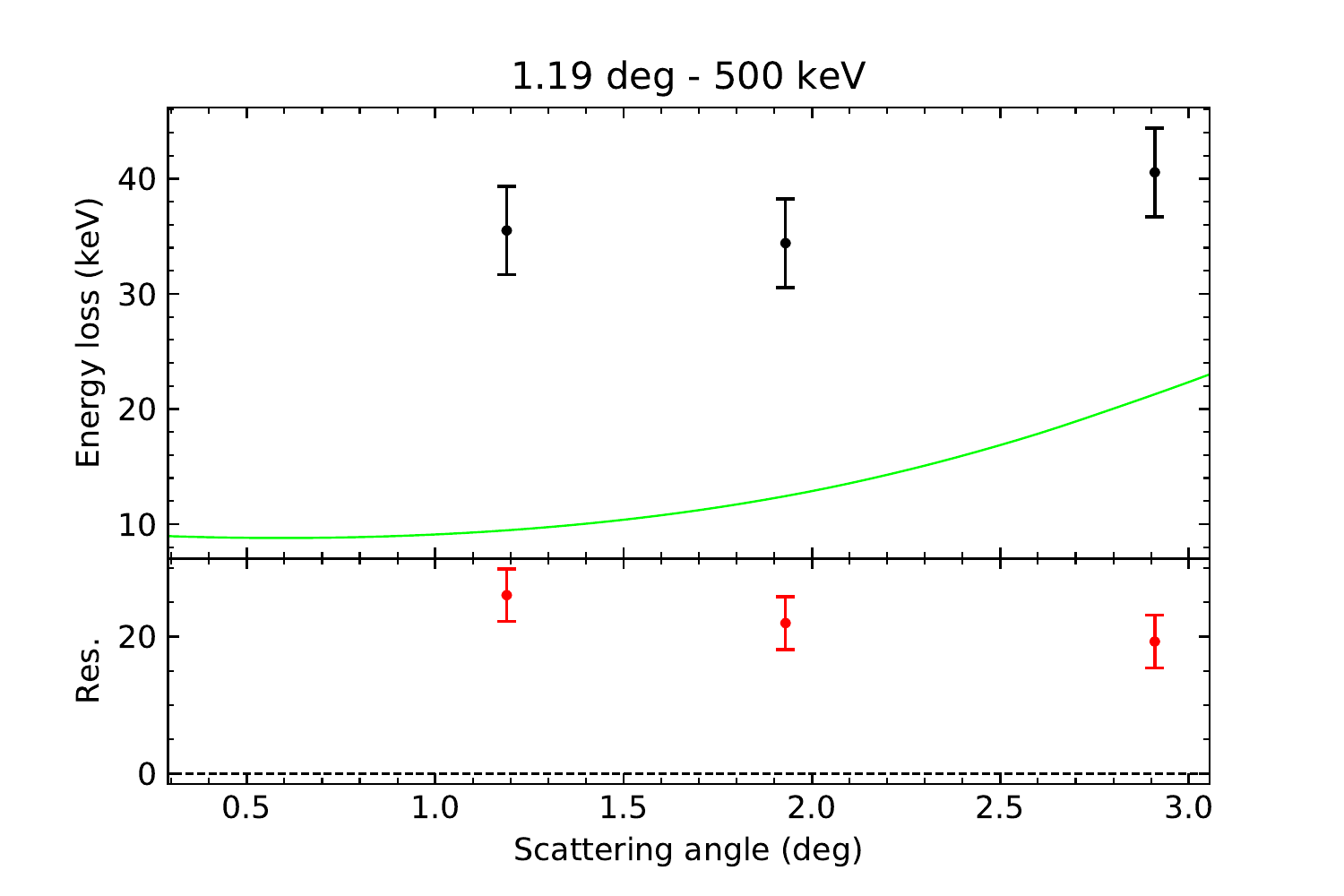}}
\caption{As before, for the incident energy of 500 keV.}
\label{fig:ene_all_500}
\end{figure}

\begin{figure}[ht]
\centering
   {\includegraphics[width=.48\textwidth,height=.23\textheight]{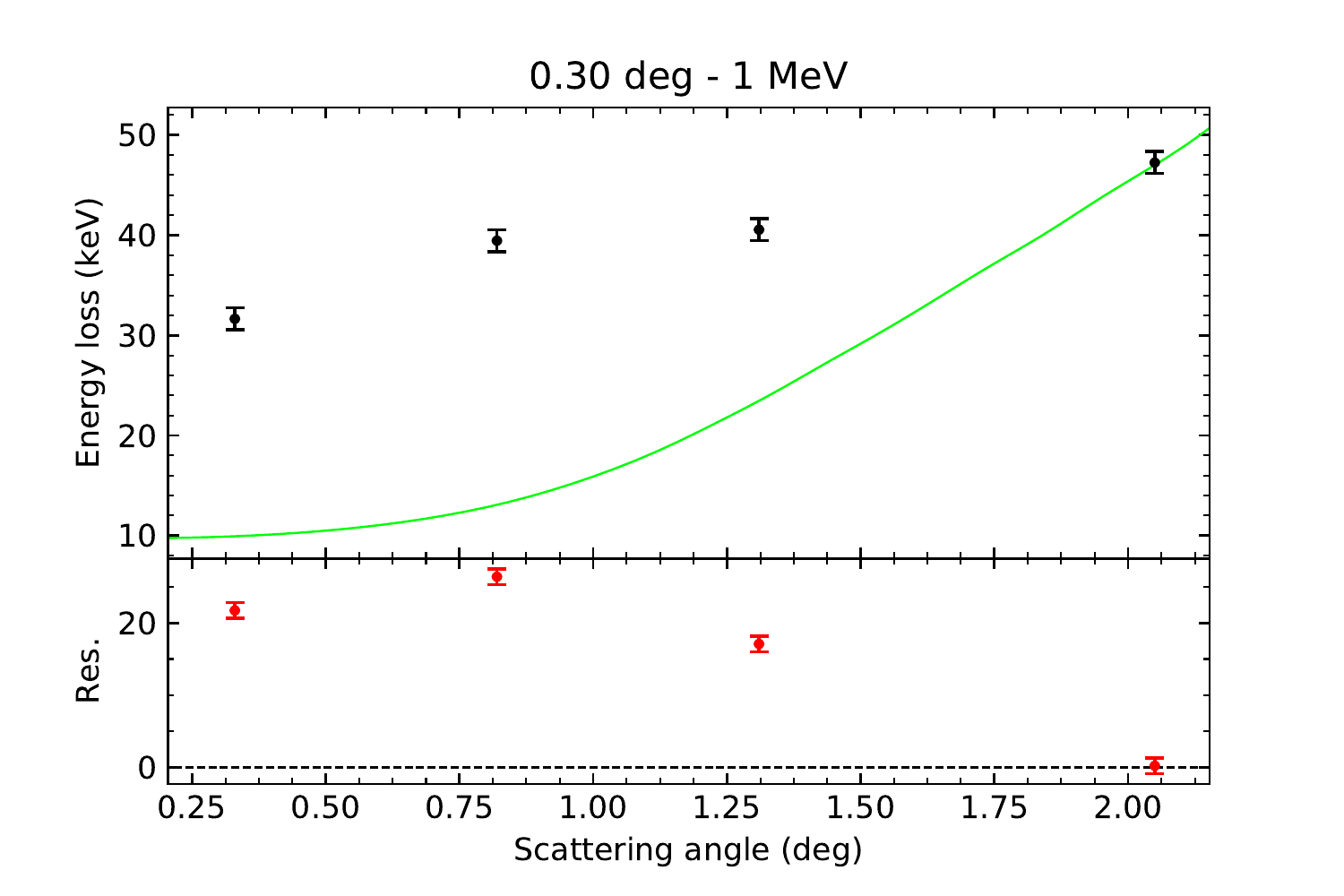}} 
   {\includegraphics[width=.48\textwidth,height=.23\textheight]{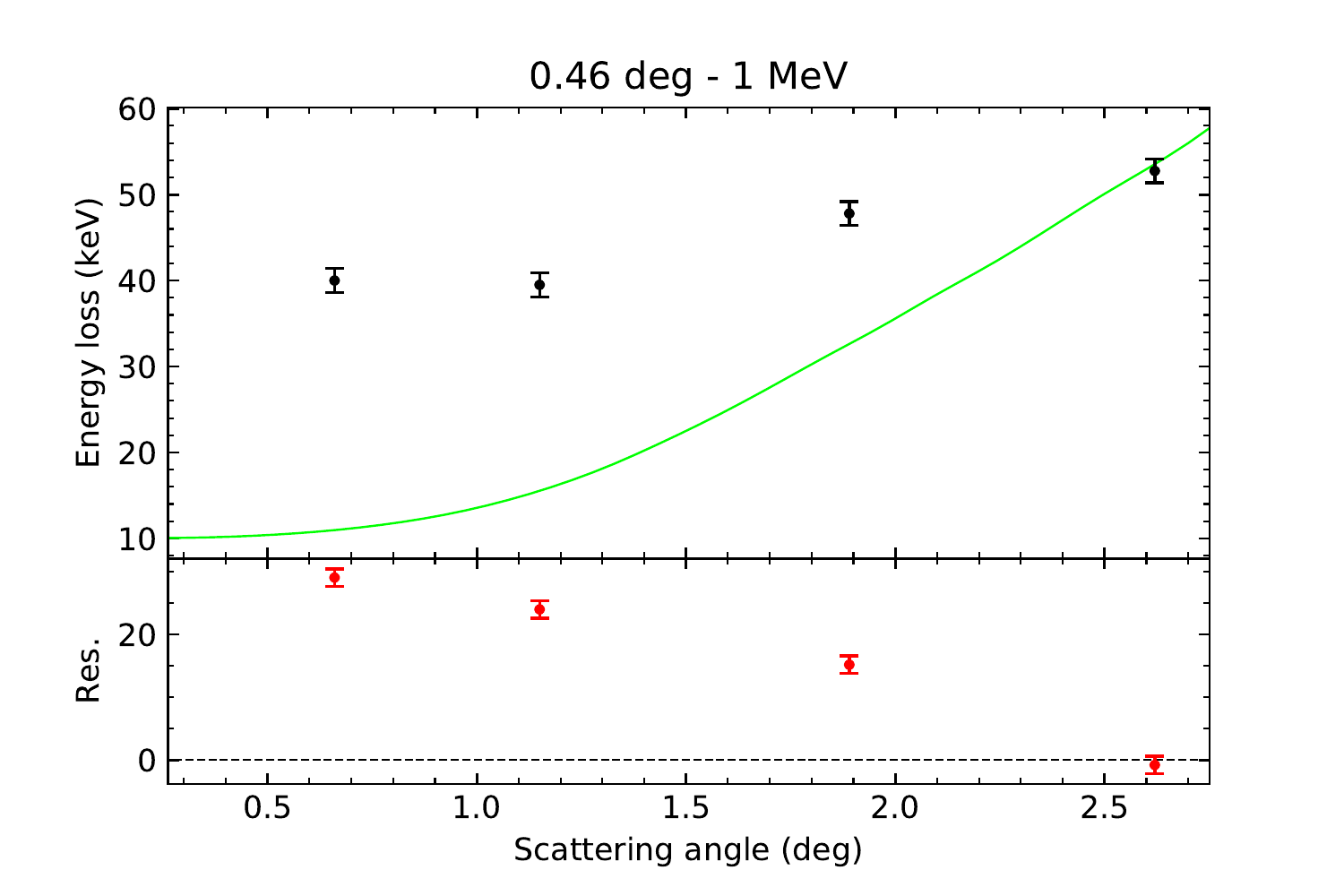}} 
   {\includegraphics[width=.48\textwidth,height=.23\textheight]{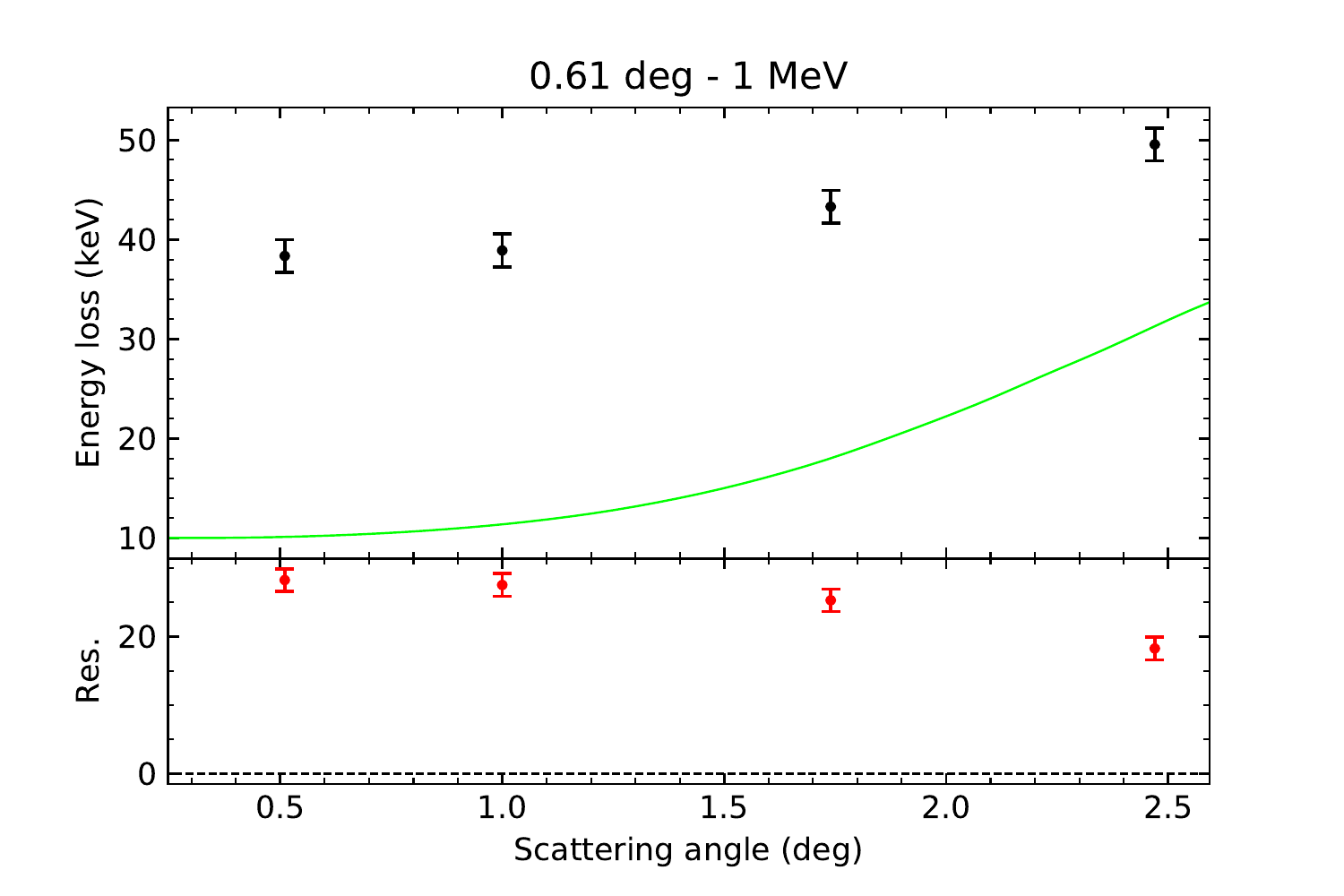}}
   {\includegraphics[width=.48\textwidth,height=.23\textheight]{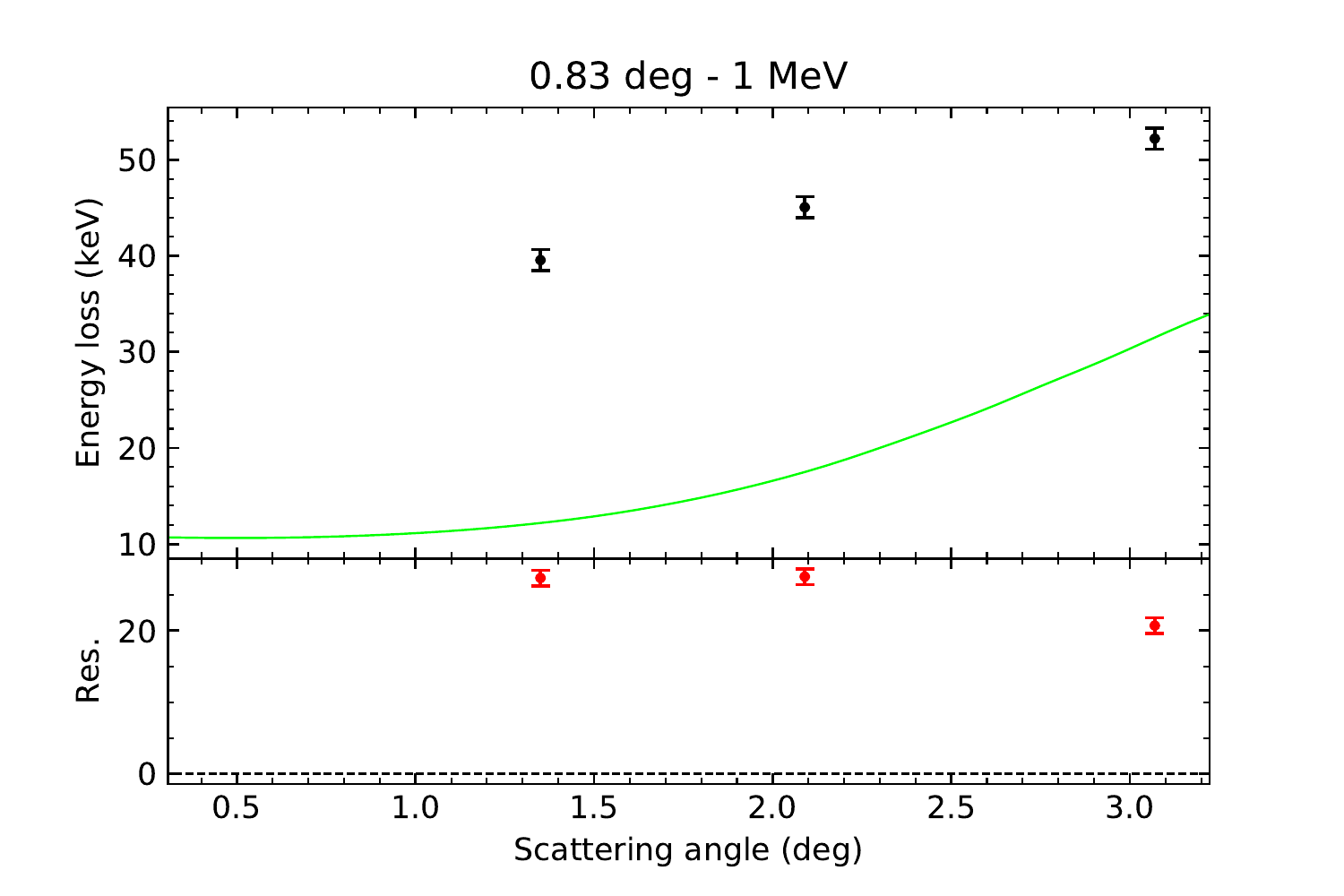}}
\caption{As before, for the incident energy of 1 MeV.}
\label{fig:ene_all_1000}
\end{figure}

\end{document}